\documentclass[11pt]{article}
\usepackage[font=small,labelfont=bf]{caption}
\usepackage{todonotes}
\usepackage{graphicx}
\usepackage[numbers,sort&compress]{natbib}
\usepackage{graphicx,epsfig}
\usepackage{amsmath}
\usepackage{graphicx,overpic}%
\usepackage{amsfonts,amsmath}%
\usepackage{subcaption}
\usepackage{color}
\usepackage{amsmath,bm}
\usepackage{amsthm}
\usepackage{booktabs}
\usepackage{threeparttable}
\usepackage{multirow}
\usepackage{url}
\usepackage{amssymb}
\usepackage{authblk}

\usepackage{lineno}
\usepackage [autostyle, english = american]{csquotes}
\MakeOuterQuote{"}
\usetikzlibrary{shapes.geometric, arrows}

\let\oldtabular\tabular
\renewcommand{\tabular}{\small\oldtabular}
\usepackage{tikz}
\usetikzlibrary{shapes.geometric,arrows}
\tikzstyle{startstop}=[rectangle,rounded corners,minimum width=3cm,minimum height=1cm,text centered, draw=black,fill=red!30]
\tikzstyle{io}=[trapezium, trapezium left angle=70, trapezium right angle=110, minimum width=3cm,minimum height=1cm, text centered, text width=3.5cm, draw=black, fill=yellow!30]
\tikzstyle{process}=[rectangle, minimum width=3cm,minimum height=1cm,text centered, draw=black,text width=3.5cm,fill=blue!30]
\tikzstyle{comments}=[rectangle, minimum width=3cm,minimum height=1cm,text centered, draw=black,text width=3.5cm,fill=green!30]
\tikzstyle{decision}=[diamond,align=center, minimum width=3cm,minimum height=1cm,text centered, draw=black,fill=green!30]
\tikzstyle{arrow}=[thick,->,>=stealth]
\tikzstyle{dashedarrow}=[dashed,->,>=stealth]

\usepackage[title]{appendix}

\def\beq{\begin{equation}}
\def\eeq{\end{equation}}
\def\bea{\begin{eqnarray}}
\def\eea{\end{eqnarray}}

\def\bu{\mathbf{u}}

\def\bu{\mathbf{u}}

\newcommand\etal{\mbox{\textit{et al.}}}
\def\vuko{Vuk{\v{c}}evi{\'c}}
\usepackage{listings}

\providecommand{\keywords}[1]
{
  \small
  \textbf{\textit{Keywords:}} #1
}

\title
{
\bf Spectral Wave Explicit Navier-Stokes Equations for wave-structure interactions using two-phase Computational Fluid Dynamics solvers}

\author[1]{\small Zhaobin LI} 
\author[1]{Benjamin Bouscasse\thanks{Corresponding author: benjamin.bouscasse@ec-nantes.fr}}
\author[1]{Guillaume Ducrozet}
\author[1]{Lionel Gentaz}
\author[1]{David Le Touz\'e}
\author[1]{Pierre Ferrant}
\affil[1]{\small \'Ecole Centrale Nantes, LHEEA Lab.\ (ECN and CNRS), Nantes, France}
\date{}%
\begin{document}
\maketitle
\vspace{-3\baselineskip}
\begin{abstract}
This paper proposes an efficient potential and viscous flow decomposition method for wave-structure interaction simulation with {\color{black} single-phase potential flow wave models} and two-phase Computational Fluid Dynamics (CFD) solvers. {\color{black} The potential part - represents the incident waves - is solved with spectral wave models; the viscous part - represents the complementary perturbation on the incident waves - is solved with the CFD solver. This combination keeps the efficiency and accuracy of potential theory on water waves and the advantage of two-phase CFD solver on complex flows (wave breaking, flow separation, etc.).} The decomposition strategy is called Spectral Wave Explicit Navier-Stokes Equations (SWENSE) \cite{ferrant2003potential}, originally proposed for single-phase CFD solvers. Firstly, this paper presents an extension of the SWENSE method for two-phase CFD solvers.  Secondly, an accurate and efficient method to interpolate potential flow results obtained by the High Order Spectral (HOS) wave model on CFD mesh is proposed. The method is able to reduce the divergence error of the interpolated velocity field to meet the CFD solver's needs without reprojection.  Implemented within OpenFOAM\textsuperscript{\textregistered}, these  methods are tested by three convincing verification, validation and application cases, considering incident wave propagation, high-order loads on a vertical cylinder in regular waves, and a Catenary Anchor Leg Mooring (CALM) buoy in both regular and irregular waves. Speed-ups between 1.71 and 4.28 are achieved with the test cases. {\color{black} The wave models and the interpolation method are released open-source to the public.} 
\end{abstract}



\section{Introduction}
The accurate prediction of wave-structure interactions is of vital importance in ship hydrodynamics and ocean engineering.  For ocean-going vessels, the wave-induced loads are essential for analyzing ships' seakeeping property, the resistance in waves, and the structural integrity in extreme sea states.  In ocean engineering,  the calculation of wave-structure interaction helps to optimally design offshore structures, for example, oil and gas facilities or marine renewable energy devices.

Traditionally, water waves and wave-structure interactions are addressed with single-phase potential theory (PT) with the perfect fluid and irrotational flow assumption. \textcolor{black}{ The majority of wave models - from the simplest 1st order Stokes wave theory \cite{stokes} to more complex fully nonlinear theories   \cite{rienecker1981fourier,DUCROZET2016OpenSourceOcean} - and a large number of numerical solvers for wave-structure interactions - using the Boundary Element Method (BEM) \cite{yeung1974,newman2002boundary,babarit2015NEMOH}, the Finite Element Method (FEM) \cite{ma2006FNPTFEM}, the Finite Difference Method (FDM) \cite{bingham2007FDM}, the Finite Volume Method (FVM) \cite{mehmood2015fvmPotential}, the Harmonic Polynomial Cells (HPC) \cite{ShaoHPC}, etc - are developed based on PT. Thanks to the irrotational flow assumption, the velocity vector of flow (3 variables in 3D cases) can be represented by a scalar (the velocity potential), hence the computational complexity is much reduced. PT solvers are often considered to be more efficient than Navier-Stokes (NS) equation based solvers. However, due to the irrotational assumption, PT fails in complex scenarios, such as flow separation around structures and wave breaking.}

\textcolor{black}{
In contrast, Computational Fluid Dynamics (CFD) solvers, based on NS equations, have a more sophisticated mathematical model and accept rotational flow.  Moreover, CFD solvers offer more possibilities to deal with the water-air interface: the flow is either treated as a single-phase problem containing only water with a moving free surface boundary \cite{CARRICA2006fdm,DIMASCIO2007levelSet,reliquet2013lvlset} or modeled as a two-phase flow containing both water and air with an interface \cite{iafrati2001level,isisCFDDES,JZang2014OF}. The latter makes CFD solvers advantageous for violent free surface deformations, such as wave breaking, wave impact, etc.
Numerous validations have proved that viscous CFD solvers are able to provide high-fidelity results for a wide variety of marine and offshore applications \cite{stern2013computational}. However, CFD solvers demand  higher computational cost than PT solvers. A recent blind test \cite{ransley2019blind} reveals a comparison between several CFD and PT solvers on a wave-structure interaction with an intermediate wave steepness (without wave breaking) and concludes that even the quickest CFD code is 1.5 orders of magnitude slower than a FEM-PT solver in that non-violent wave case. But the authors of \cite{ransley2019blind} also admit that CFD solvers may win the comparison when the non-breaking and irrotational conditions of PT are violated, i.e., when the wave becomes larger.}

\textcolor{black}{
To summarize, {\color{black}each method has} advantages and limits. If they can be coupled appropriately, one can expect the resulting method to benefit from {\color{black}both sides}. In the literature, this idea has been explored by many researchers. Among others, two main coupling strategies are commonly used: domain decomposition (DD) and functional decomposition (FD).}

DD splits the computational domain into a potential region and a viscous region, and in each region, uses the appropriate solver with the best efficiency and accuracy. From a physical point of view, the complex interaction, e.g., viscous effects and violent free-surface deformation, appears near the structure only; in the far-field, the viscous effects can often be neglected, allowing the use of potential theory. For this reason, the computational domain can be split into a viscous inner sub-domain plus an irrotational outer sub-domain. Information is exchanged on the common boundary, either in a two-way interactive fashion or a one-way forcing manner. {\color{black}
For two-way coupling: the most common practice is to couple a PT solver and a Reynolds Averaged Navier-Stokes Equation (RANSE) solver. If the structure does not induce wave breaking, both PT and RANSE solvers can be single-phase solvers as in \cite{campana1995viscous,chen1999rans,zhang2013coupling}.  When the waves near structure are more violent, two-phase RANSE solvers are more suitable, but in the far-field single-phase PT solver can still be used, as shown in \cite{SiddiquiHPC,lu2017overlapping}. In extreme violent free surface deformation case (such as dam breaking), both PT and RANSE solvers are two-phase, as in \cite{colicchio2006bemlevelSet}. }The two-way coupling reduces the size of the viscous computational domain, but it requires iterations between the PT and the CFD solver, which in return needs extra efforts.  The alternative one-way manner sends information from the PT to the viscous CFD solver only.  The PT solver considers the wave propagation in the far-field until the inner zone. The inner CFD zone then uses this information as wave making boundary conditions. This method has been applied for shoaling and breaking wave problem in near-shore areas \cite{guignard1999computation,lachaume2003modeling,grilli2004numerical}, calculation of wave force on offshore structures \cite{christensen2009transfer,paulsen2014efficient,JZang2014OF}. Several authors \cite{jacobsenFuhrmanFredsoe2012,ihFOAM} proposed a general one-way coupling method to combine potential wave theories and two-phase viscous solvers in OpenFOAM\textsuperscript{\textregistered}, which has been widely used by the offshore and coastal engineering community. However, the one-way coupling needs a larger computational domain to avoid wave reflection \cite{li2019comparison}.


The second category, FD, splits the
total flow problem into (i) an irrotational part to be solved with PT solvers and (ii) a complementary part to be solved by CFD codes \cite{dommermuth1993laminar}.
Since such a decomposition is not unique, multiple choices exist in the literature, which can be classified into two categories, according to the complexity of the irrotational part. \textcolor{black}{One can first use PT solvers to obtain an irrotational solution of the wave-structure interaction and then correct the solution with a rotational part calculated with a CFD solver solving a complementary equation, as proposed in   \cite{kim2005complementary,edmund2013velocity,rosemurgy2016velocity}. However, a more simple decomposition is to use the PT solver for the incident waves only and solve all the complementary phenomena by the CFD solver, as proposed in  \cite{ferrant2003potential}.} The FD methods usually allow the CFD solver to use coarse mesh in the far-field to reduce the computational cost, since the interesting zone of the complementary phenomena locates often near the structure only. 

The Spectral Wave Explicit Navier-Stokes Equations (SWENSE) method presented in this paper belongs to the FD category and it decomposes the total solution into an incident wave part and a complementary part.  The essential benefits of the method are: (i) the fast and accurate incident wave simulation and (ii) that it allows the use of coarse mesh in the far-field region while keeping the near-field accuracy. A refined mesh is required only near the structure where the complementary field needs to be solved accurately.
The original SWENSE method is proposed only for single-phase CFD solvers.  It has been successfully applied to calculate wave force on vertical cylinder \cite{licalculation}, on a Catenary Anchor Leg Mooring Buoy \cite{monroy2010phd}, and ship resistance in waves \cite{reliquet2013lvlset}. \textcolor{black}{In the single phase scenario,} the use of coarser mesh reduced the CPU time of the SWENSE method by \textcolor{black}{one order of magnitude} compared with conventional NS solvers for an equivalent accuracy \cite{luquet2007simulation}. 

\textcolor{black}{
Recently, the SWENSE method has been extended for two-phase CFD solvers  \cite{vukvcevic2016decomposition}. However, reference  \cite{vukvcevic2016decomposition} adopts a FD strategy different from the original SWENSE method, which leaves the pressure field undecomposed. Consequently, the use of coarse mesh in the far-field may introduce errors to the incident pressure field and also to the incident waves, losing the advantage of the SWENSE method. This point will be discussed in detail in Sect. \ref{sect:compareWtihVuko}.}

\textcolor{black}{
In this paper, we present a novel extension of the SWENSE method for two-phase viscous CFD solvers, which keeps all advantages of the original method, i.e. the ability to use a coarse mesh in the far-field to enhance the efficiency. Moreover, coupling two-phase CFD solvers with SWENSE enables the method to deal with violent free surface deformations, such as breaking waves near the structure.}
Secondly, the paper also proposes an accurate and convenient reconstruction method to interpolate the results of High-Order Spectral (HOS) wave models onto CFD mesh. {\color{black}The method can reduce the divergence error of the interpolated velocity field to meet the CFD solver's need without reprojection.} Implemented in OpenFOAM\textsuperscript{\textregistered}, the validity and efficiency of the proposed methods are assessed thanks to three convincing cases, including the simulation of progressive regular waves, the calculation of high order loads on a vertical cylinder in regular waves, and a Catenary Anchored Leg Mooring (CALM) buoy in regular and irregular waves. 


The rest of the paper is organized as follows. Section \ref{sect:equations} derives the two-phase SWENS governing equations {\color{black} and compares them with those of reference  \cite{vukvcevic2016decomposition}.}   
Section \ref{Section:IncidentWave} presents the fully non-linear spectral wave models for incident wave modeling 
and proposes the reconstruction technique to interpolate the results of HOS wave models to CFD mesh. Section \ref{sect:OpenFOAM} details the numerical discretization and an implementation example of a two-phase SWENSE solver in OpenFOAM\textsuperscript{\textregistered} \cite{openFoam}. The validation and application cases are shown in Sect.\ \ref{sect:validationAndApplication}.

\section{Governing equations of the two-phase SWENSE method\label{sect:equations}}

\subsection{SWENSE decomposition}



The SWENSE method \cite{ferrant2003potential,ferrant2008fully} defines three notions in a wave-structure interaction:

\begin{enumerate}
\item Total field: the total field represents the real flow including the incident waves, the scattered waves and viscous effects caused by the wave-structure interaction. It is assumed that the flow is governed by the incompressible (NS) equations.

\item Incident field: the incident part concerns the propagation of the incident waves in the computational domain without structures. The viscosity is neglected. The flow is described by the Euler equations. It is further assumed that the flow is irrotational so that incident waves can be solved by PT wave models.

\item Complementary field: the complementary part represents the difference between the total field and the incident field. This field is generated due to the presence of structures in the computational domain and the viscosity of the fluid. The complementary variables are governed by the SWENS equations.
\end{enumerate}

With these notions, a primitive field of the flow $\chi$ (pressure, velocity) is decomposed into an incident part $\chi_{I}$ and a complementary part $\chi_{C}$ (see Eqn.\ \ref{eq:changeofVariable}). $\chi_{I}$ is explicitly given by PT wave models; $\chi_{C}$ is to be calculated by the CFD solver implementing the SWENS equations. In the following contents, variables without subscript denote the total field, subscript $I$ denotes the incident part, and subscript $C$ denotes the complementary part.

\begin{equation}
\chi=\chi_{I}+\chi_{C}\label{eq:changeofVariable}
\end{equation}

\begin{figure}[ht]
\centering
\includegraphics[width=10cm]{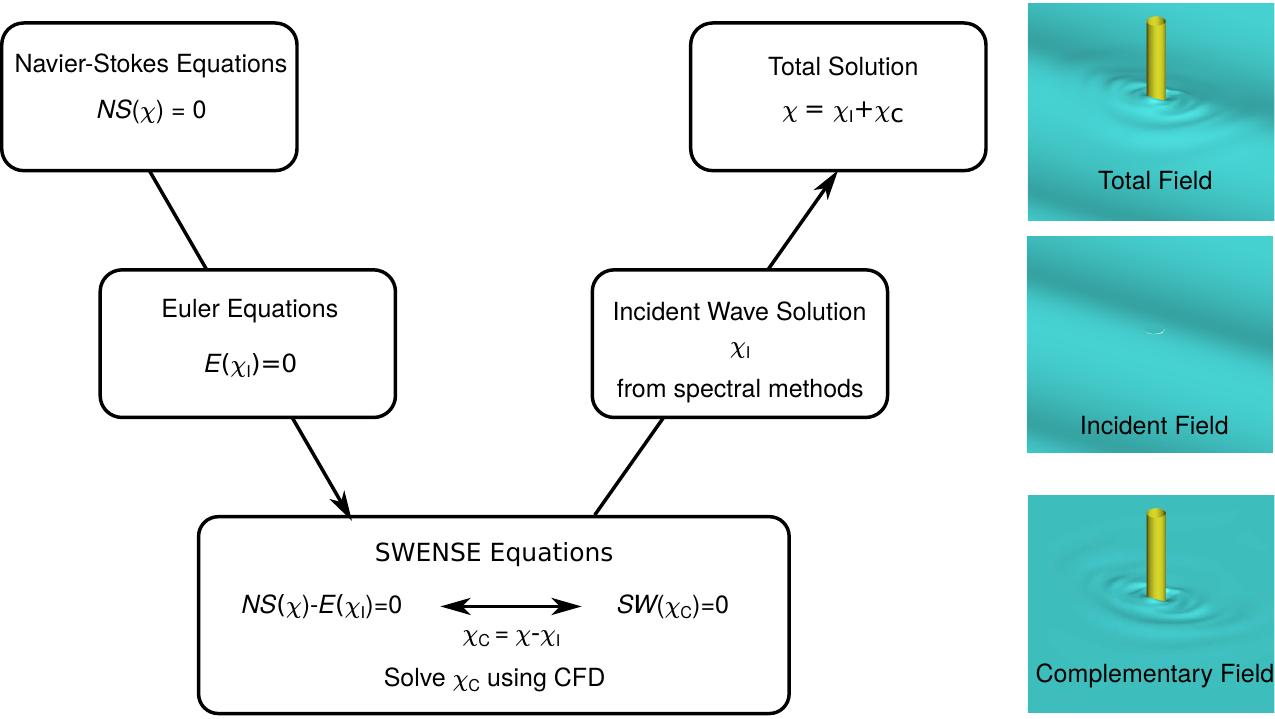}
\caption{The functional decomposition in the SWENSE method\label{fig:swenseMethod}}
\end{figure}

In Fig.\ \ref{fig:swenseMethod}, a V-shape illustrates how the SWENS equations are derived with the functional decomposition and how the total field is reconstructed.  
The left half of this V-shape shows the derivation procedure of the SWENS equations. Assuming the Navier-Stokes equations (the first row) and the Euler equations (the second row) have the same definition zone, we can subtract the Euler equations from the NS equations. The SWENS equations (the third row) is obtained by writing the remainder with the variable $\chi_{C}$.  On the right half, the complementary field $\chi_{C}$ is added to the incident solution $\chi_{I}$ to reconstruct the total solution $\chi$. 

\subsection{Derivation of two-phase SWENS equations}

The two-phase SWENS equations are derived following the SWENSE decomposition procedure in Fig.\ \ref{fig:swenseMethod} with the two-phase NS equations and the Euler equations.

\subsubsection{Two-phase incompressible Navier-Stokes equations}


%
%

In the present work, the immiscible air-water flow with a deformable common interface is considered as incompressible and viscous. The definition zone contains both water and air.

\begin{figure}[h]
\centering
\includegraphics[width=7cm]{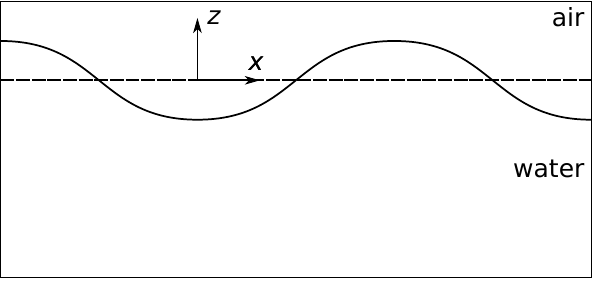}
\caption{The definition zone of the two-phase NS equations: both water and air are considered.}
\label{fig:VOFMethod}
\end{figure}

{\color{black}

The Volume-of-Fluid  (VOF) method \cite{HIRT1981VOF} is used for interface capturing. The method is chosen among others for its advantage of mass conservation.
The VOF field $\alpha$ is used to represent the volume fraction of water. It is transported by the total velocity field $\textbf{u}$ with the following equation:
\begin{equation}
    \frac{\partial \alpha}{ \partial t} +  \nabla. (\bu\alpha )+ \nabla.(\bu_r\alpha(1-\alpha)) = 0 \label{eqn:alphaTransportSwense}
\end{equation}
where $\bu=(u,v,w)$ is the fluid velocity, and $\nabla .(\alpha(1- \alpha)\bu_r)$ is an artificial compression term to avoid the interface smearing with $\bu_r$  the compression velocity \cite{albadawi2014assessment,rusche2003computational}}. {\color{black}A sharp interface is approximated with $\alpha = 0.5$. The density $\rho$ and the molecular viscosity $\mu$ are phase-dependent and are defined as:   
\begin{equation}
\rho =
\begin{cases}
\rho_w & \alpha>0.5 \\
\rho_a &\alpha \leqslant 0.5
\end{cases}
\label{eqn:rhoEquation}
\end{equation}
and
\begin{equation}
\mu =
\begin{cases}
\mu_w & \alpha>0.5 \\
\mu_a &\alpha\leqslant 0.5
\end{cases}
\label{eqn:partial}
\end{equation}
with subscriptions $w$ and $a$ representing water and air respectively. This kind of definition has been used in the literature \cite{abadie2010numerical,jasak2019cfd}. Despite its simplicity, it shows adequate accuracy in the present work, while other more accurate techniques (such as PLIC \cite{scardovelli1999direct}) can be considered in the future.
}

The flow is described by incompressible NS equations.
\begin{equation}
    \nabla . \textbf{u} = 0 \label{eqn:NSContituity}
\end{equation}
{\color{black}
\begin{equation}
    \frac{\partial \textbf{u}}{ \partial t} + \textbf{u} . \nabla \textbf{u} = - \frac{\nabla p }{\rho} + \textbf{g} +  \frac{\nabla.\left((\mu + \mu_t)\left(\nabla\bu+\nabla\bu^T\right)\right)}{\rho} \label{eqn:NSnonConservative}
\end{equation}
}where $p$ is the total pressure; $\textbf{g}$ is the gravitational acceleration; {\color{black} $\mu_t$ represents the turbulent viscosity, obtained from Reynolds averaged turbulence models with the Boussinesq assumption.} Eqn.\ \eqref{eqn:NSnonConservative} neglects the surface tension since it is small in classical marine and ocean engineering applications.

%
%

\subsubsection{Euler equations for incident water waves}

\begin{figure}[h]
\centering
\includegraphics[width=7cm]{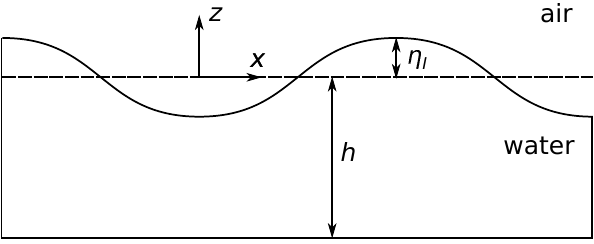}
\caption{The definition zone of the single-phase Euler equations: only water is considered. Water waves are modeled as a single-phase problem with a deforming free surface.
\label{fig:EulerEquation}}
\end{figure}

The incident waves are modeled as a single-phase flow.  The water free surface is considered as a moving boundary (see Fig.\ \ref{fig:EulerEquation}).
and  is defined by a time-dependent free-surface elevation function, as follows
\begin{equation}
 z = \eta_I(\widetilde{\textbf{x}},t)
\end{equation}
where $\widetilde{\textbf{x}} = (x,y)$. It assumes $\eta_I$ is a single-valued function, and the application is consequently limited to non-breaking incident waves.

Neglecting the viscosity of water, the flow beneath the free surface is modeled with the incompressible Euler equations, as follows,

\begin{eqnarray}
\nabla.\textbf{u}_{I} &=& 0 \label{eqn:EulerContinuity} \\
\frac{\partial \textbf{u}_{I}}{ \partial t} +\textbf{u}_{I}. \nabla \textbf{u}_{I} &=& - \frac{\nabla p_{I} }{\rho_I} + \textbf{g}  \label{eqn:EulerNonconservative}
\end{eqnarray}
where the subscript $I$ represents "\textit{incident}" field. The density $\rho_I $ is equal to the water density since the equations are defined only in water.
The boundary conditions are:

\begin{align}
\frac{\partial \eta_I}{\partial t } + u_I \frac{\partial \eta_I}{\partial x} + v_I \frac{\partial \eta_I}{\partial y} - w_I &= 0 \quad at \quad z = \eta_I(\widetilde{\textbf{x}},t)  &\text{ on the free surface}\\
p_I &= 0 \quad at \quad z = \eta_I(\widetilde{\textbf{x}},t)&\text{ on the free surface} \\
w_I &= 0 \quad at \quad z = -h  &\text{ on the seabed}
\end{align}
with $\textbf{u}_I = (u_I,v_I,w_I)$ and $\widetilde{\textbf{x}} = (x,y)$.

{\color{black}\noindent\textbf{Extension of the definition zone}}

It is worth noting that the definition zone of Euler equations is different from that of two-phase NS equations. Consequently, the former must be extended to apply the same SWENSE decomposition in both water and air. The extension technique will be presented in Sect.\ \ref{sect:extensionAirVelocity}. Herein, please just assume the Euler equations and their solution have the same definition zone as that of the two-phase NS equations, i.e., in the whole air-and-water domain.

\subsubsection{Two-phase Spectral Wave Explicit Navier-Stokes Equations}


The two-phase SWENS equations are derived by subtracting the Euler equations from the two-phase Navier-Stokes equations.

\noindent\textbf{Continuity Equation}

Using the continuity equation of NS equations (Eqn.\ \ref{eqn:NSContituity}) minus the that of Euler equations (Eqn.\ \ref{eqn:EulerContinuity}), yields:
\begin{equation}
  \nabla.(\textbf{u}-\textbf{u}_I)=0 \label{eqn:SwContinuity1}
\end{equation}
With the notation of $\textbf{u}_C=\textbf{u}-\textbf{u}_I$, the continuity equation using the complementary variable reads:
\begin{equation}
  \nabla.\textbf{u}_C=0 \label{eqn:SwContinuity2}
\end{equation}

\noindent\textbf{Momentum Equation}

The momentum equation should also be derived by subtracting the momentum equation of the Euler equations (Eqn.\ \ref{eqn:EulerNonconservative}) from that of the Navier-Stokes equations (Eqn.\ \ref{eqn:NSnonConservative}). However, the direct use of Eqn.\  \eqref{eqn:EulerNonconservative} causes stability problems in the air (see Appendix \ref{section:directSubtractionProblem}). For this reason, Eqn.\  \eqref{eqn:EulerNonconservative} is modified by introducing a pressure term $p_I^*$:
\begin{equation}
p_I^* = \rho \frac{p_I}{\rho_I} \label{eqn:pIStar}
\end{equation}
where $\rho$ is the density of the two-phase flow and $\rho_I$ is equal to the water density.
Eqn.\  \eqref{eqn:EulerNonconservative} written in its modified version by using $p_I^*$ reads:
\begin{equation}
\frac{\partial \textbf{u}_{I}}{ \partial t} +\textbf{u}_{I}. \nabla \textbf{u}_{I} = - \frac{\nabla p_I^* }{\rho} + \frac{p_I}{\rho_I}\frac{\nabla\rho}{\rho} + \textbf{g}  \label{eqn:modifiedEuler}
\end{equation}

Subtracting Eqn.\  \eqref{eqn:modifiedEuler} from Eqn.\  \eqref{eqn:NSnonConservative} and using the notation of $p_C=p-p_I^*$ the two-phase SWENSE momentum equation written with the complementary variables is obtained as:
{\color{black}
\begin{equation}
\frac{\partial \textbf{u}_C}{ \partial t} + \textbf{u} . \nabla \textbf{u}_C + \textbf{u}_C . \nabla \textbf{u}_{I}
= -\frac{\nabla p_C}{\rho} - \frac{p_I}{\rho_I}\frac{\nabla\rho}{\rho} +
\frac{\nabla. \left((\mu+\mu_t)\left(\nabla\bu+\nabla\bu^T\right)\right)}{\rho} \label{eqn:twophaseSwMomentNotSimplyfied}
\end{equation}

We now simplify the viscosity term, by separating the molecular and turbulent terms. Looking at the molecular terms first, we can decompose it into incident and complementary terms.

\begin{equation}
   \nabla. \left(\mu\left(\nabla\bu+\nabla\bu^T\right)\right)
  = \nabla. \left(\mu\left(\nabla\bu_I+\nabla\bu_I^T\right)\right) + \nabla. \left(\mu\left(\nabla\bu_C+\nabla\bu_C^T\right)\right)
\end{equation}

Continue to simplify the incident part using $\nabla.\bu_I=0$ and $\nabla \times \bu_I=0$, since the incident velocity field is incompressible and irrotational. we have
\begin{equation}
  \begin{split}
    \nabla. \left(\mu\left(\nabla\bu_I+\nabla\bu_I^T\right)\right) &=2\nabla. \left( \mu\nabla\bu_I^T \right)\\
    &=2 \left(\mu\nabla\left(\nabla.\bu_I\right)+ \nabla \bu_I.\nabla \mu \right) \\
    &=2 \nabla \bu_I.\nabla \mu.
  \end{split}
\end{equation}

The turbulence viscosity can be also simplified in the same way, the resulting equation is:

\begin{equation}
\frac{\partial \textbf{u}_C}{ \partial t} + \textbf{u} . \nabla \textbf{u}_C + \textbf{u}_C . \nabla \textbf{u}_{I}
= -\frac{\nabla p_C}{\rho} - \frac{p_I}{\rho_I}\frac{\nabla\rho}{\rho} +
\frac{\nabla. \left((\mu+\mu_t)\left(\nabla\bu_C+\nabla\bu_C^T\right)\right)}{\rho} + \frac{2\nabla \bu_I.\nabla \mu}{\rho} +\frac{2\nabla \bu_I.\nabla \mu_t}{\rho}
\label{eqn:twophaseSwMomentNotSimplyfied2}
\end{equation}

It worth noting that in Eqn. \eqref{eqn:twophaseSwMomentNotSimplyfied2}, two source terms have non-zero values only on the air-water interface, i.e., $\displaystyle \frac{p_I}{\rho_I}\frac{\nabla\rho}{\rho}$ and $\displaystyle \frac{2\nabla \bu_I.\nabla \mu}{\rho}$. 
The first suggests a restoring force proportional to the density gradient. The second represents a viscous dissipation at the interface, which is proportional to the molecular viscosity gradient. The second term is neglected since it is several orders of magnitude smaller. 
Moreover, the turbulence stress on the incident wave part is also neglected $\displaystyle \left( \frac{2\nabla \bu_I.\nabla \mu_t}{\rho}\right)$, because we assume the incident waves provided by PT solver is accurate for the aspect of application and should not interact with the turbulence viscosity.

The final simplified momentum equation reads,

\begin{equation}
\frac{\partial \textbf{u}_C}{ \partial t} + \textbf{u} . \nabla \textbf{u}_C + \textbf{u}_C . \nabla \textbf{u}_{I}
= -\frac{\nabla p_C}{\rho} - \frac{p_I}{\rho_I}\frac{\nabla\rho}{\rho} +
\frac{\nabla. \left((\mu+\mu_t)\left(\nabla\bu_C+\nabla\bu_C^T\right)\right)}{\rho}
\label{eqn:twophaseSwMoment}
\end{equation}
}

{\color{black}
\noindent\textbf{VOF Equation}
 
The VOF field is not decomposed as other variables because it is challenging to consider the boundedness in the decomposition, i.e., it is not straightforward to define an incident VOF field $\alpha_I$ and a complementary VOF field $\alpha_C$ and to keep $0 \le \alpha_I+\alpha_C \le1$.  Instead, the total VOF field is transported with the reconstructed total field $\bu = \bu_I + \bu_C $ using Eqn.\ \eqref{eqn:alphaTransportSwense}. The authors are aware that in this step (transporting VOF field with a given velocity field), the proposed method introduces numerical errors as large as convectional two-phase VOF NS solvers, and thus may restrict the accuracy of the present method when using coarse mesh.  However, the mesh requirement of an accurate VOF convection is often not as demanding as for the momentum equations (see Appendix \ref{sect:compareLSVOF}): using 15 cells per wave length can already provide adequate accuracy. Decomposition approaches can be used on other approaches such as the Decomposed Level-Set (DLS) \cite{vukvcevic2016decomposition,reliquet2013lvlsetphd}.  Appendix \ref{sect:compareLSVOF} also provides a comparison between VOF and DLS. 

%
%
}

{\color{black}
\subsection{Boundary conditions}

The boundary conditions (BCs) of the complementary field are derived from the boundary conditions of the total fields, with the same physical significance. Table \ref{tab:SWENSE_BC} shows some commonly used BCs in the SWENSE method with their equivalence in Navier-Stokes Equation (NSE).

For far-field and the non-slip wall boundary condition, the BCs of NSE and SWENSE are mathematical equivalent. The pressure BCs are not given, as they should be calculated to ensure the fixed velocity BC. The far-field BC is applied to the outer boundaries, assuming that the fluid motion is equal to the incident waves, i.e., the complementary field vanishes because the wave amplitude decay with $1/\sqrt{r}$ ($r$ is the distance from the structure). The no-slip wall BC is used on the body. For a fixed body, the total velocity on the wall $\bu = 0$, thus in SWENSE $\bu_C = -\bu_I$.

The atmosphere boundary is the upper limit of the computational domain in two-phase flow simulation, which is connected to the atmosphere. In NSE, such boundary condition is imposed by the atmosphere pressure ($p=0$) and assumes the gradient of velocity $\bu$ is equal to zero. The strict equivalence is $p_C=-p_I^*$. However, the $p_I^*$ is small by definition in the air (Eqn. \eqref{eqn:pIStar}). For simplicity, we impose $p_C = 0$ and assume the gradient of $\bu_C$ is zero.

\renewcommand{\arraystretch}{1.8}
\begin{table}[h!]
\centering
  {%
  \caption{Boundary conditions of SWENSE compared with Navier-Stokes Equations\label{tab:SWENSE_BC}}
  \begin{tabular}{lll}
  \hline
   & NSE & SWENSE \\ \hline
  \multirow{2}{*}{Far-field} & $\alpha = \alpha_I$ & $\alpha = \alpha_I$ \\
   & $\mathbf{u} = \mathbf{u}_I$ & $\mathbf{u}_C = \mathbf{0}$ \\ \hline
  \multirow{2}{*}{No-slip wall} & $\displaystyle{\frac{\partial \alpha}{\partial n} = 0} $ & $\displaystyle{\frac{\partial \alpha}{\partial n} = 0} $ \\
   & $\mathbf{u} = \mathbf{0}$ & $\mathbf{u}_C = -\mathbf{u}_I$ \\ \hline
  \multirow{3}{*}{Atmosphere} & $\displaystyle{\frac{\partial \alpha}{\partial n} = 0} $ & $\displaystyle{\frac{\partial \alpha}{\partial n} = 0} $ \\
   & $ p = 0 $ & $ p_C = 0 $ \\
   & $ \displaystyle{\frac{\partial \mathbf{u}}{\partial n} = \mathbf{0}} $ & $ \displaystyle{\frac{\partial \mathbf{u}_C}{\partial n} = \mathbf{0}} $ \\ \hline
  \end{tabular}
  }
\end{table}
\renewcommand{\arraystretch}{1.0}
}

\subsection{Discussion\label{sect:swenseEquationsAbility}}

\subsubsection{Relation with NS equations\label{sect:analogyWithNS}}

The given SWENSE equation can regress to the NS equations when the incident solution is set to be calm water, i.e., Eqn.\ \eqref{eqn:twophaseSwMoment} in the two-phase SWENSE momentum equation is equivalent to Eqn.\ \eqref{eqn:NSnonConservative} in the two-phase NS Equations, if the incident wave part is explicitly given as no waves.

To prove that, one can let the incident wave field be zero, i.e.:
\begin{eqnarray}
\textbf{u}_I = 0
\end{eqnarray}
and thus the incident pressure is equal to the hydrostatic pressure:
\begin{equation}
\frac{p_I}{\rho_I}=-gz
\end{equation}

Thus,  Eqn.\  \eqref{eqn:twophaseSwMoment} can be written as:
{\color{black}
\begin{equation}
  \frac{\partial \textbf{u}_C}{ \partial t} + \textbf{u} . \nabla \textbf{u}_C
  = -\frac{\nabla p_C}{\rho}+gz\frac{\nabla\rho}{\rho} +
  \frac{\nabla. \left((\mu+\mu_t)\left(\nabla\bu_C+\nabla\bu_C^T\right)\right)}{\rho} \label{eqn:SWENSENOWAVE}
\end{equation}
}

{\color{black}
Transforming this equation with the relations below:
\begin{eqnarray}
\mathbf{u}_C &=& \mathbf{u} -  \mathbf{u}_I =  \mathbf{u} \\
p_C &=& p - p_I^* = p + \rho gz
\end{eqnarray}
yields 
%
%
%
Eqn.\ \ref{eqn:NSnonConservative}, showing that the SWENSE momentum equation (Eqn.\ \ref{eqn:twophaseSwMoment}) can regress to the two-phase NS equation (Eqn.\ \ref{eqn:NSnonConservative}) if there are no incident waves.}


\subsubsection{Keeping incident solution with coarse mesh \label{sect:accurateIncidentWaves}}

The second property is to preserve exactly the incident wave solution throughout the domain when no structure is present, regardless of the mesh used.

In the case of pure incident wave simulation, the initial values of the complementary fields are zero, i.e.:
\begin{eqnarray}
\mathbf{u}_C (\mathbf{x},t=0) &=& 0 \\
p_C(\mathbf{x},t=0) &=& 0
\end{eqnarray}

The second term on the R.H.S of Eqn.\  \eqref{eqn:twophaseSwMoment} is equal to zero in the entire computational domain, 
\begin{equation}
    \frac{p_I}{\rho_I}\frac{\nabla\rho}{\rho}= 0
\end{equation}
because
\renewcommand{\arraystretch}{1.8}
\begin{table}[h!]
\centering
\label{tab:sourceTermIsZero}
\begin{tabular}{|c|c|c|}
\hline
 & At the incident free-surface & elsewhere     \\ \hline
$p_I$ & 0                            & non-zero  \\ \hline
$\nabla\rho$ & non-zero                & 0              \\ \hline
$\displaystyle\frac{p_I}{\rho_I}\frac{\nabla\rho}{\rho}$  & 0 & 0 \\\hline
\end{tabular}
\end{table}
\renewcommand{\arraystretch}{1.0}

Using the above conditions, Eqn.\  \eqref{eqn:twophaseSwMoment} is simplified to:
\begin{equation}
    \frac{\partial \textbf{u}_C}{ \partial t} = 0
\end{equation}
showing that complementary velocity fields $\textbf{u}_C$ remains zero. In this condition, the pressured field $p_C$ also remains zero after solving the pressure Poisson equation. As a result,
Eqn.\ \eqref{eqn:twophaseSwMoment} preserves the kinematics of the incident waves.


{\color{black}
\subsubsection{Main differences from \vuko\ \etal\ \cite{vukvcevic2016decomposition} \label{sect:compareWtihVuko}}

Another two-phase SWENSE method has been introduced early by \vuko \etal\  \cite{vukvcevic2016decomposition}  with different characteristics with the method proposed here. These differences are presented as follows. 

The first and the fundamental difference is that in reference \cite{vukvcevic2016decomposition}, terms containing incident wave information are not simplified with known relations from PT.
 For example, the continuity equation used in reference \cite{vukvcevic2016decomposition} reads,
 \begin{equation}
   \nabla.\textbf{u}_C=-\nabla.\textbf{u}_I. \label{eq:vukoContinuity}
 \end{equation}
 Although the R.H.S. is equal to zero in the wave theory, reference \cite{vukvcevic2016decomposition} evaluates it by the CFD solver, to consider the errors coming from:
 \begin{itemize}

 \item the spectral solution techniques;
   \item the interpolation of incident wave velocity on the CFD mesh.
 \end{itemize}

Reference \cite{vukvcevic2016decomposition} emphasizes the necessity to correct these errors with the complementary field. However, in the present work, we choose to drop off these errors and cancel out the R.H.S. term with the theoretical relation $\nabla.\textbf{u}_I=0$ for incompressible flow.
The reasons for this choice are:

\begin{itemize}
\item The PT methods are accurate for water wave problems. The numerical error resides in a converged spectral potential flow result is often negligible. Compensating this error by CFD solvers is not beneficial because this correcting step may introduce larger numerical errors itself.
\item The interpolation of incident field on CFD mesh is improved in this work (see Section \ref{sectionReconstructionMethod}), so that the numerical error related to the interpolation procedure is drastically reduced.
\item The numerical evaluation of the incident quantities requires fine CFD discretization to be accurate. This is not consistent with the objective of the SWENSE method, i.e., to use coarse mesh for the incident wave propagation problem.
\end{itemize}

Similarly, the momentum equation (Eqn.\ \ref{vukoMomentum}) in reference \cite{vukvcevic2016decomposition} is not simplified either with the analytic relation provided by the Euler equations. It reads,
\begin{equation}
\frac{\partial \textbf{u}_C}{ \partial t} + \textbf{u} . \nabla \textbf{u}_C
-\nu \nabla^2 \textbf{u}_C= -\frac{\partial \textbf{u}_I}{ \partial t} - \textbf{u} . \nabla \textbf{u}_I
+ \nu \nabla^2 \textbf{u}_I - \beta \nabla p_d \label{vukoMomentum}
\end{equation}
where $\nu$ is the kinematic viscosity and $\beta = 1/\rho$. This treatment also prevents the use of coarse mesh for incident wave propagation, because coarse mesh (even only used in the far-field) generates errors on the R.H.S. of Eqn.\ \eqref{vukoMomentum} and results in spurious complementary fields that interfere with incident waves. In contrast, the present work cancels out these terms with analytical relations so that the numerical errors can be avoided, as shown in Sect.\ \ref{sect:accurateIncidentWaves}.

The second difference is that reference \cite{vukvcevic2016decomposition} uses the Decomposed Level Set (DLS) method to capture the interface, instead of the standard VOF (not decomposed) here. Appendix \ref{sect:compareLSVOF} compares both methods on a pure convection case, i.e., the interface capturing function is transported by the incident wave velocity. The results demonstrate that the DLS is not more advantageous than the VOF in the present second-order accurate Finite Volume framework.
}
%
%

\section{Incident wave modeling}\label{Section:IncidentWave}

This section describes how to obtain the incident wave solution by potential flow solvers (spectral wave models) and how to interpolate it onto the CFD mesh.

\subsection{Potential flow theory and spectral solution techniques}

By further assuming the incident wave velocity is irrotational, the Euler equations can be solved efficiently and accurately \cite{dean1991waterwaves} with PT. Such an assumption allows to reduce the number of unknowns in the Euler equation, by defining a scalar field $\phi$ (called velocity potential or potential), so that:
\begin{equation}
    \textbf{u}_I = \nabla \phi \label{eqn:velocityPotential}
\end{equation}
Using $\phi$, the Euler equations are written in an equivalent form, as follows,
\begin{align}
\Delta\phi  &= 0 &\text{ Laplace's equation} \label{eqn:laplacian}\\
\frac{\partial \eta_I}{\partial t } + \frac{\partial \phi}{\partial x} \frac{\partial \eta_I}{\partial x} + \frac{\partial \phi}{\partial y} \frac{\partial \eta_I}{\partial y} - \frac{\partial \phi}{\partial z} & = 0 \quad at \quad z = \eta_I(\textbf{x},t) &\text{ Free surface kinematic condition} \label{eqn:FSKC}\\
\frac{\partial \phi}{\partial t} + \frac{1}{2}(\nabla\phi)^2 + gz  &= 0 \quad at \quad z = \eta_I(\textbf{x},t) &\text{ Free surface dynamic condition} \label{eqn:FSDC}\\
\frac{\partial \phi}{\partial z}  &= 0 \quad at \quad z = -h  &\text{ Seabed boundary condition} \label{eqn:seabed}
\end{align}

In the present work, spectral methods are used to solve this non-linear potential flow problem with high accuracy and efficiency. These methods decompose the free surface elevation $\eta_I(\widetilde{\mathbf{x}},t)$ and the velocity potential $\phi(\widetilde{\mathbf{x}},z,t)$ using a set of basis functions. An example in a 2D uni-directional wave case is shown as follows:
\begin{align}
  \eta_I(x,t) &= \sum_i A_i^{\eta}(t)\psi_i(x) \label{eqn:whatIsSpectralMethod1}\\
  \phi(x,z,t) &= \sum_i A_i^{\phi}(t) \frac{\cosh(k_i(z+h))}{\cosh(k_ih)}\psi_i(x)
  \label{eqn:whatIsSpectralMethod}
\end{align}
where $A_i$ are the modal amplitudes, $k_i$ are the basis wave numbers, and $\psi_i(x)$ are the horizontal basis functions (sine and cosine functions or their complex exponential equivalents). With these basis functions, Eqn.\ \eqref{eqn:whatIsSpectralMethod} satisfy automatically the Laplace's equations and the seabed boundary condition (Eqns.\ \ref{eqn:laplacian} and \ref{eqn:seabed}). The free surface boundary conditions (Eqns.\ \ref{eqn:FSKC} and \ref{eqn:FSDC}), discretized on uniformly distributed points on the free surface, are used to establish a linear system to determine the modal amplitudes $A_i^{\eta}(t)$ and $A_i^{\phi}(t)$.

In this work, two fully non-linear spectral methods are adopted. The first is based on the stream function waves theory \cite{rienecker1981fourier,CN_stream} for 2D regular waves. The second is the High-Order Spectral (HOS) method \cite{DUCROZET2016OpenSourceOcean,ducrozet2012hosNWT} for 2D/3D arbitrary waves in open seas or in experimental wave tanks.

\subsubsection{Regular waves: stream function theory}
For 2D regular nonlinear waves, the algorithm proposed by Rienecker \& Fenton  \cite{rienecker1981fourier} based on the stream function theory is adopted \cite{CN_stream,cnSteamGithub}. This algorithm solves 2D progressing regular waves over a horizontal seabed for a wide range of depths, amplitudes and wavelengths. The free surface elevation and the velocity potential are decomposed with basis functions as shown in Eqns.\ \eqref{eqn:whatIsSpectralMethod1} and \eqref{eqn:whatIsSpectralMethod}. The basis function $\psi_i(x)$ is replaced by $\psi_i(x-ct)$, with $c$ the phase velocity to make the modal amplitudes $A_i^{\eta}$ and $A_i^{\phi}$ independent of time. These amplitudes are solved from a linear system established with the kinematic and the dynamic boundary conditions (Eqns.\ \ref{eqn:FSKC} and \ref{eqn:FSDC}). The velocity field $\textbf{u}_I$ is calculated with Eqn.\ \eqref{eqn:velocityPotential}, and the pressure field is obtained with the Bernoulli equation. The reader can find more information in references \cite{rienecker1981fourier}.

%

\subsubsection{Arbitrary waves: High-Order Spectral method}

The High-Order Spectral (HOS) method is widely used for the study of arbitrary waves wave propagating in open domains \cite{west1987new,dommermuth1987high}. It considers the full non-linearity of the free surface and exhibits high efficiency and accuracy thanks to its pseudo-spectral formalism. Two open-source solvers developed at LHEEA Lab.\ (Ecole Centrale Nantes and CNRS), \textit{HOS-Ocean} \cite{DUCROZET2016OpenSourceOcean,HOSOceanGitHub} and \textit{HOS-NWT} \cite{ducrozet2012hosNWT,HOSNWTGitHub}, are used to solve arbitrary incident waves in open domains and experimental wave tanks, respectively.

For simplicity, only \textit{HOS-Ocean} in a 2D case is presented as an example. The reader is referred to references \cite{DUCROZET2016OpenSourceOcean,ducrozet2012hosNWT} for more information.

The HOS wave model solves an unsteady wave propagation problem with the following equations derived from the free surface boundary conditions (Eqns.\ \ref{eqn:FSKC} and \ref{eqn:FSDC}) to describe the time evolution of the free surface elevation $\eta_I$ and the velocity potential $\tilde{\phi}$.

\begin{eqnarray}
\frac{\partial \eta_I}{\partial t} &=& \left(1+\left(\frac{\partial \eta_I}{\partial x}\right)^2\right)w_I - \frac{\partial \tilde{\phi}}{\partial x} .\frac{\partial \eta_I}{\partial x}  \label{eqn:hosetaDt}\\
\frac{\partial \tilde{\phi}}{\partial t}&=& -g\eta_I - \frac{1}{2}\left(\frac{\partial \tilde{\phi}}{\partial x}\right)^2 + \frac{1}{2}\left(1+\left(\frac{\partial \eta_I}{\partial x}\right)^2\right)w_I^2 \label{eqn:hosphiDt}
\end{eqnarray}
where $x$ is the horizontal coordinate, $\tilde{\phi}$ is the free surface velocity potential defined as follows:
\begin{equation}
   \tilde{\phi}(x,t) = \phi\left(x,z=\eta_I(x,t),t\right)
\end{equation}
$\widetilde{\phi}$ is decomposed in the spectral domain with the following equation:
\begin{equation}
\tilde{\phi}(x,t) = \sum_{i=1}^{N} A_{i}^{\tilde{\phi}} (t) \exp(jk_{i}x) \label{eqn:HOSphiDecompose}
\end{equation}
where $j$ is the unit imaginary number $j^2=-1$. The free surface elevation $\eta_I$ is decomposed with Eqn.\ \eqref{eqn:whatIsSpectralMethod1}. Such a decomposition allows the efficient calculation of $\eta_I$ and $\widetilde{\phi}$ and their spatial derivatives on a series of uniformly distributed HOS calculation points with the Fast Fourier Transform (FFT). 
%

\begin{figure}[ht]
\centering
\includegraphics[width=0.6\textwidth]{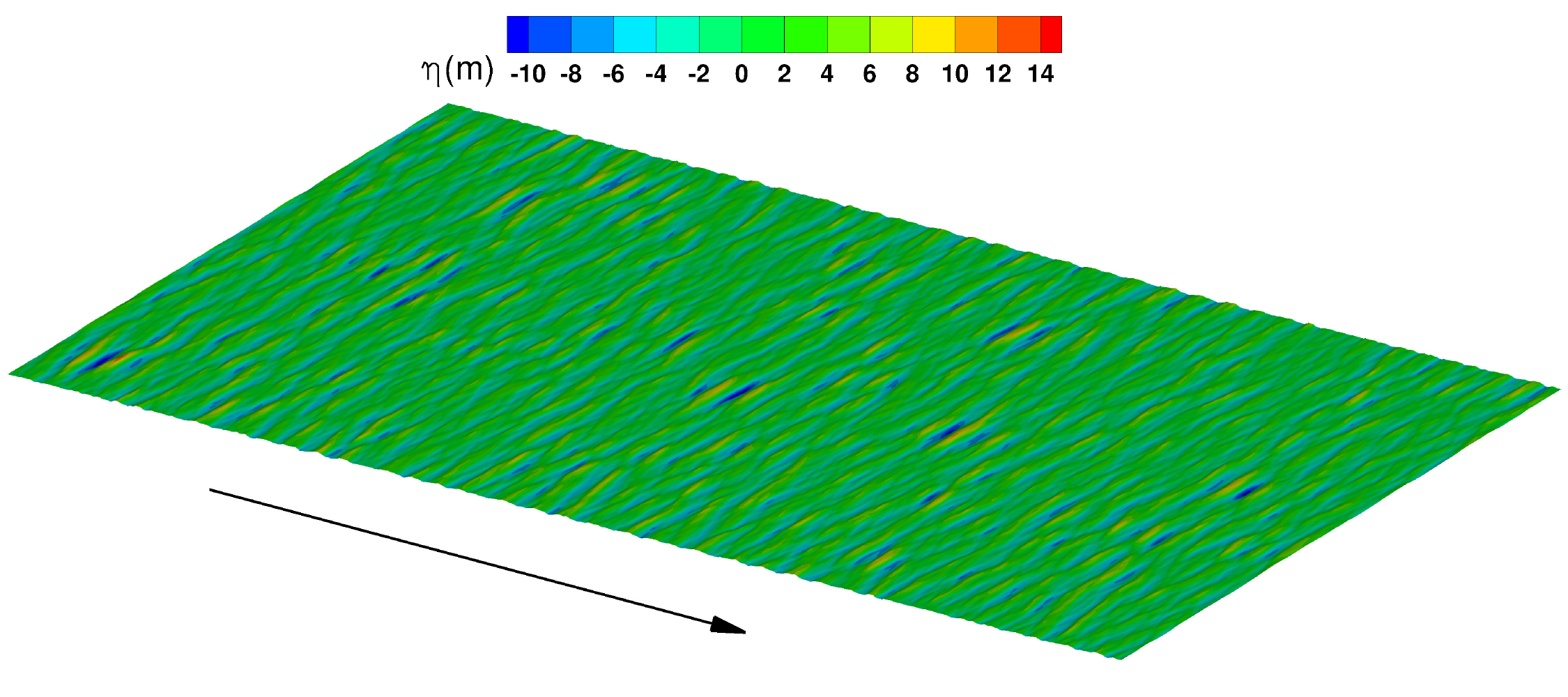}
\caption{Free-surface elevation of a \textit{HOS-Ocean} simulation with a JONSWAP spectrum. The size of computational domain is about 50 $\text{ km}^2$. The picture is reproduced from the work of Ducrozet \etal\  \cite{DUCROZET2016OpenSourceOcean}.  Reprinted with permission, copyright Elsevier\textsuperscript{\textregistered}. \label{fig:HOS-OCEAN}}
\end{figure}

Figure \ref{fig:HOS-OCEAN} gives an example of an irregular sea state simulation conducted with \textit{HOS-Ocean} \cite{HOSOceanGitHub}. The computational domain and time duration are made to be large enough for investigating the occurrence of extreme wave events. The irregular sea state is generated using a Joint North Sea Wave Project (JONSWAP) spectrum. 
The size of the domain is $(Lx,Ly)=(40\lambda_p,20\lambda_p)\approx50 \text{ km}^2$.

Besides \textit{HOS-Ocean}, the second solver, \textit{HOS-NWT} (Numerical Wave Tank) \cite{HOSNWTGitHub,ducrozet2012hosNWT}, is used to reproduce wave evolution in experimental test basins. The basis functions and the numerical schemes are adapted to reproduce a rectangular wave tank with a 3D wave-maker, an absorbing beach, and perfectly reflective side walls.

\subsection{Reconstruction of incident wave information on the CFD mesh}\label{sectionReconstructionMethod}

The reconstruction step transforms the results of spectral wave models onto the CFD mesh. For example, in a 2D case, the velocity potential $\phi$ is stored as the modal amplitudes $A_i^{\phi}(t)$ in Eqn.\ \eqref{eqn:whatIsSpectralMethod}. With these modal amplitudes, the potential $\phi$ can be reconstructed and the velocity can be further calculated by evaluating the derivatives of the potential (Eqn. \ref{eqn:velocityPotential}) as follows:
\begin{align}
  u_I(x,z,t) = \sum_{i=1}^{N} A_i^{\phi}(t)j k_i \frac{\cosh(k_i(z+h))}{\cosh(k_ih)}\psi_i(x) \\
  w_I(x,z,t) = \sum_{i=1}^{N} A_i^{\phi}(t) k_i \frac{\sinh(k_i(z+h))}{\cosh(k_ih)}\psi_i(x)
\end{align}

The velocity can be obtained by substituting the coordinates $(x,z)$ and the time $t$ into the above equations. This reconstruction method is referred to as "analytical evaluation" method hereafter. The computational cost at each cell location and at each timestep is proportional to the number of Fourier components. For the stream function wave theory, the number of Fourier components are $N \approx O(10)$, whereas the HOS method typically needs 100 to 1000 Fourier components in a single direction for an irregular wave train. As a result, the analytical evaluation is very time-consuming for HOS waves. Instead, an interpolation method is used: the modal results are first transferred on a series of space points with the Inverse Fast Fourier Transform (IFFT) algorithm and then interpolated to the CFD cells. In this section, the analytical evaluation and the interpolation method are explained in the first two subsections. The third subsection focuses on the reconstruction error and its consequences. The last subsection proposes a new interpolation technique to improve the interpolation accuracy. For simplicity, only the velocity field is shown as an example. The pressure field can be reconstructed similarly.

\subsubsection{Analytical evaluation for the stream function wave theory}
For regular waves obtained with the stream function wave theory, the analytical evaluation method is applied since the number of Fourier components is small. For example, a regular wave with a moderately large steepness ($ka = 0.24$) converges with only 9 Fourier components ($N = 9$).

\subsubsection{Interpolation for the HOS waves}

The interpolation procedure \cite{ducrozet2007modelisation} to reconstruct the HOS results on CFD cells firstly reconstructs the wave information on a coarse rectangular HOS grid via IFFT and then interpolates it on the CFD mesh. Compared to the analytical evaluation, the efficiency is much improved by the IFFT algorithm. This procedure can be divided into two steps.

\begin{enumerate}
  \item IFFT: This step translates the spectral information in the spatial domain via IFFT.
  After the IFFT, the wave information is available on uniformly spaced points. The number of points is equal to the number of Fourier modes ($N$) used in the HOS computation. The distance between two points is equal to the shortest wave length of the basis function ($2 \pi/k_N$). Take a 2D case as an example:
  \begin{itemize}
  \item For the free surface elevation ($\eta_I(x,t)$), IFFT transforms $A_i^{\eta}$ to $N$ uniformly distributed points in the $x$ direction (see Figure \ref{fig:hosModeToSpace}).

\begin{figure}[ht]
\centering
    \includegraphics[width=0.7\textwidth]{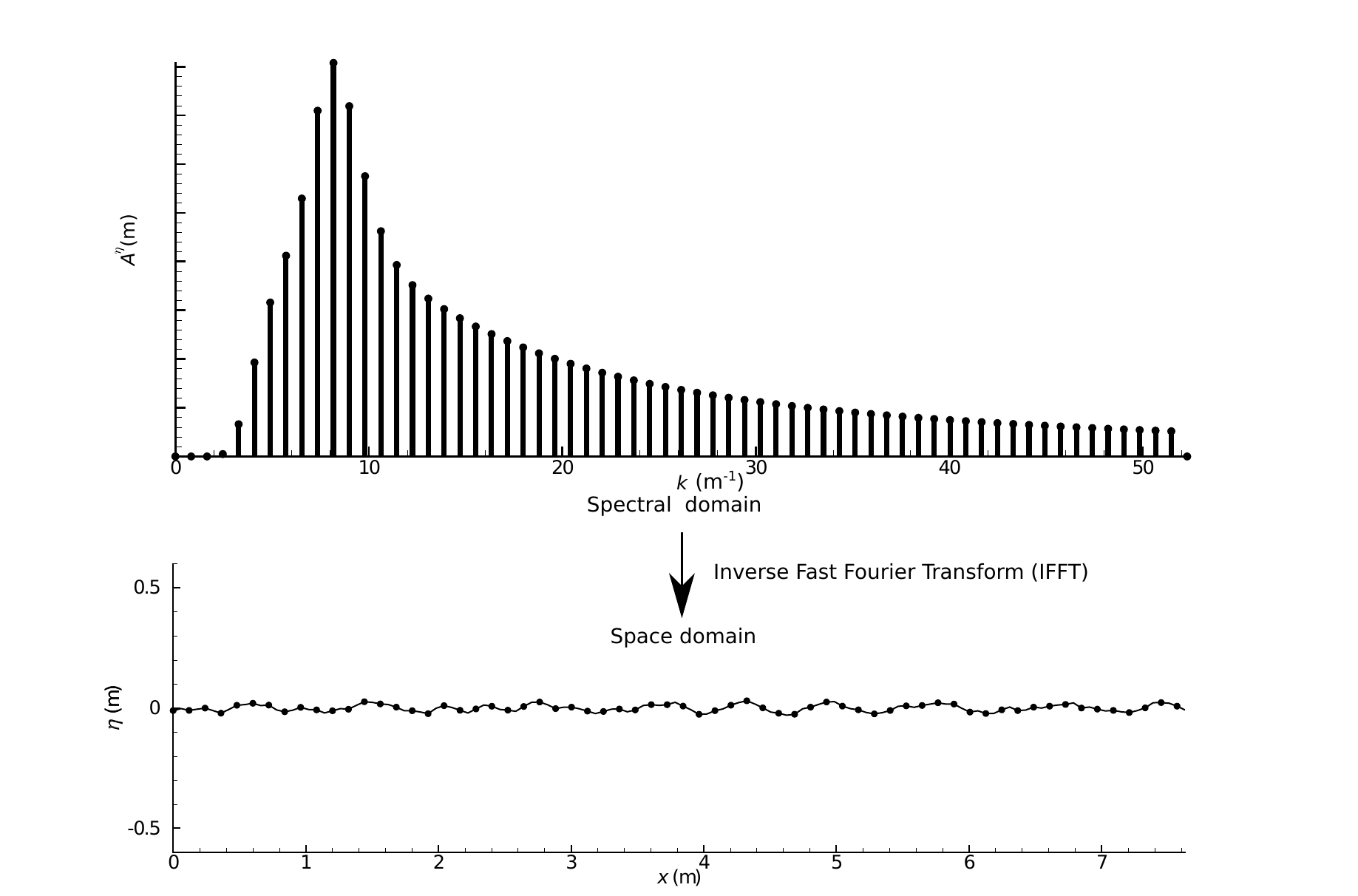}
    \caption{Reconstruction of the incident wave solution on the HOS grid. Top plot: free surface elevation ($\eta_I$) is stored in spectral domain as amplitudes of its Fourier components. Bottom plot: the modal information is transformed in the spatial domain with the Inverse Fast Fourier Transform (IFFT) on uniformly-spaced points.}
    \label{fig:hosModeToSpace}
\end{figure}
  \item The velocity field ($\bu_I(x,z,t)$) is reconstructed similarly. But since the velocity field also depends on the vertical direction ($z$), several IFFTs are used. Each  IFFT transforms $A_i^{\phi}$ to the velocity at $N$ horizontally uniformly distributed points with a given vertical position $z$.  After this reconstruction, the velocity field is available on a coarse rectangular grid (referred to as the "HOS grid" hereafter) with several horizontal layers. Each horizontal layer contains $N$ uniformly distributed points. Note the vertical position of these layers can be given arbitrarily, so that a vertical refinement is possible. In contrast, horizontal refinement cannot be  done directly since the number of points $N$ is equal to the number of modes in the HOS calculation.
\end{itemize}

  \item Interpolation: the incident wave information is interpolated from the HOS grid onto the CFD mesh, as the HOS grid is usually different from a CFD mesh. For example, a typical HOS simulation uses 10 points per peak wave length, while CFD solvers usually use around 100 cells. Figure \ref{fig:hosCFDWholeMesh} interpolated a comparison of a typical HOS grid (on the left) and a CFD mesh (on the right). The figure is colored by the velocity magnitude. The result on the CFD mesh is interpolated from the HOS grid behind it with the cubic spline scheme.
  \begin{figure}[h]
      \centering
        \includegraphics[width=0.45\textwidth]{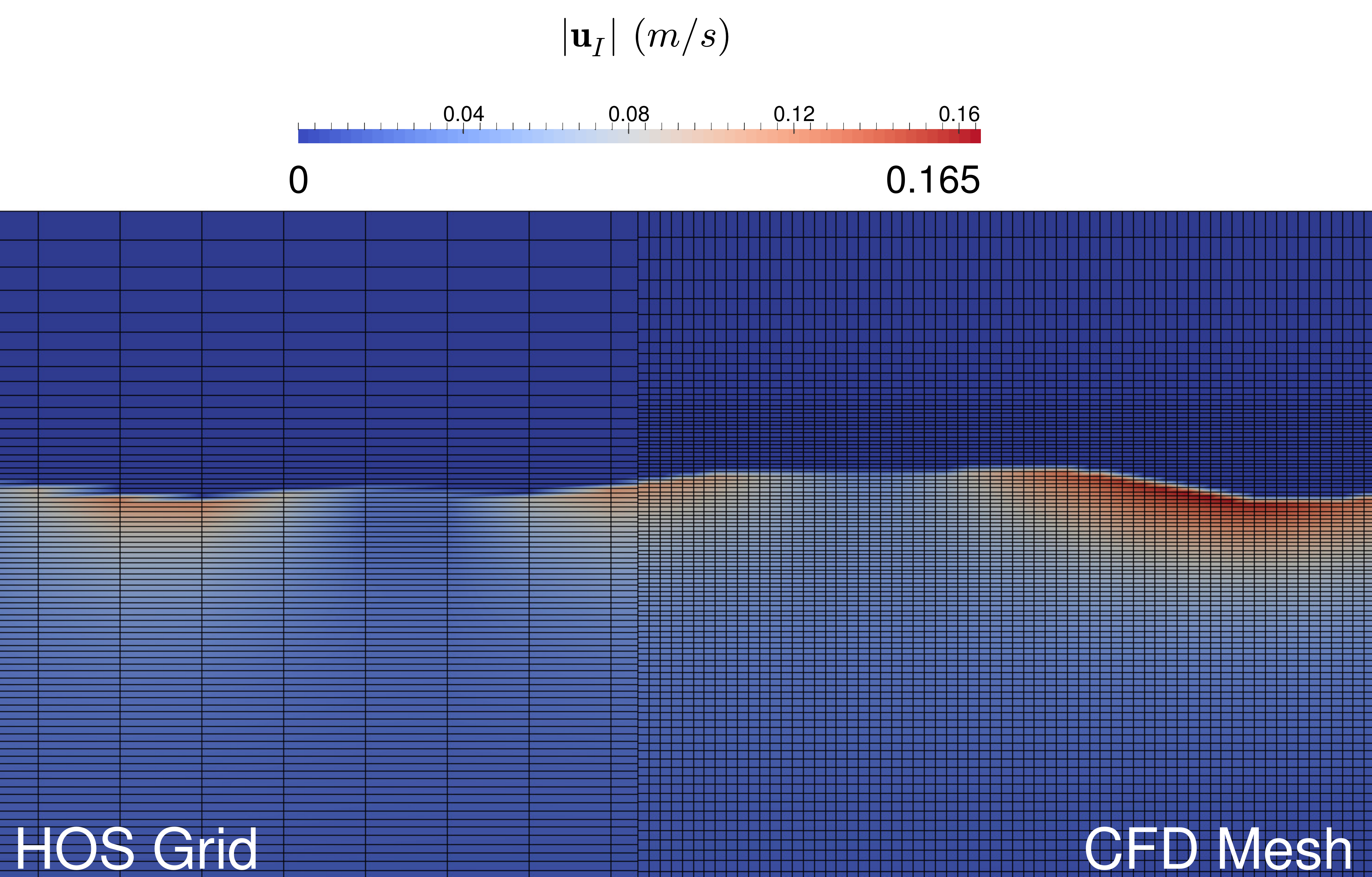}
        \caption{Comparison of a typical HOS grid having the same horizontal resolution used in the HOS simulation (left) and the mesh of a CFD solver (right).}
        \label{fig:hosCFDWholeMesh}
  \end{figure}
\end{enumerate}

\begin{figure}[h]

\begin{subfigure}{\textwidth}
    \centering
  \includegraphics[width=0.7\textwidth]{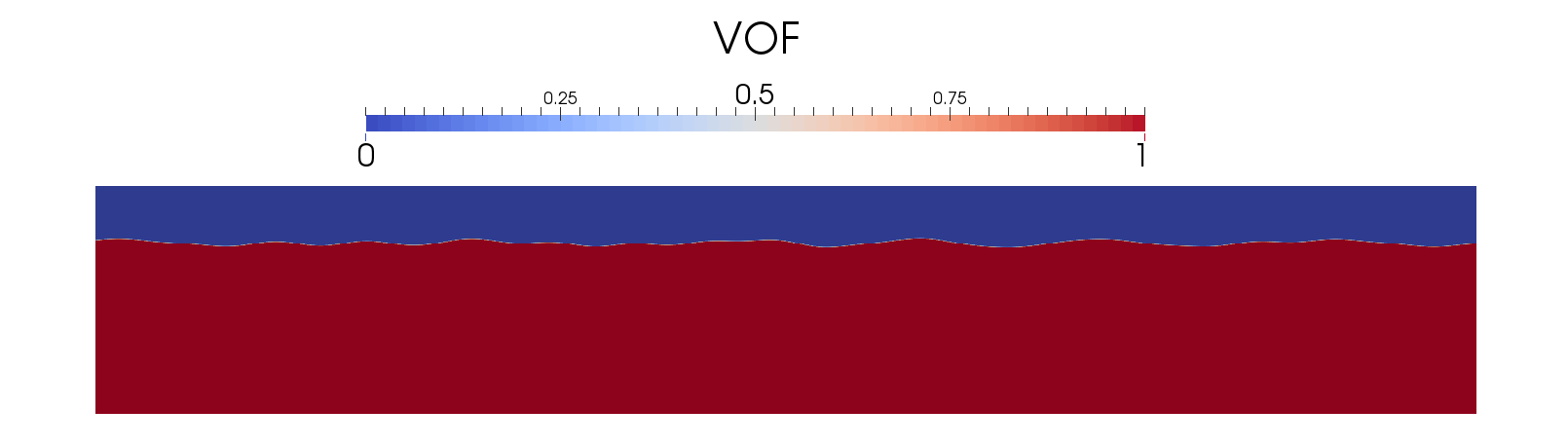}
  \caption{Volume of Fluid Field \label{fig:interpolatedVOFField}}
\end{subfigure}

\centering
\begin{subfigure}{\textwidth}
    \centering
  \includegraphics[width=0.7\textwidth]{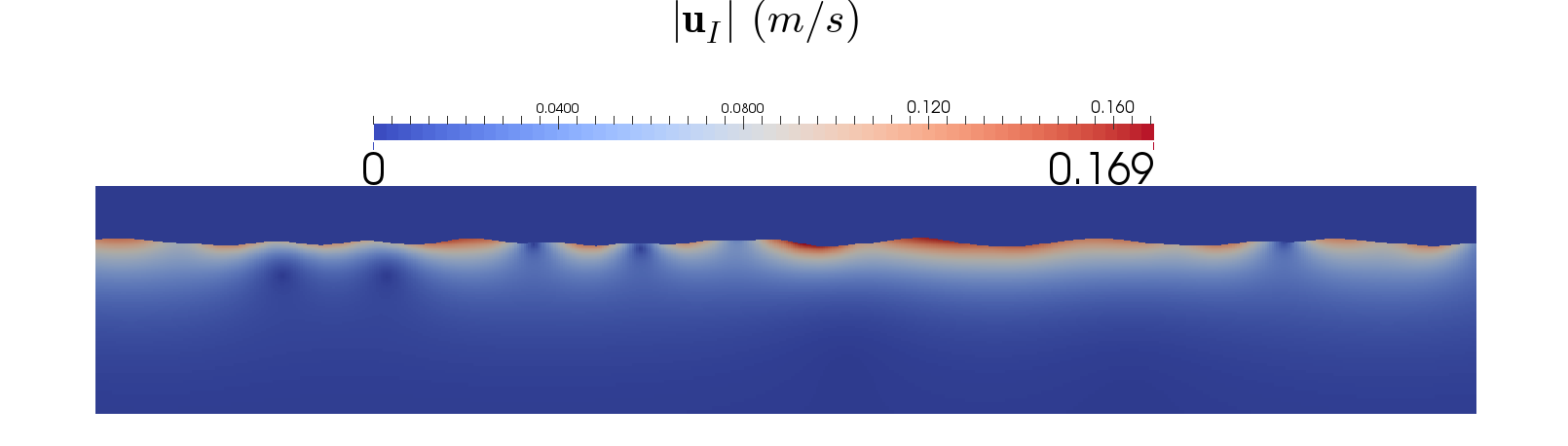}
  \caption{Interpolated velocity field under the free surface \label{fig:interpolatedVelocityField}}
\end{subfigure}

\caption{Irregular wave field interpolated from the HOS grid onto the CFD mesh.}
\label{fig:IrregularWaveReconstruction}

\end{figure}

An example of the interpolation result is shown in Fig.\ \ref{fig:IrregularWaveReconstruction}. The VOF field (on the top) is reconstructed from the free surface elevation ($\eta_I$). The velocity field is shown at the bottom. The waves are simulated by \textit{HOS-Ocean} (JONSWAP wave spectrum with $T_p =  0.7s, H_s = 0.028 m, \gamma = 3.3 $). The HOS simulation domain contains 10 peak wave lengths ($L_x=10\lambda_p$) with 128 Fourier components ($N=128$).

\subsubsection{Reconstruction error: velocity divergence}

This section analyzes the reconstruction error, represented by the velocity divergence ($\nabla. \bu_I $).  The physical significance of the velocity divergence is the volume change rate. Its unit is $s^{-1}$. Having $\nabla. \bu_I = a (s^{-1}) $ means that the fluid will have a relative volume change of $a$ in one second. If the reconstruction is accurate, the velocity divergence is equal to zero because the flow is incompressible.

The VOF method is very sensitive to the velocity divergence because the boundedness of the VOF field requires a divergence free velocity field. Such an error finally leads to instabilities in the simulation.

In the present work, the velocity field is reconstructed on Finite Volume mesh. The divergence is then calculated by:
\begin{equation}
\nabla.\bu_I=\sum_f \bu_{I,f}.\textbf{A}_f/\Delta V
\end{equation}
where $f$ represents the cell faces, $\bu_{I,f}$ represents the face-averaged incident velocity, $\mathbf{A}_f$ the surface vector of the cell face $f$, and $\Delta V$ the volume of the cell. It may contain numerical errors coming from:
\begin{enumerate}
  \item Approximating the face-averaged value by the face-center value ($\bu_{I,f} \approx \bu_I(\mathbf{x}_{cf})$), where $\mathbf{x}_{cf}$ is the face center coordinates.
  \item Evaluating $\bu_I(\mathbf{x}_{cf})$ by interpolation in the interpolation method.
\end{enumerate}

\begin{figure}
  \centering
   \includegraphics[width=0.7\textwidth]{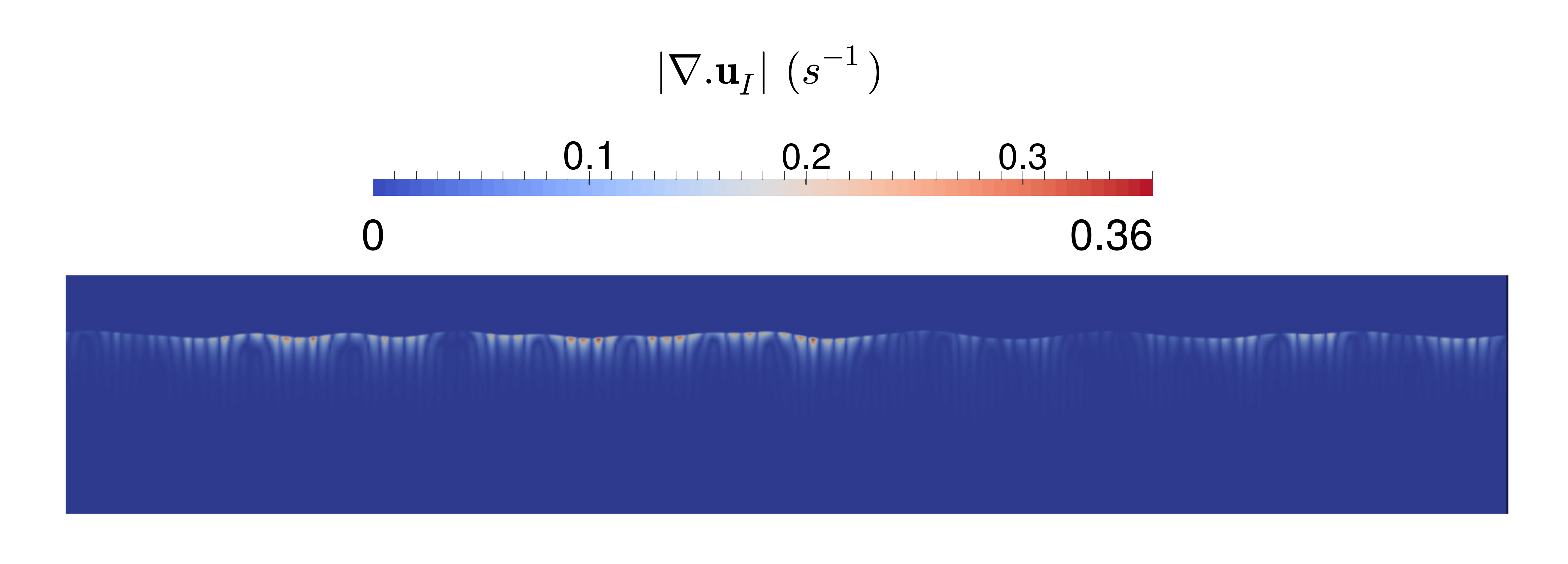}
   \caption{Divergence error of an interpolated HOS velocity field on a CFD mesh: HOS grid with 128 points. {\color{black}The error suggests the necessity to improve the interpolation accuracy.} \label{fig:DivergenceErrorHOS}}
\end{figure}

\begin{figure}
  \centering
   \includegraphics[width=0.72\textwidth]{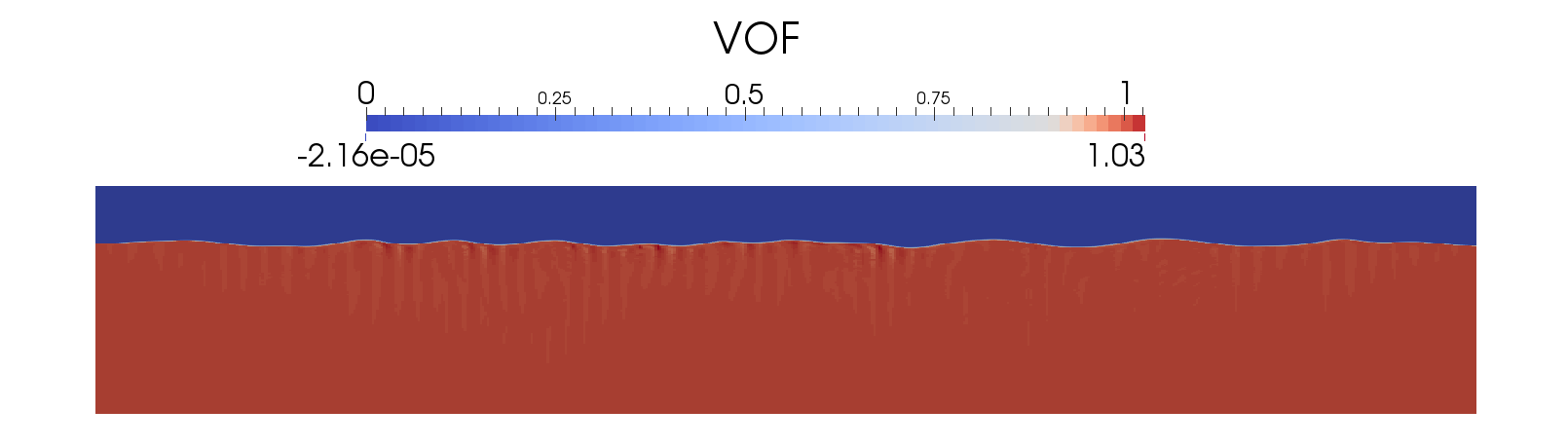}
   \caption{VOF field convected by the irregular wave field interpolated from the HOS method after one peak period. {\color{black}The unbounded VOF field suggests the necessity to improve the interpolation accuracy.} \label{fig:VOFErrorHOS}}
\end{figure}

The first source of error is due to the use of the second-order Finite Volume Method, which affects both the analytical evaluation and the interpolation method and is often negligible. In contrast, the second source of error is usually much larger.  Figure  \ref{fig:DivergenceErrorHOS} shows the divergence error of the interpolated velocity field of Fig.\ \ref{fig:IrregularWaveReconstruction}. The CFD mesh is discretized with $(\Delta x, \Delta z)=(\lambda_p/100,H_s/20)$.  The VOF field transported by this velocity field after one wave period is shown in Fig.\ \ref{fig:VOFErrorHOS}. Its values are no more bounded by 0 and 1. {\color{black}This problem suggests the necessity to improve the interpolation accuracy. }

\subsubsection{Improvement of the interpolation accuracy}


The main source of inaccuracy in the interpolation is the large space interval between HOS points. Refining the HOS grid is definitively helpful to reduce the interpolation error. However, using a grid as refined as a CFD mesh directly in the HOS computation is not feasible, since it introduces too many very short waves and may make the computation unstable \cite{ducrozet2017applicability}. To overcome this difficulty, we propose a refinement at the postprocessing stage. The method takes any HOS simulation result as input and allows to refine the grid to a user-defined level.

\begin{figure}[hp]
\centering
    \includegraphics[width=0.85\textwidth]{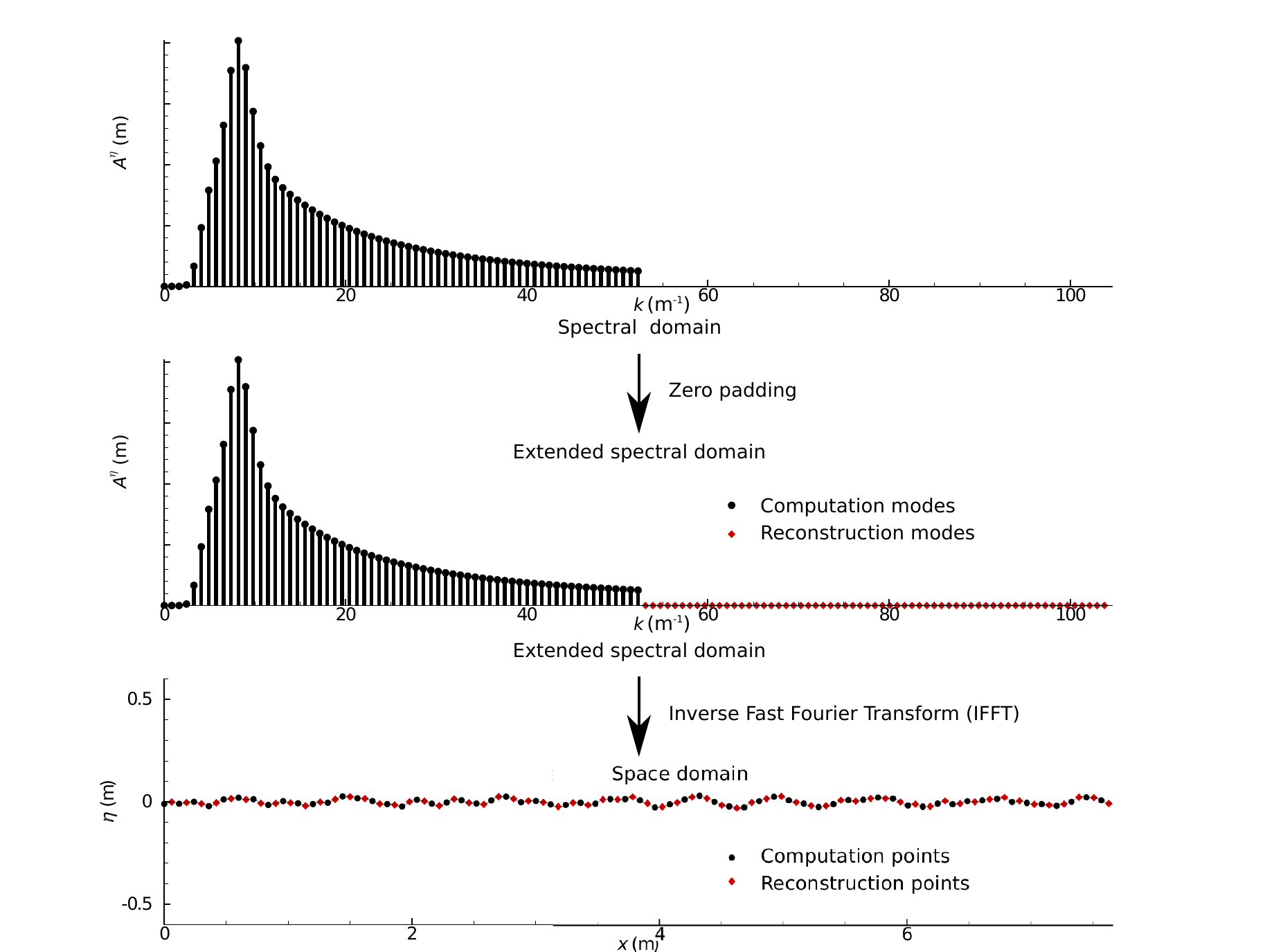}
    \caption{Zero-padding procedure for increasing the spatial resolution of the HOS results. The original spectrum of the HOS solution (on the top) is extended by adding extra modes of zero amplitude at the end (in the middle). It results in extra points in the space domain, and thus the HOS grid is refined.}
    \label{fig:hosModeToSpaceRefinement}
\end{figure}

This refinement is achieved with a zero-padding step in the spectral domain  before the IFFT (see Fig.\ \ref{fig:hosModeToSpaceRefinement}). After reading the amplitudes of Fourier components from an HOS result, the method extends the spectrum by adding extra modes with zero amplitude at the end. According to the relation between the spectral and the spatial domain in the FFT algorithm, additional Fourier components result in extra spatial points, i.e., if the spectrum is extended $n$ times, the distance between two HOS points can be reduced to $1/n$. Note that the factor $n$ has to be an integer to end up with the original spatial points plus $n-1$ additional points between two original points. In this way, the incident result is available on a finer grid and the interpolation error is reduced consequently.  Although more points are added, the efficiency is still greatly enhanced compared to the analytical evaluation, since this method still relies on the IFFT.

\begin{figure}[hp]
  \centering
    \includegraphics[width=0.6\textwidth]{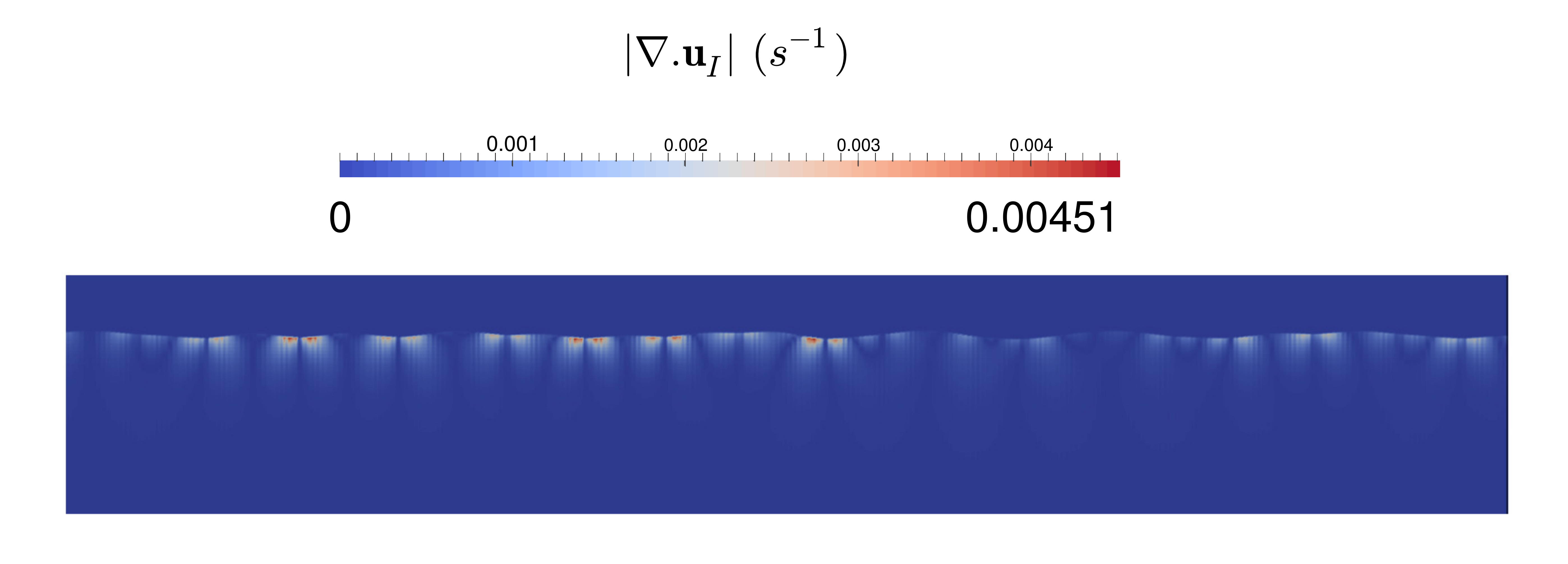}
    \caption{Divergence error of an interpolated HOS velocity field on a CFD mesh with the zero-padding improvement: HOS grid refined with a factor of 4.}
    \label{fig:IrregularWaveDivVelocityDirectAndRefinementComparison}
\end{figure}

\begin{figure}[hp]
	\centering
	\includegraphics[width=0.6\textwidth]{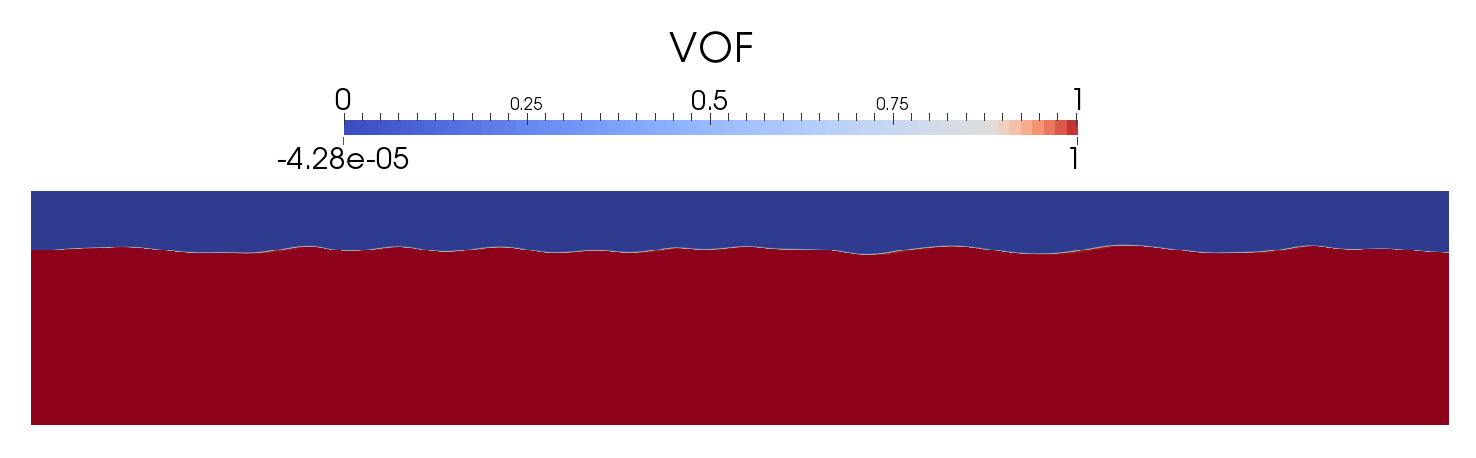}
	\caption{{\color{black} VOF field convected by the irregular wave field interpolated from the HOS method after one peak period with the zero-padding improvement: HOS grid refined with a factor of 4.}}
	\label{fig:IrregularWaveVOFAfterRefinement}
\end{figure}

To demonstrate the improvement in the interpolation accuracy, the divergence of the HOS velocity reconstructed by the zero-padding method is shown in Fig.\ \ref{fig:IrregularWaveDivVelocityDirectAndRefinementComparison}. The same CFD mesh and the same HOS simulation results are used as in Fig.\ \ref{fig:DivergenceErrorHOS}. The zero-padding method transfers the modal information on 512 reconstruction points by extending the Fourier modes by a factor of 4. Compared to Fig.\ \ref{fig:DivergenceErrorHOS} the divergence error is reduced by 2 orders of magnitudes. This level of accuracy appears to be enough in our experience to convect the VOF field in a stable way (compare Fig.\ \ref{fig:IrregularWaveVOFAfterRefinement} to  Fig.\ \ref{fig:VOFErrorHOS}).

\subsubsection{Code availability}

This reconstruction technique is included in the open-source library \textit{Grid2Grid}  \cite{2018arXiv180100026C,Grid2GridGitHub}. This library is developed by the LHEEA research department (Ecole Centrale Nantes and CNRS) to connect any CFD solver with \textit{HOS-Ocean} \cite{DUCROZET2016OpenSourceOcean,grilli2004numerical} and \textit{HOS-NWT} \cite{ducrozet2012hosNWT,HOSNWTGitHub}, two open-source HOS wave solvers. The library is published with the GNU General Public License v3.0 and can be downloaded from the url: \url{https://github.com/LHEEA/Grid2Grid}.

\subsection{Extension of incident wave solution in the air \label{sect:extensionAirVelocity}}

\begin{figure}[ht!]
 \centering
 \begin{subfigure}[b]{0.45\textwidth}
   {\includegraphics[trim={0 5cm 0 0cm},clip,width=\textwidth]{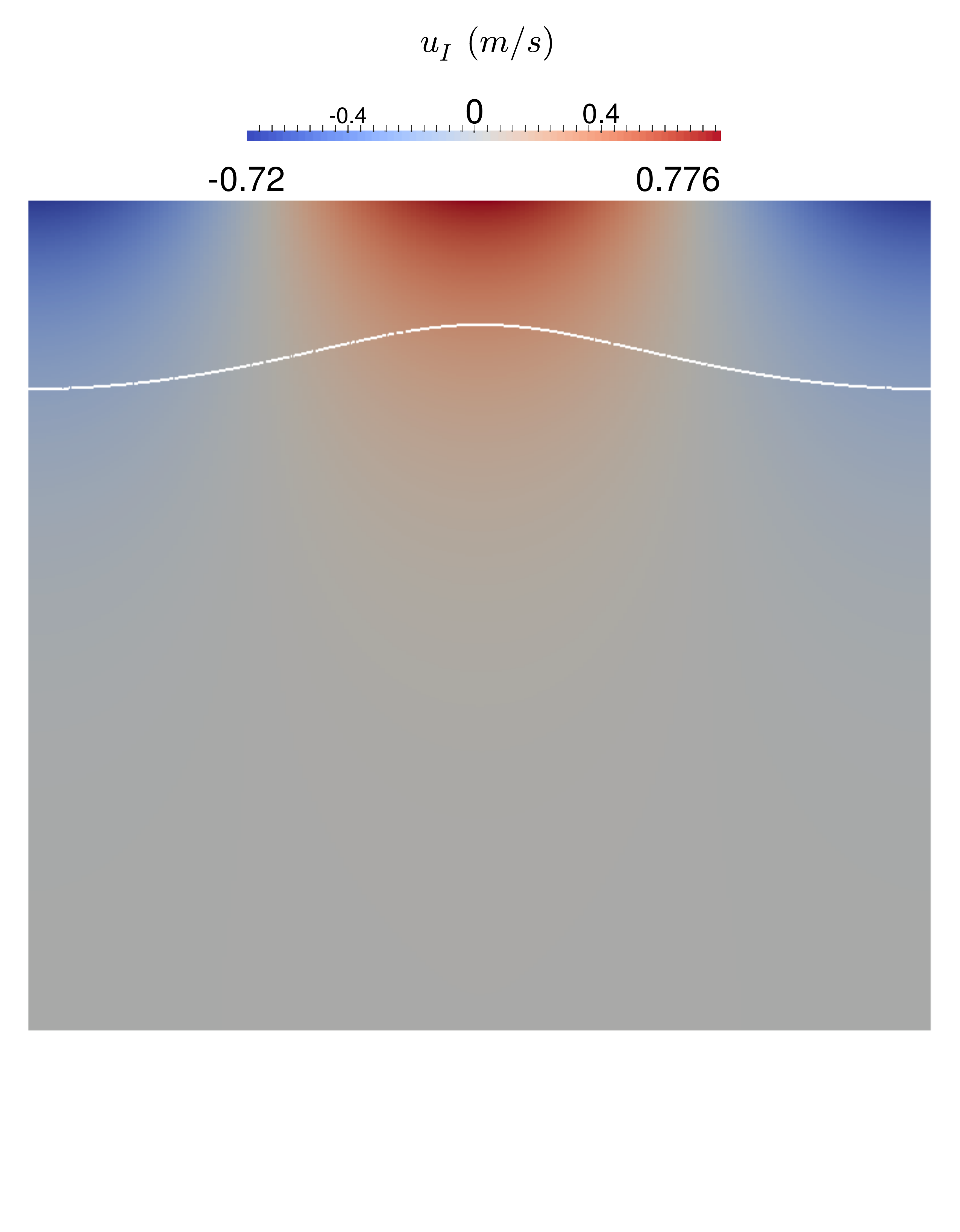}
   \caption{Horizontal velocity}}
 \end{subfigure}
 \begin{subfigure}[b]{0.45\textwidth}
   {\includegraphics[trim={0 5cm 0 0cm},clip,width=\textwidth]{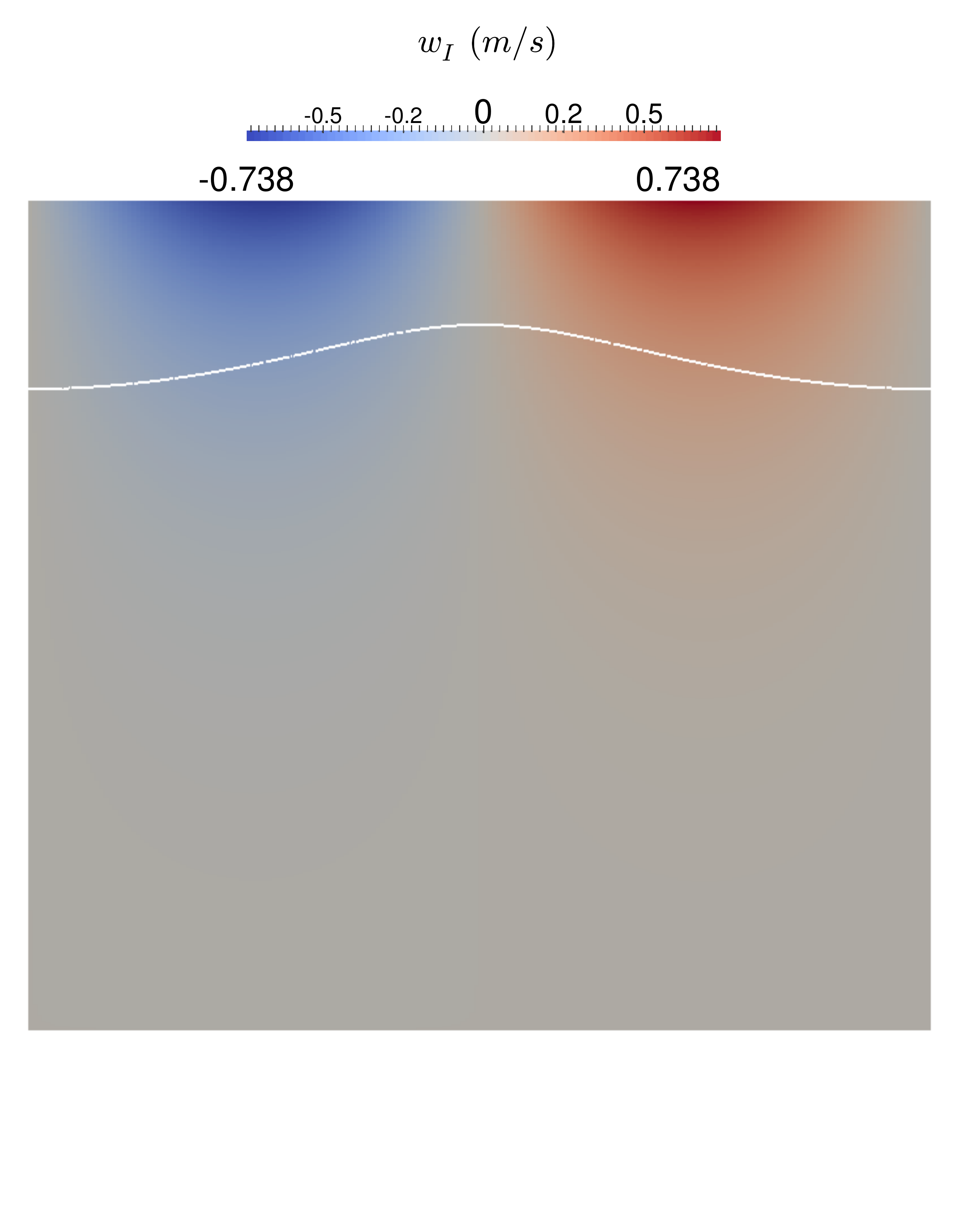}
   \caption{Vertical velocity}}
 \end{subfigure}
 \caption{Extended velocity field calculated by the stream function wave theory. The incident free surface is represented by the white line. ($T = 0.7s, H = 0.06m, h = 0.6m$)}
 \label{fig:streamfunctionExtended}
\end{figure}

The two-phase SWENSE method requires the incident solution both in water and in air. However, the wave model only defines the incident solution under the incident free surface position. To obtain the information above the free surface, the incident solution is extended by using Eqn.\ \eqref{eqn:whatIsSpectralMethod}, which gives an incident "solution" even above the free surface. Fig.\ \ref{fig:streamfunctionExtended} gives an example of the extended velocity field of a regular wave. The white line represents the free surface position of the incident wave. It is worth noting that the extended solution still satisfies the Euler equations (Eqn.\  \ref{eqn:EulerContinuity} and \ref{eqn:EulerNonconservative}). This property of the spectral method helps the SWENSE method to adjust the definition zone of the incident wave, so that the same functional decomposition can be used in the entire CFD computational domain.

This direct extension results in a smooth transition across the free surface, which is beneficial numerically. The drawback of such an extension is an oscillating "\textit{incident wind}" field in the air. {\color{black} By using the boundary conditions given in Tab. \ref{tab:SWENSE_BC}, this oscillatory wind remains in the total field. Although it is possible to correct this unrealistic phenomenon, (for example, by using no-slip BC on the top), such correction requires extra computational cost in the air, which goes against the objective to reduce CPU time.}  In the present work, we accept this oscillatory wind as a side effect of the SWENSE decomposition in the air, without any special treatment, since in most classical wave-structure interaction problems, the air-effects are often very weak, as it will be verified in Sect.\ \ref{sect:validationAndApplication}.

\newpage
\section{Implementation in OpenFOAM\label{sect:OpenFOAM}}
{\color{black}The present work uses the open-source CFD package OpenFOAM\footnote{https://openfoam.org/} to implement the proposed method. OpenFOAM uses the second-order Finite Volume method (FVM) with unstructured polyhedral meshes. All the variables are located at the cell center. Its native two-phase solver, \textit{interFoam}, adopts the VOF method \cite{HIRT1981VOF} and the Multi-Dimensional Limiter for Explicit Solution (MULES) algorithm \cite{deshpande2012evaluatingInterFoam}. The PIMPLE algorithm \cite{PIMPLE}, which combines the Pressure Implicit with Splitting of Operator (PISO) and the Semi-Implicit Method for Pressure-Linked Equations (SIMPLE) \cite{versteeg2007introduction}, is used to obtain converged results of the velocity-pressure-VOF coupling at each time step. The readers are referred to \cite{deshpande2012evaluatingInterFoam,ubbink1999jcp} for more details of \textit{interFoam}. The marine hydrodynamics community of OpenFOAM has developed codes based on \textit{interFoam} by adding wave modeling techniques, \textit{e.g.}, \textit{waves2Foam} \cite{jacobsenFuhrmanFredsoe2012}, \textit{foamStar} \cite{monroy2016}, etc.}

The proposed two-phase SWENSE method is implemented on top of \textit{foamStar} \cite{monroy2016} developed by Bureau Veritas and Ecole Centrale Nantes. The new solver is named as \textit{foamStar-SWENSE}. The only difference between the two solvers is that the NS equations in \textit{foamStar} are replaced by the SWENS equations in \textit{foamStar-SWENSE}.

\subsection{Equations discretization}
This section briefly describes how the governing equations are discretized. Standard schemes in OpenFOAM are used, as listed in the end of the section. More details of these schemes can be found in the literature \cite{Moukalled2016FVMOpenFOAM,ferzigerPericComputational}.

For convenience, the sign $[\cdot]$ is used to indicate that the terms enclosed are treated implicitly. Otherwise, the terms are evaluated explicitly.

\subsubsection{VOF equation}
The discretization of Eqn.\ \eqref{eqn:alphaTransportSwense} is standard as in \textit{interFoam}:
\begin{equation}
    \left[\frac{\partial \alpha}{ \partial t}\right] +  \left[\nabla.(\textbf{u}\alpha)\right] + {\color{black}\nabla .\left(\alpha(1-\alpha)\bu_r\right)}= 0 \label{eqn:DiscretizedAlphaTransportSW}
\end{equation}
The MULES algorithm {\color{black}\cite{deshpande2012evaluatingInterFoam}} is used to keep the boundedness of the $\alpha$ field.

\subsubsection{Momentum equation}
Eqn.\ \eqref{eqn:twophaseSwMoment}, rewritten in its conservative form, is discretized as follows:
\begin{equation}
\left[\frac{\partial \textbf{u}_C}{ \partial t}\right] + [\nabla.(\textbf{u}\otimes \textbf{u}_C)]+  \nabla. (\textbf{u}_C \otimes \textbf{u}_{I})
-\frac{[\nabla.(\mu_{eff} \nabla\textbf{u}_C)] + (\nabla.(\mu_{eff} \nabla\textbf{u}_C)^T)}{\rho} = -\frac{\nabla p_C}{\rho} - \frac{p_I}{\rho_I}\frac{\nabla\rho}{\rho} \label{eqn:DiscretizedTwophaseSwMoment}
\end{equation}
where $\mu_{eff} = \mu + \mu_t$.
The time derivative, the convection, and the diffusion terms of the $\textbf{u}_C$ are discretized implicitly. The rest terms are explicitly evaluated.

Note that the R.H.S. terms are evaluated with a reconstruction operation from face flux \cite{fvcReconstruction, fvcReconstruction2}, to introduce a pseudo-staggered grid setup and to avoid checker-board pressure oscillations that may occur on co-located grids. For example, $\displaystyle{\frac{\nabla p_C}{\rho}}$ is evaluated at the cell face $\displaystyle{\frac{(\nabla p_C)_f}{\rho_f}}$ and reconstructed to the cell center.  The pressure $(\nabla p_C)_f$ at the cell face is directly calculated with the pressure at the two neighbor cell centers. The face density $\rho_f$ is interpolated as:
\begin{equation}
  \rho_f =\frac{\rho_w|\Delta x_w| + \rho_a|\Delta x_a|}{|\Delta x_w + \Delta x_a |}
  \label{eqn:interfaceRhoInterpolation}
\end{equation}
where $|\Delta x_w|$ and $|\Delta x_a|$ denote the distance of cell center to the interface of water cell and air cell respectively. This interpolation scheme mimics the Ghost Fluid Method for two-phase incompressible flow \cite{lalanne2015computation}. $\displaystyle{\frac{p_I}{\rho_I}\frac{\nabla\rho}{\rho}}$ is evaluated similarly.

Eqn.\ \eqref{eqn:DiscretizedTwophaseSwMoment}, written in a semi-discretized form for a cell $P$ reads:

\begin{equation}
a_{P}\mathbf{u}_{C,P}+\sum_{N}a_{N}\mathbf{u}_{C,N}= \mathbf{S}-\frac{\nabla p_C}{\rho} -\frac{p_I}{\rho_I}\frac{\nabla\rho}{\rho} \label{eqn:algebraicMomentumEquationSW}
\end{equation}
where the subscript $P$ and $N$ represent the value of the current cell and the values of its neighbors. The coefficients of the discretized system are $a_P$ and $a_N$. $\mathbf{S}$ represents the source terms in the discretized equation except for the complementary pressure term and the interface term.


In OpenFOAM, the solution step of Eqn.\  \eqref{eqn:algebraicMomentumEquationSW} is called the momentum prediction. It provides a momentum conserving velocity field $\textbf{u}_C$. However, the result of the velocity prediction does not guarantee the incompressibility of the field. A correction is necessary via the pressure equation step.

\subsubsection{Pressure equation}

The discretized pressure equation in the SWENSE solver is derived in the same way in as in standard incompressible flow OpenFOAM solvers. The semi-discretized form of the momentum equation Eqn.\ \ref{eqn:algebraicMomentumEquationSW}, as follows,
\begin{equation}
    a_P\mathbf{u}_{C,P} = H(\mathbf{u}_C) - \frac{\nabla p_C}{\rho} -\frac{p_I}{\rho_I}\frac{\nabla\rho}{\rho}
    \label{eqn:semi-discretizedMonemtumEquationSW}
\end{equation}
where
\begin{itemize}
\item $a_p\mathbf{u}_{C,P}$ is the diagonal contribution of Eqn.\  \eqref{eqn:algebraicMomentumEquationSW},
\item $\displaystyle{- \frac{\nabla p_C}{\rho}}$ is the contribution of the complementary pressure gradient,
\item $\displaystyle{-\frac{p_I}{\rho_I}\frac{\nabla\rho}{\rho}}$ is the contribution of interface density gradient,
\item $H(\mathbf{u}_{C,P})$ is the off-diagonal contribution of the matrix and the source terms $\mathbf{S}$ in Eqn.\  \eqref{eqn:algebraicMomentumEquationSW}:
\begin{equation}
    H(\mathbf{u}_C) = - \sum_{N}a_{N}\mathbf{u}_{C,N} + \mathbf{S}
\end{equation}
\end{itemize}

From Eqn.\  \eqref{eqn:semi-discretizedMonemtumEquationSW}), the complementary velocity at the center of the cell is:

\begin{equation}
  \mathbf{u}_{C,P} = \frac{H(\mathbf{u}_C)}{a_P} - \frac{1}{a_P}\frac{\nabla p_{C}}{\rho}-\frac{1}{a_P}\frac{p_I}{\rho_I}\frac{\nabla\rho}{\rho}
\end{equation}

Interpolating this value to the face center, yields:
\begin{equation}
    \mathbf{u}_{C,f}=\overline{\mathbf{u}_{C}} = \overline{\left(\frac{H(\mathbf{u}_C)}{a_P}\right)} -\overline{\left(\frac{1}{a_P}\right)} \frac{(\nabla p_C)_f}{\rho_f} -\overline{\left(\frac{1}{a_P}\right)} \frac{p_{I,f}}{\rho_I}\frac{(\nabla\rho)_f}{\rho_f} \label{eqn:RhieChow}
\end{equation}
where the $\overline{\,\cdot\,}$ symbol denotes the value on the cell face, which is linearly interpolated from the cell center of both sides.

Substituting Eqn.\ \eqref{eqn:RhieChow} into Eqn.\ \eqref{eqn:SwContinuity2}, the discretized pressure equation is written as:
\begin{equation}
    \left[\nabla.\left(\overline{\left(\frac{1}{a_P}\right)} \frac{(\nabla p_C)_f}{\rho_f}\right)\right] = \nabla. \left(\overline{\left(\frac{H(\mathbf{u}_C)}{a_p}\right)} -\overline{\left(\frac{1}{a_P}\right)} \frac{p_{I,f}}{\rho_I}\frac{(\nabla\rho)_f}{\rho_f} \right) \label{eqn:discretizedPressureSW}
\end{equation}

This Poisson equation is used to determine the complementary pressure field and to correct the complementary velocity flux. At last, the flux is used to reconstruct the cell-centered $\mathbf{u}_C$.

{\color{black}
\subsubsection{Discretization Schemes}

The discretization schemes are listed in Tab. \ref{tab:discretizationSchemes}. The time derivative is discretized with second-order Crank-Nicolson scheme and blended with first-order Euler implicit scheme to compromise between the accuracy and the stability. Spatial schemes are also 2nd-order in general, but first-order schemes will be used locally to avoid over-shoots (or under-shoots).

\begin{table}[h!]

\centering
\caption{The discretization schemes in foamStar-SWENSE \label{tab:discretizationSchemes}}
\resizebox{\textwidth}{!}{%
\begin{tabular}{lll}
  \hline
 & Term & Discretization scheme \\
 \hline
\textbf{Time derivative} & default & Crank Nicolson (2nd-order) blended with Euler Implicit (1st-order) \\
\textbf{Gradient} & default & cellLimited leastSquares (2nd-order with 1st order limiter) \\
 & $\nabla \alpha$ & Gauss linear (2nd-order) \\
\textbf{Divergence} & default & Gauss linear (2nd-order) \\
 & $\nabla \cdot (\bu \otimes \bu_C) $ & Gauss linearUpwindV (2nd-order with 1st-order limiter) \\
 & $\nabla \cdot (\bu \alpha) $ & Gauss vanLeer01 (2nd-order with 1st-order limiter) \\
\textbf{Laplacian} & default & Gauss linear limited correct 0.5 (2nd-order with 1st-order limiter) \\
\textbf{Interpolation} & default & linear (2nd-order) \\
 & $\rho_{c \rightarrow f}$ & linear and interface density interpolation (Eqn. \eqref{eqn:interfaceRhoInterpolation}) (2nd-order) \\
\textbf{Surface normal gradient} & default & limited corrected 0.5 (2nd-order with 1st-order limiter) \\
 \hline
\end{tabular}%
}
\end{table}

}

\subsection{Prevention of wave reflections with the relaxation zone technique}
\begin{figure}[ht!]
\centering
\includegraphics[width=9cm]{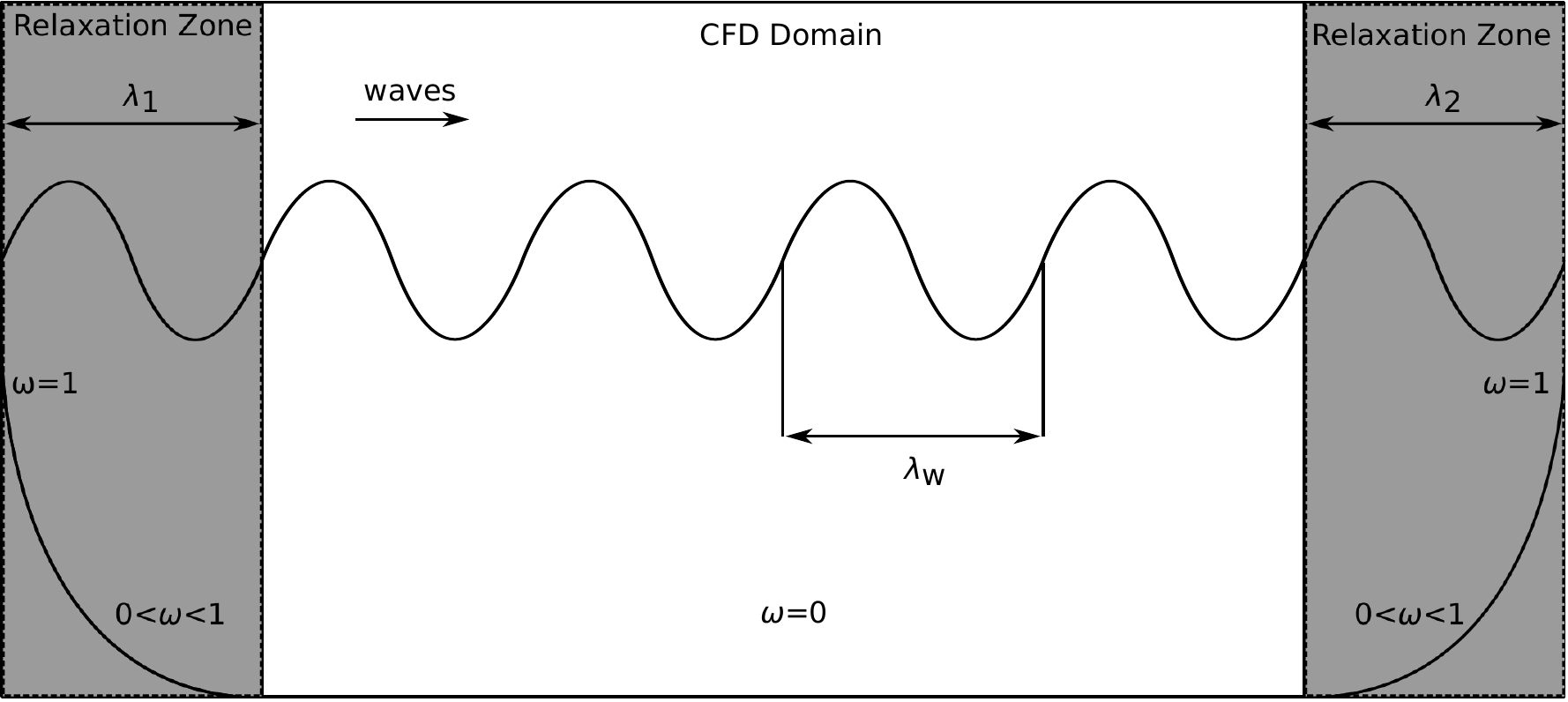}
\caption{Relaxation zone technique principle} \label{fig:relaxationZone}
\end{figure}

A relaxation zone technique is used to prevent wave reflections at the computational domain boundaries. It defines regions where the computed value is gradually blended to the target value using a space-dependent weight function $\omega$ as shown in Fig.\ \ref{fig:relaxationZone}. For a given quantity $\chi$, the relaxed value $\chi_{relax}$ in these regions is defined as the linear combination of the numerical solution $\chi_{CFD}$ and the target value $\chi_{target}$, as follows:
\begin{equation}
\chi_{relax}=\omega\chi_{target}+(1-\omega)\chi_{CFD}  \label{eqn:relaxationZone}
\end{equation}
This relaxation step is applied at the end of each time step after the solution of CFD solver has converged. The relaxed value is then used in the further simulation.

The weight function is defined in a way such that it is equal to 1 on the boundary and 0 in the full CFD domain. The present implementation follows reference \cite{monroy2016}. The weight function reads,
\begin{equation}
\omega = -2 x^3 + 3x^2 \label{eqn:weightFunction}
\end{equation}
where $x$ is the non-dimensional distance to the inner boundary of the relaxation zone. $x = 1$ corresponds to the outer boundary of the relaxation zone. $x = 0$ corresponds to the interface between the pure CFD and the relaxation zone.

In the relaxation zones, the complementary velocity field $\mathbf{u}_C$ is gradually attenuated to zero, and the VOF field $\alpha$ is blended to the target value of incident wave field $\alpha_I$, as follows:

\begin{eqnarray}
  \mathbf{u}_{C,relax} &=& \omega\mathbf{0}+(1-\omega)\mathbf{u}_{C,CFD} \label{eqn:swUcRelax} \\
  \alpha_{relax} &=& \omega\alpha_{I} + (1-\omega)\alpha_{CFD} \label{eqn:swVOFRelax}
\end{eqnarray}
since the SWENSE method assumes that the complementary waves {\color{black}vanish} in the far-field. 
The pressure is not relaxed by this method as it is solved implicitly at each step.

\subsection{Solution algorithm}
\begin{figure}[hp!]
\centering
\scalebox{0.8}
{
\begin{tikzpicture}[node distance=2cm]
\node (newTimeStep)[startstop]{Start new time step};
\node (updateIncidentProperites) [process,below of=newTimeStep] {Update Incident waves $\mathbf{u}_I$ and $p_I$};
\node (pimpleStart) [startstop,below of=updateIncidentProperites]  {Start $\alpha-\mathbf{u}_C-p_C$ coupling (PIMPLE)};
\node (Grid2Grid)[io,right of = updateIncidentProperites,xshift=4cm]{Map the incident wave solutions};
\node (reconstructVelocity) [process,below of=pimpleStart] {Reconstruction $\mathbf{u} = \mathbf{u}_I + \mathbf{u}_C$};
\node (solveAlpha) [process,below of=reconstructVelocity] {Solve for $\alpha$ (Eqn.\ \ref{eqn:DiscretizedAlphaTransportSW})};
\node (updateProperties) [process,below of=solveAlpha] {Update $\rho$, $\nu$, and $p_I^*$ (Eqns.\ \ref{eqn:rhoEquation} and \ref{eqn:partial} and \ref{eqn:pIStar})};

\node (pisoStart) [startstop,right of=updateProperties,xshift=4cm] {Start $\mathbf{u}_C-p_C$ coupling (PISO)};
\node (solveU) [process, below of=pisoStart] {Solve for $\mathbf{u}_C$ (Eqn.\ \ref{eqn:DiscretizedTwophaseSwMoment})};
\node (solveP) [process, below of=solveU] {Solve for $p_C$ (Eqn.\ \ref{eqn:discretizedPressureSW})};
\node (correctU) [process, below of=solveP] {Update velocity $\mathbf{u}_C$ };
\node (pisoConverge) [decision,below of=correctU,yshift=-1cm] {Max iteration \\ reached? };
\node (turbulence) [process, left of=pisoConverge,xshift=-4cm] {Solve for turbulence};
\node (Relaxation) [process, below of=turbulence] {Relaxation Zones (Eqn.\  \ref{eqn:swUcRelax} and \ref{eqn:swVOFRelax})};
\node (pimpleConverge) [decision,below of=Relaxation,yshift=-1.5cm] {PIMPLE \\ converged?};
\node (nextTime) [startstop,right of=pimpleConverge,xshift=4cm] {Next time step};
\draw [arrow] (newTimeStep)--(updateIncidentProperites);
\draw [arrow] (updateIncidentProperites)--(pimpleStart);
\draw [arrow] (pimpleStart)--(reconstructVelocity);
\draw [arrow] (reconstructVelocity)--(solveAlpha);
\draw [arrow] (solveAlpha)--(updateProperties);
\draw [arrow] (solveAlpha)--(updateProperties);
\draw [arrow] (updateProperties)--(pisoStart);
\draw [arrow] (pisoStart)--(solveU);
\draw [arrow] (solveU)--(solveP);
\draw [arrow] (solveP)--(correctU);
\draw [arrow] (correctU)--(pisoConverge);
\draw [arrow] (pisoConverge)--node[anchor=south]{yes}(turbulence);
\draw [arrow] (turbulence)--(Relaxation);
\draw [arrow] (Relaxation)--(pimpleConverge);
\draw [arrow] (pimpleConverge)--node[anchor=south]{yes}(nextTime);

\draw [dashedarrow] (pimpleConverge.west) --node[anchor=south]{no}([xshift=-1cm]pimpleConverge.west) |-(reconstructVelocity.west);
\draw [dashedarrow] (pisoConverge.east) --node[anchor=south]{no}([xshift=1cm]pisoConverge.east) |-(solveP.east);

\draw[dashedarrow](Grid2Grid)--(updateIncidentProperites);

\end{tikzpicture}
}

\caption{Solution algorithm of the SWENSE method} \label{fig:SolutionAlgorithmSW}
\end{figure}
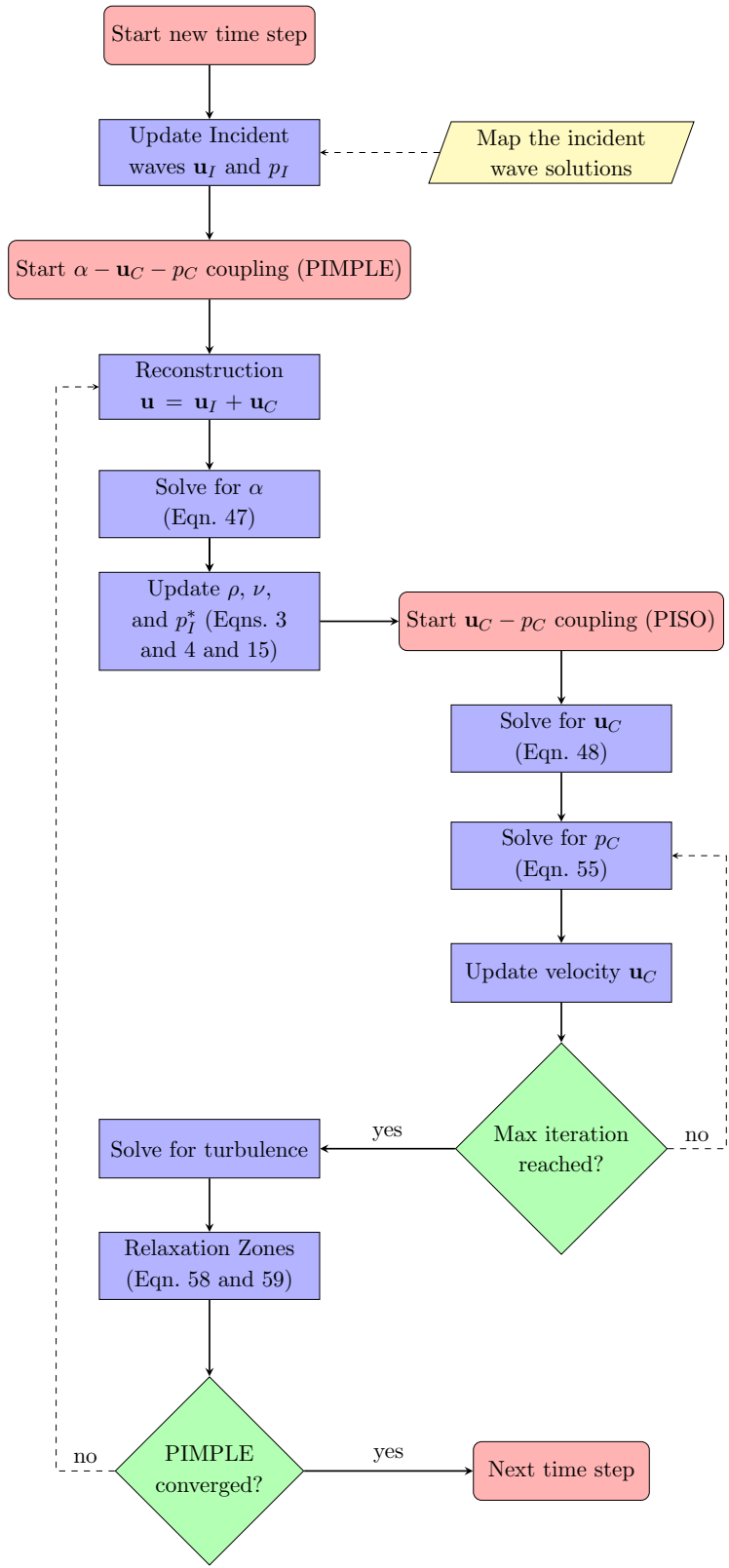

The structure of the SWENSE solver is provided in the flowchart of Fig.\  \ref{fig:SolutionAlgorithmSW}.
\begin{enumerate}
  \item At the beginning of each time step, incident wave properties $\textbf{u}_I$ and $p_I$ are updated from potential wave solvers and mapped on the CFD mesh. The total velocity is reconstructed with $\bu = \bu_I + \bu_C $.
  \item Solve for the VOF field $\alpha$ (Eqn.\  \ref{eqn:DiscretizedAlphaTransportSW}); update the fluid properties with the new VOF field (Eqn.\  \ref{eqn:rhoEquation} and \ref{eqn:partial}). Update the modified incident wave pressure ($p_I^*$) with the new density field (Eqn.\ \ref{eqn:pIStar}).
  \item In the PISO loop, the complementary velocity field is solved from Eqn.\  \eqref{eqn:DiscretizedTwophaseSwMoment} at first (the momentum prediction step). Secondly, the complementary pressure $p_C$ is solved from Eqn.\  \eqref{eqn:discretizedPressureSW}. The flux of the complementary velocity is corrected by the pressure field $p_C$.  After the correction, the flux is used to reconstruct the complementary velocity field $\textbf{u}_C$ at the cell center. The PISO loop iterates until {\color{black} reaching the maximum iteration number (user defined value, set to 6 in the present cases)}.
  \item The solution is then blended to the target values in the relaxation zones to attenuate the complementary waves in the far-field (Eqns.\  \ref{eqn:swUcRelax} and \ref{eqn:swVOFRelax})
  \item Outer nonlinear iterations are made to achieve the convergence of the VOF, the velocity, and the pressure before stepping to the next time. {\color{black}The convergence criterion is that the residual of $\bu_C$ is reduced 3 orders of magnitude.}
\end{enumerate}

\section{Validation and Application\label{sect:validationAndApplication}}

In this section, the two-phase SWENSE method is validated on three test cases:
\begin{itemize}
\item Progressive waves: This test case consists simply in a simple regular wave propagation in a periodic domain. No structure is present. Different discretizations are used to check whether a SWENSE solver can keep incident waves accurate with coarse discretizations. A comparison with the conventional NS solver \textit{foamStar} is also provided.
\item High-order wave loads on a vertical cylinder: This case validates the method on the calculation of the wave force on a fixed structure with simple geometry. The horizontal wave force on the cylinder is recorded. Up to fourth harmonic components are extracted and compared with experimental data and numerical results of potential flow solvers. A systematic convergence study is also conducted.
\item Fixed Catenary Anchored Leg Mooring (CALM) buoy with a heave-damping skirt in regular and irregular waves: This configuration is used to assess the method with a complex geometry and includes violent free surface deformation and viscous effects representing a real ocean engineering application. The results of \textit{foamStar-SWENSE} are compared with experimental data. For the regular wave case, comparative simulations conducted with \textit{foamStar} are used to demonstrate the efficiency and accuracy of the proposed method.
\end{itemize}

\subsection{Progressive waves\label{sect:progressiveWaves}}

Good quality of the incident waves is the first requirement for wave-structure interaction simulations. NS solvers (e.g., \textit{foamStar}), which simulate the wave system with a direct method, need fine meshes to propagate waves accurately. Instead, incident wave propagation in SWENSE solvers is much easier because the incident waves are explicitly given by wave models. In a SWENSE solver, maintaining the accurate incident waves is nothing more than keeping the complementary field is equal to zero. Although such a property of the proposed two-phase SWENSE method has been demonstrated theoretically in Sect.\ \ref{sect:swenseEquationsAbility}, numerical errors may exist in CFD simulations. Therefore, this case tests the real behavior of  \textit{foamStar-SWENSE}, taking all the probable numerical errors into account.

The case simulates regular wave propagation in a 2D periodic domain. Periodic boundary conditions at inlet and outlet are used to get rid of the wave generation and absorption issues and to focus on the wave propagation. The case is intentionally designed with a large computational domain and a long time duration to increase the numerical errors, in order to test the stability and the accuracy of the method.

\textit{foamStar-SWENSE} and \textit{foamStar} are tested and compared  with several identical spatial and temporal discretizations.

\subsubsection{Test case setup}

Regular waves with a moderately large wave steepness $(ka = 0.22)$ are simulated. The wave characteristics are listed in Tab.\ \ref{tab:WaveInputParemeteres}. The stream function wave theory \cite{rienecker1981fourier,CN_stream} is used to generate the reference solution and provide the incident wave information to \textit{foamStar-SWENSE}.

\begin{table}[ht]
\begin{center}
\caption{Progressive wave test case: wave parameters}
  \begin{tabular}{l r}
    \toprule
    Parameter & Value \\
    \midrule
    Wave period $ (T) $& $  0.70 \,\text{s} $ \\
    Wave height $ (H=2a) $& $ 0.058 \,\text{m} $ \\
    Water depth $ (h) $& $  0.60 \, \text{m} $ \\
    Wave steepness $ (ka) $& $  0.22 $ \\
    Relative water depth $(kh)$&  $4.7$ \\
    \bottomrule
  \end{tabular}
  \label{tab:WaveInputParemeteres}
\end{center}
\end{table}

\begin{figure}[ht]
  \begin{center}
  \includegraphics[width=0.4\paperwidth]{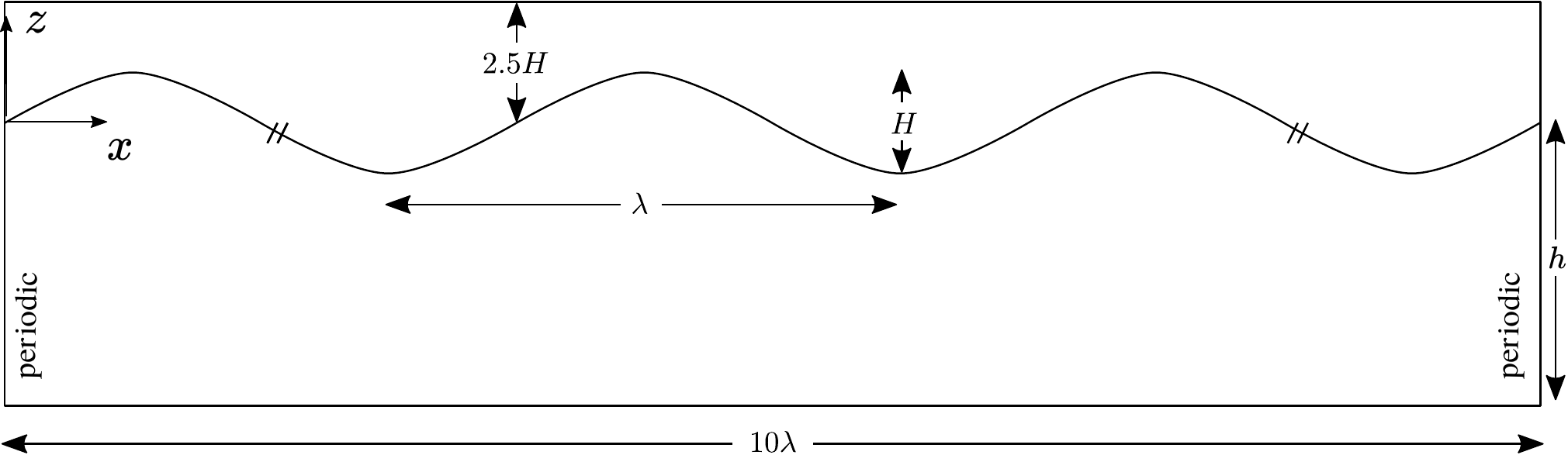}
  \end{center}
  \caption{Progressive waves test case: computational domain}
  \label{fig:WaveNumericalSettings}
\end{figure}

The simulations use a two-dimensional rectangular computational domain (see Fig.\ \ref{fig:WaveNumericalSettings}). The waves travel from the left to the right. Periodic boundary conditions are applied on the left and the right boundaries. The origin of the coordinate system is located at the left of the computational domain on the free surface position at rest. The axis $x$ points right and the axis $z$ points up. The length of the computational domain is equal to ten wave lengths ($\lambda$), and the height of the computational domain is equal to the water depth ($h$) plus 2.5 times the wave height ($H$). The simulation time is equal to 20 wave periods.

Three sets of space and time resolutions are used (see Tab.\ \ref{tab:NumericalSettings}). The medium discretization corresponds to the typical mesh used by two-phase VOF solvers for the wave-structure interaction problems.

\begin{table}[h!]
\begin{center}
\caption{Progressive waves test case: numerical resolutions}
  \begin{tabular}{l c c c}
    \toprule
    Parameter & Coarse & Medium & Fine \\
    \midrule
    Mesh size $ (\Delta x, \Delta z) $ & $\lambda/50,H/10$ & $ \lambda/100,H/20 $  & $\lambda/200,H/40$ \\
    Time step $ (\Delta t) $ & $  T/200 $ & $  T/400 $ & $  T/800 $ \\
    \bottomrule
  \end{tabular}
  \label{tab:NumericalSettings}
\end{center}
\end{table}

For \textit{foamStar-SWENSE}, the initial values of the complementary velocity field $\bu_C$ and the complementary pressure field $p_C$ are set to zero; the VOF field is set according to the stream function wave theory. For \textit{foamStar}, the initial values of the velocity field $\textbf{u}$, pressure field $p_d$, and the VOF field $\alpha$ are set according to the stream function wave theory.

\subsubsection{Numerical results}

\begin{figure}[h!]
\begin{center}
	\includegraphics[width=0.8\textwidth]{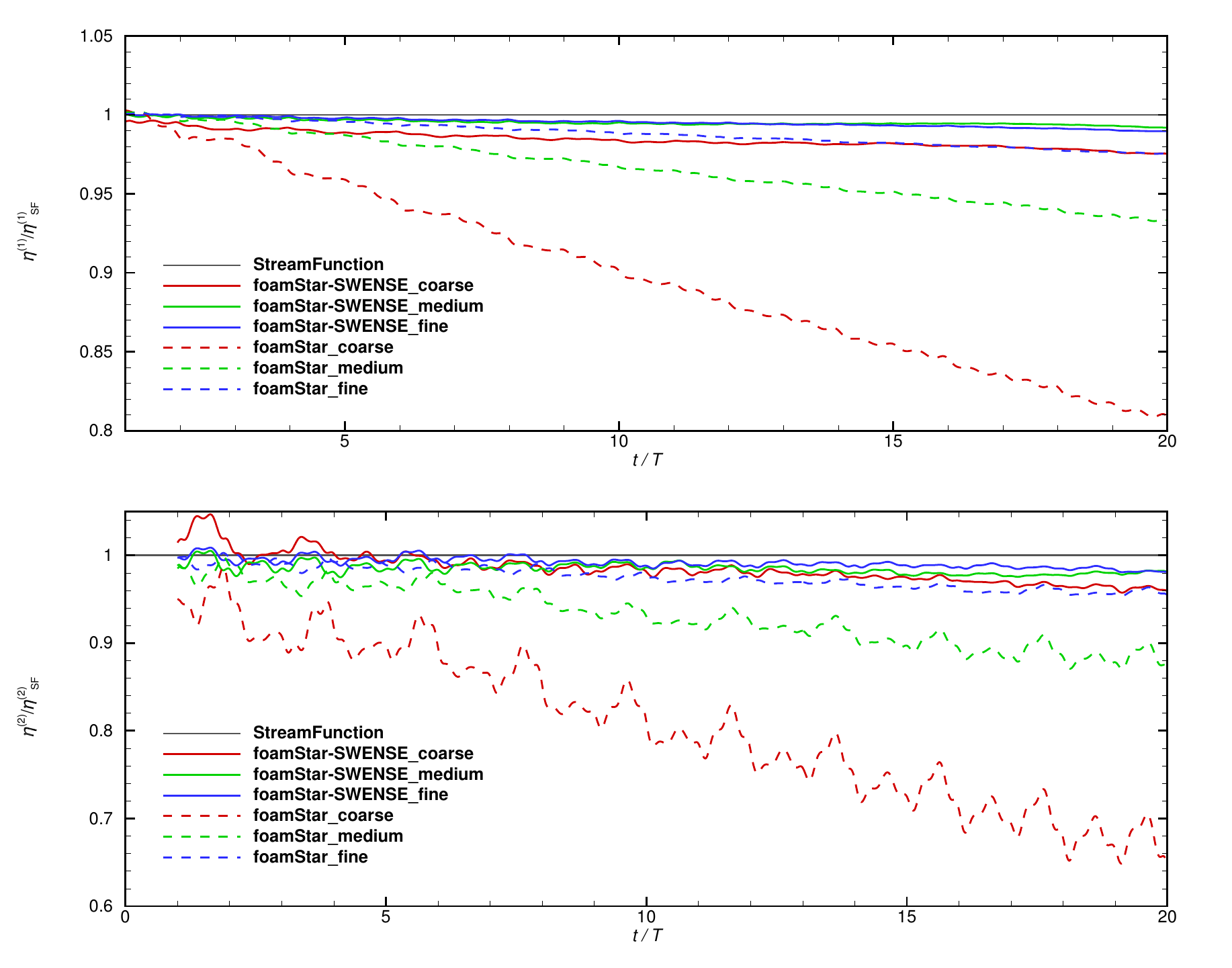}
\end{center}

\caption{Progressive waves test case: Harmonic amplitudes of the free surface elevation at the center of the domain. Top plot: first harmonic amplitude. Bottom plot: second harmonic amplitude.}
\label{fig:waveHarmonics}

\end{figure}

The free surface elevation at the center of the domain is measured. Figure \ref{fig:waveHarmonics} plots the time evolution of the first and second harmonics. The vertical axis is normalized by the reference stream function value; the horizontal axis is non-dimensionalized  by the wave period. The errors on the first harmonic are listed in Tab.\ \ref{tab:firstHarmonicAmplitudes} every five periods.

\begin{table}[h]
\caption{Progressive waves test case: Relative errors on the first harmonic amplitude of the free surface elevation at the center of the computational domain}
\centering
\begin{tabular}{cccccc}
  \toprule
                                 &        & 5T       & 10T      & 15T      & 20T      \\
                                 \hline
                                 \multirow{3}{*}{\textit{foamStar-SWENSE}} & coarse & 1.11\%  & 1.61\%  & 1.80\%  & 2.45\%  \\
                                                                  & medium & 0.27\%  & 0.47\%  & 0.53\%  & 0.81\%  \\
                                                                  & fine   & 0.17\% & 0.38\% & 0.66\% & 1.03\% \\ \hline
\multirow{3}{*}{\textit{foamStar}}        & coarse & 4.10\%  & 9.89\%  & 14.50\%  & 18.97\%  \\
                                 & medium & 1.30\%  & 3.33\%  & 4.90\%  & 6.68\%  \\
                                 & fine   & 0.44\%  & 1.19\%  & 1.75\%  & 2.45\% \\
\hline
\end{tabular}
\label{tab:firstHarmonicAmplitudes}
\end{table}

\textit{foamStar-SWENSE}: Figure \ref{fig:waveHarmonics} reveals that the first and the second harmonic amplitudes are well kept for the entire 20 wave periods, with all the three discretizations. The fine and the medium discretizations give very close results; the coarse discretization produces results with slightly larger errors, but still simulates the waves rather accurately. Tab.\ \ref{tab:firstHarmonicAmplitudes} confirms the above observations. The three meshes have relative errors of 2.45\%, 0.81\%, and 1.03\% at  $t=20T$.

\textit{foamStar}: Figure \ref{fig:waveHarmonics} suggests a more remarkable wave damping throughout the 20 periods with all the discretizations. The loss of wave amplitude is of 18.97\%, 6.68\% and 2.45\% for the coarse, medium and fine meshes at $t=20T$.

%

\subsubsection{Discussion}

\noindent \textbf{Numerical errors in two-phase SWENSE method}

{\color{black} Although Sect. \ref{sect:accurateIncidentWaves} proves that the SWENSE method keeps the waves equal to the incident solution in theory, the simulation results show a difference. One reason for this contradiction is the error created when approximating the interface position by the contour of  $\alpha = 0.5 $. Theoretically, the complementary fields are kept zero in a pure incident wave propagation case because the free surface position (the density jump) coincides with the $p_I = 0$ contour (incident free surface position). However, the present implementation defines the density jump at $\alpha = 0.5 $, where $p_I$ is slightly different from 0. An error in  $\displaystyle\frac{p_I}{\rho_I}\frac{\nabla\rho}{\rho} $ of the momentum equation (Eqn.\ \ref{eqn:twophaseSwMoment}) 
is introduced and generates a tiny spurious complementary field near the free surface.} {\color{black}
Even with refinement, this error cannot be fully reduced to zero. In the future, better definition of the density jump can be considered with more advanced interface capturing methods to improve the result. For now, we satisfy with the method since it provides more accurate results than the NS solver, as shown below.
}

%

\noindent \textbf{Advantage in keeping incident waves over conventional solvers}


In Fig.\ \ref{fig:waveHarmonics}, the results of \textit{foamStar} show the well known numerical wave damping problem of two-phase NS solvers \cite{monroy2017,choi2018} and the wave damping is more severe with coarser discretizations. {\color{black} For this reason, the International Towing Tank Conference (ITTC) suggests more than 80 grids per wavelength when using 2nd-order CFD codes for wave problems \cite{ittcCFDGuideline}.} 
For \textit{foamStar-SWENSE}, the results are very close to the reference value, even with the coarse discretization. Moreover, 
\textit{foamStar-SWENSE} with the coarse mesh gives results as good as \textit{foamStar} with the fine mesh, showing that \textit{foamStar} needs 4 times more refined discretization to achieve the same accuracy. An obvious efficiency gain of the SWENSE method is confirmed.

\subsection{High-order wave forces on a vertical cylinder}


High-order wave force on cylinders is a classical problem in offshore engineering. Despite their small amplitudes, the high-order wave forces may cause sudden structural vibration (the so-called "ringing" phenomenon), since their frequencies are close to the natural frequency of the structure. This problem has been addressed by numerous approaches in the literature, including asymptotic analytical solutions \cite{faltinsen1995nonlinear,malenica1995third}, experiment \cite{huseby2000experimental}, and numerical simulations with fully nonlinear potential flow approaches \cite{ferrant1996computation,ShaoHPC}. Therefore, these well-established reference data make this case very suitable for validation purposes.

The test case contains two following parts:
\begin{itemize}
  \item A convergence study in a rather steep wave condition ($ka=0.24$).

  \item Validation of the method with simulations covering eight different wave steepnesses ($0.6<ka<0.24$).
\end{itemize}

\subsubsection{Test case setup}

\begin{figure}[ht]
  \begin{center}
  \includegraphics[width=0.6\paperwidth]{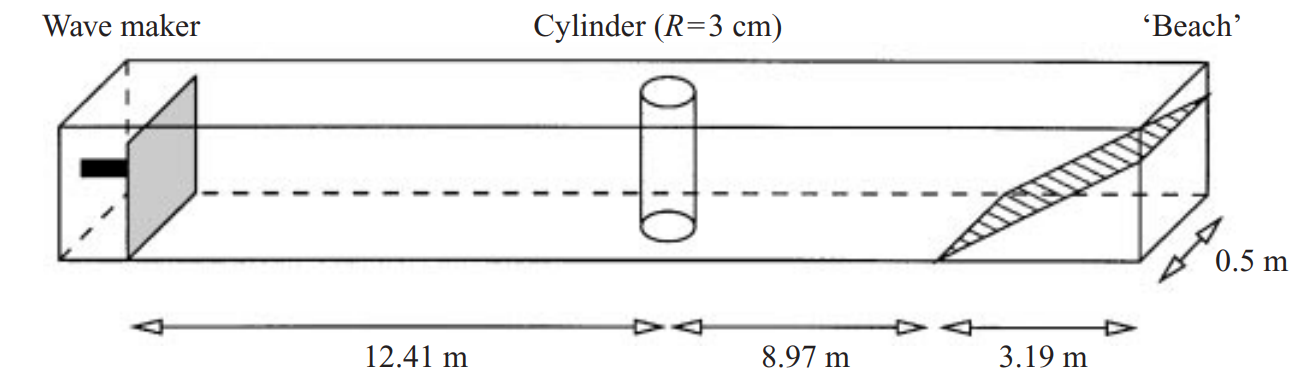}
  \end{center}
  \caption{High-order wave forces test case: The experiment setup \cite{huseby2000experimental}. Reprinted with permission, copyright Cambridge University Press.}
  \label{fig:GridCylinderInWavesExp}
\end{figure}

The simulation reproduces the experiment in \cite{huseby2000experimental}, where a thin cylinder is exposed to regular waves in deep water (see Fig.\ \ref{fig:GridCylinderInWavesExp}). The cylinder has a radius of $R=0.03$ m, being fixed in the water tank of water depth $h=0.6$ m. The incident wave frequency $f$ is equal to $1.425 $ Hz. Different wave amplitudes $(a)$ are used in the experiment and the data are available for a series of wave steepnesses in the range of $ka\in [0.03,0.24]$.

\begin{figure}[ht]
  \begin{center}
  \includegraphics[width=0.4\paperwidth]{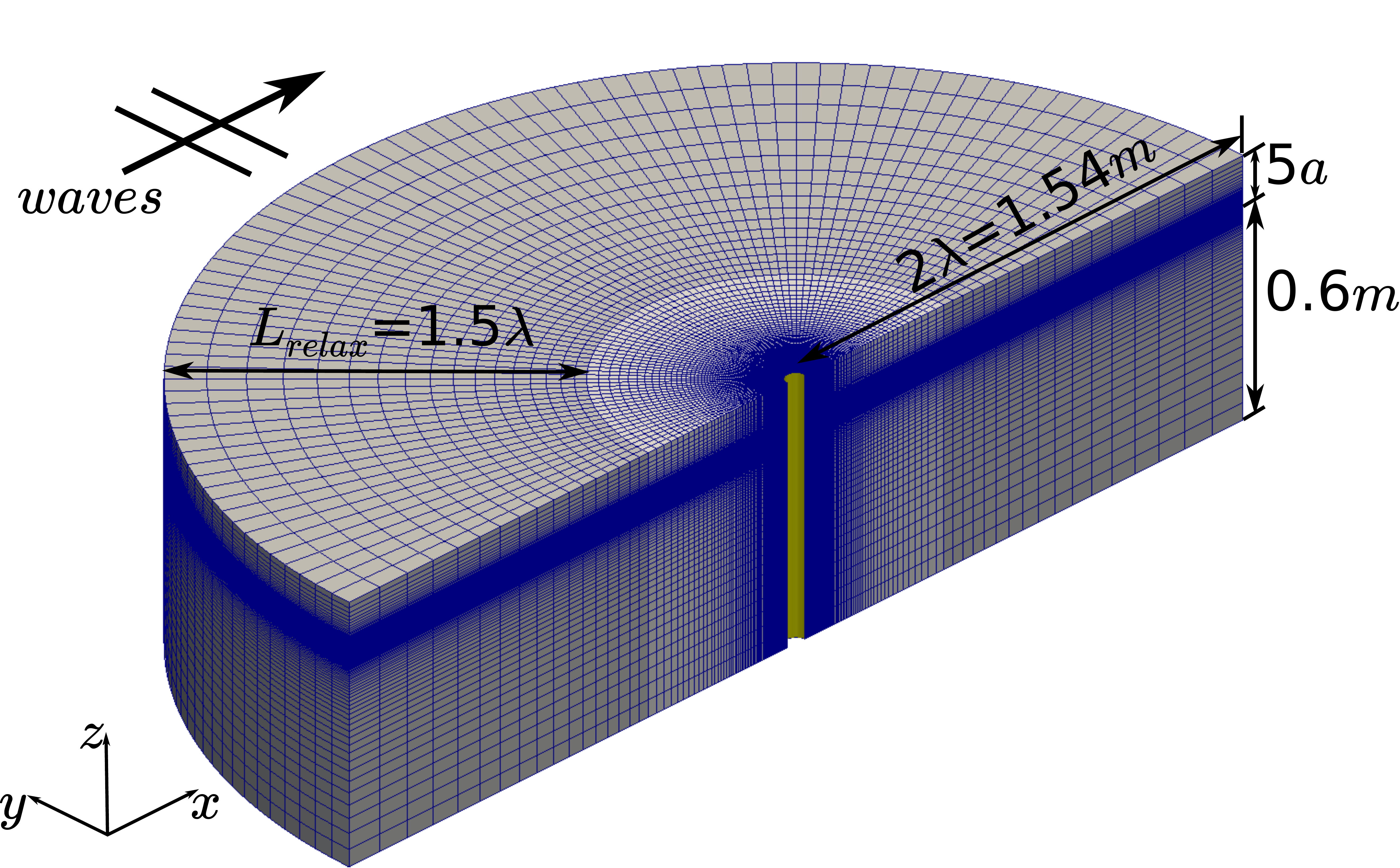}
  \end{center}
  \caption{High-order wave forces test case: Cylindrical mesh used for the simulations}
  \label{fig:GridCylinderInWaves}
\end{figure}

For convenience, a cylindrical mesh is used by the simulation, instead of modeling the entire experimental wave tank (see Fig.\ \ref{fig:GridCylinderInWaves}). The cylinder is located at the center. At far-field, the complementary field is damped to avoid reflections. This configuration represents an ideal experimental condition, where no wave reflects back from the boundaries. This kind of mesh is also used by the potential flow solvers producing the reference data \cite{ferrant1996computation,ShaoHPC}. 

The cylindrical mesh is also accurate and efficient for the SWENSE method: The mesh is fine near the center, which helps to capture the complementary fields accurately near the structure; Coarse mesh in the far-field helps to reduce the computational cost. Please note that  NS solvers cannot use such a mesh, since coarse far-field mesh results in inaccurate incident waves.


A longitudinal symmetry plane is used. The origin of the coordinate system coincides with the cylinder's center-line and is located at the still water level, axis $x$ points to the incident wave propagation direction and axis $z$ points upward. Along $z$ axis, the computational domain extends from the tank bottom at $z=-h$ until $z=5a$ in the air. In $z$ direction, the mesh is uniform near the air-water interface $z\in (-1.5a,1.5a)$.  The cell size increases gradually out of this refined zone. The domain’s radius is equal to 2 wave lengths $(L_r = 2\lambda)$. The mesh is refined near the cylinder and gradually enlarged along the radius direction. In the far-field, a relaxation zone with a length of $L_{relax}=1.5\lambda$ is used to absorb the complementary field, leaving a pure CFD zone with one wave length diameter. {\color{black} The setup of the relaxation zone follows the study in \cite{paulsen2014efficient}, which shows a relaxation zone of at least 1$\lambda$ long locating at least 1/6$\lambda$ away from the structure is sufficient.} The setup is summarized in Tab.\ \ref{tab:cynlinderCaseParemeteres}.

{\color{black}
The maximum Reynolds number $Re = 1.55\times 10^4$ when the wave crest reaches the cylinder, indicating the flow is between the laminar and transitional regime with a Keulegen-Carpenter number $KC = 3.01$. As a result, no turbulence model is used. No-slip wall boundary condition is applied on the cylinder with under-resolved boundary layer. It is acceptable since the viscous force is very small compared to the pressure contribution.
}

\begin{table}[ht]
\begin{center}
  \caption{Parameters for test case cylinder in waves}  \label{tab:cynlinderCaseParemeteres}
  \begin{tabular}{l r}
    \toprule
    Parameter & Value \\
    \midrule
    Cylinder radius $ (R) $  & $0.03 \, m $ \\
    Water depth $ (h) $& $  0.60 \,m $ \\
    Wave frequency $ (f)$& $ 1.425 \,Hz $ \\
    Domain size $ (L_{r} \times L_{\theta} \times L_z) $ & $ 2\lambda \times 180^{o}  \times (h+5a)$ \\
    Relaxation Zone Length $(L_{relax}) $ & $1.5\lambda $ \\
    \bottomrule
  \end{tabular}
\end{center}
\end{table}
%

\subsubsection{Convergence study\label{sect:convergenceStudy}}

A systematic convergence study with six sets of temporal and spatial resolutions is conducted. The time step and the mesh size---in radial, tangential, and vertical directions---are changed simultaneously. The discretization details are summarized in Tab.\ \ref{tab:cylinderConvergenseDiscretization}. This study uses the steepest wave case ($ka=0.24$) in the experiment.

\begin{table}[ht]
\begin{center}
  \caption{High-order wave forces test case: Spatial and temporal resolutions for the convergence study}
  \label{tab:cylinderConvergenseDiscretization}
  \begin{tabular}{cccccc}
  \hline
  \multirow{2}{*}{Index} & \multicolumn{4}{c}{Number of cells in}                                  & Time step    \\ \cline{2-5}
                         & radial direction & tangential direction & a wave amplitude & total   & $T/\Delta t$ \\ \hline
  1                      & 50                & 40                   & 6                  & 88,000   & 200          \\
  2                      & 60                & 48                   & 7                  & 155,520  & 240          \\
  3                      & 71                & 56                   & 9                  & 246,512  & 285          \\
  4                      & 100               & 80                   & 12                 & 668,000  & 400          \\
  5                      & 140               & 112                  & 17                 & 1,881,600 & 560          \\
  6                      & 200               & 160                  & 24                 & 5,504,000 & 800          \\ \hline
  \end{tabular}

\end{center}
\end{table}

\begin{figure}[ht]
  \begin{center}
  \includegraphics[width=0.5\paperwidth]{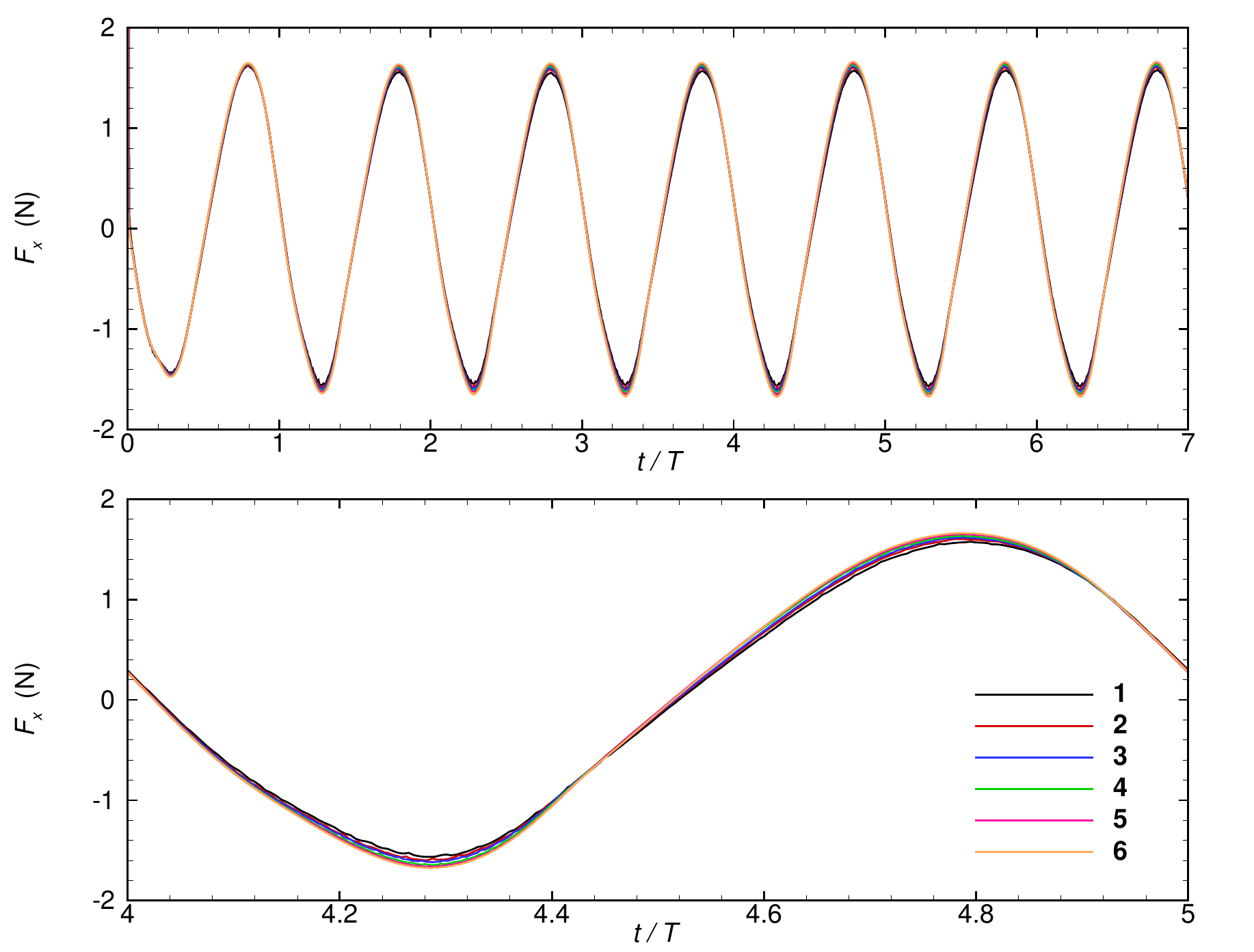}
  \end{center}
  \caption{High-order wave forces test case: Time history of the inline wave force with different mesh resolutions}
  \label{fig:convergenceForceTimeHistory}
\end{figure}

Figure \ref{fig:convergenceForceTimeHistory} shows the time histories of the inline force obtained with different discretizations. The abscissa denotes the time $t$, which is normalized by the wave period $T$. It is observed that a periodic regime appears after two wave periods. The bottom plot zooms in between the fourth and fifth periods. The first observation is that the results are very close and converge with the refinement.

For a better comparison, the time histories in the periodic regime are transformed to the frequency domain by the Fast Fourier Transform (FFT). The amplitudes of the first four harmonic components are summarized in Tab.\ \ref{tab:HarmonicAmplitudesOfForceByDifferentDiscretization}. Following \cite{huseby2000experimental}, the harmonic amplitudes in the table are normalized using:
\begin{equation}
  F_{n}^{'}=\frac{F_{n}}{\rho gR^{3}}\left(\frac{R}{a}\right)^{n}
\end{equation}

\begin{table}[ht]
\begin{center}
  \caption{High-order wave forces test case: Harmonic amplitudes of inline wave force obtained with different mesh resolutions }
  \label{tab:HarmonicAmplitudesOfForceByDifferentDiscretization}
  \begin{tabular}{ccccccc}
  \hline
  \multirow{2}{*}{Harmonic amplitude} & \multicolumn{6}{c}{Mesh}            \\ \cline{2-7}
                                      & 1     & 2     & 3     & 4     & 5     & 6     \\ \hline
  $F_{1}^{'}$                                  & 6.029 & 6.142 & 6.198 & 6.299 & 6.366 & 6.403 \\
  $F_{2}^{'}$                                  & 0.225 & 0.223 & 0.210 & 0.210 & 0.205 & 0.207 \\
  $F_{3}^{'}$                                 & 0.207 & 0.219 & 0.229 & 0.241 & 0.247 & 0.243 \\
  $F_{4}^{'}$                                 & 0.111 & 0.118 & 0.121 & 0.124 & 0.125 & 0.125 \\ \hline
  \end{tabular}
\end{center}
\end{table}

\begin{figure}[ht]
  \begin{center}
  \includegraphics[width=0.6\paperwidth]{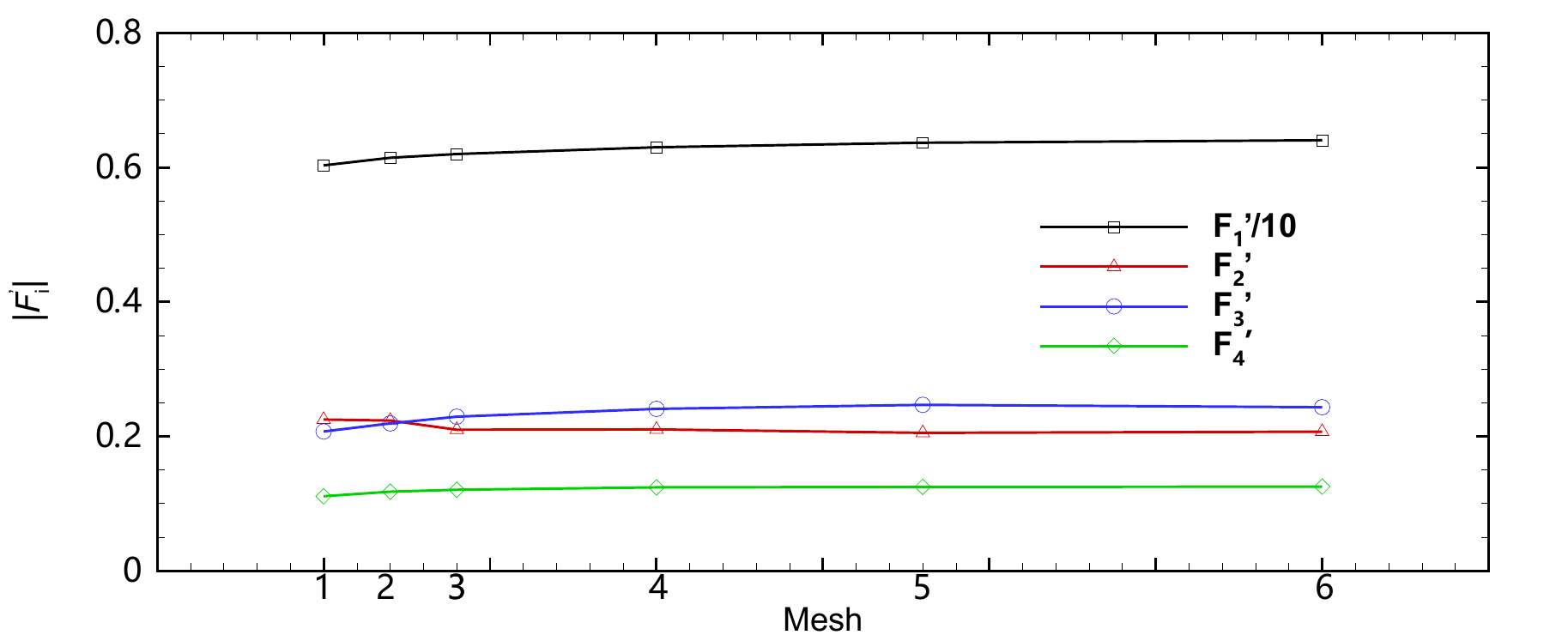}
  \end{center}
  \caption{High-order wave forces test case: Harmonic amplitudes of the inline wave force obtained with different mesh resolutions.}
  \label{fig:HarmonicAmplitudesOfForceByDifferentDiscretization}
\end{figure}

Figure \ref{fig:HarmonicAmplitudesOfForceByDifferentDiscretization} shows the variation of the normalized harmonic amplitudes with the mesh refinement. The abscissa represents the mesh 1 to 6 and is scaled by the number of cells per dimension. Again, it is observed that results with different discretizations are very close, showing that even the coarsest mesh can give a good prediction of force.

In the following, the error estimation procedure proposed by 
\cite{ecca2014procedure} is adopted to estimate the convergence rate and the converged value.  This procedure assumes that the discretization error is a power function of the mesh and time-step size, as follows:
\begin{equation}
  \epsilon_i = \phi_i - \phi_0 = \alpha (\Delta x)^{p_x} + \beta (\Delta t)^{p_t}
  \label{eqn:error}
\end{equation}
where $\epsilon_i$ is the discretization error of the simulation result $\phi_i$, while $\phi_0$ denotes the extrapolated "exact" solution of the mathematical equations. $\Delta x$ and $\Delta t$ stand for the characteristic mesh and time-step size. $p_x$ and $p_t$ are the convergence order in space and time. For each harmonic amplitude, the five parameters: $\phi_0$, $\alpha$, $p_x$, $\beta$, $p_t$ are determined by the method of least-squares, using the six discretization results.

The converged values $\phi_0$ and the convergence order are listed, in Tab.\ \ref{tab:Estimatedexactvalueandordreofaccuracy}. The results show a general convergence order between 1 and 2, which is coherent with the discretization schemes used in OpenFOAM (second order convective schemes with first order limiters).

\begin{table}[ht]
\begin{center}
  \caption{High-order wave forces test case: Estimated "exact" value and order of convergence for the first to fourth harmonic amplitudes}
  \label{tab:Estimatedexactvalueandordreofaccuracy}
\begin{threeparttable}
  \begin{tabular}{c c c c }
    \toprule
    \multirow{2}[3]{*}{Harmonic amplitude} & "Exact" Value & \multicolumn{2}{c}{Order of convergence} \\
    \cmidrule(lr){3-4}
    & ($\phi_0$) & space ($p_x$) & time ($p_t$)\\
    \midrule
    $F_{1}^{'}$ & 6.459 & 1.5 & 1.5 \\
    $F_{2}^{'}$ & 0.204 & 1.8 & 1.8 \\
    $F_{3}^{'}$ & 0.256 & 1.6 & 1.6 \\
    $F_{4}^{'}$ & 0.123 & 1.2 & 2 \\
    \bottomrule
  \end{tabular}
\end{threeparttable}

\end{center}
\end{table}


Table \ref{tab:RelativeerrorsWithDifferentDiscretizations} presents the "relative error" for different mesh resolutions. The "relative error" is defined with the following equation, according to \cite{ecca2014procedure}.
\begin{equation}
  \delta_i = \frac{|\phi_i-\phi_0|}{|\phi_0|}.
  \label{eq:relativeError}
\end{equation}

\begin{table}[ht]
\begin{center}
  \caption{High-order wave forces test case: "Relative errors" for  different mesh resolutions}
  \label{tab:RelativeerrorsWithDifferentDiscretizations}
  \begin{threeparttable}
  \begin{tabular}{ccccccc}
  \hline
  \multirow{2}{*}{Harmonic amplitude} & \multicolumn{6}{c}{"Relative errors" for different meshes (\%)} \\ \cline{2-7}
                                      & 1        & 2       & 3       & 4       & 5       & 6       \\ \hline
  $F_{1}^{'}$                                 & 6.66     & 4.90    & 4.03    & 2.48    & 1.44    & 0.87    \\
  $F_{2}^{'}$                                & 10.24    & 9.51   & 2.82   & 3.03   & 0.54    & 1.40    \\
  $F_{3}^{'}$                               & 19.11    & 14.40   & 10.51   & 5.96    & 3.64    & 5.02    \\
  $F_{4}^{'}$                              & 9.89     & 4.41 & 2.06   & 0.74  & 1.32   & 1.70   \\ \hline
  \end{tabular}
\end{threeparttable}

\end{center}
\end{table}

From Tab.\ \ref{tab:RelativeerrorsWithDifferentDiscretizations}, it is observed that:
\begin{itemize}

\item The first harmonic amplitude is predicted rather accurately even with the coarsest mesh (6.66\% error). This should be credited to the advantage of the SWENSE method and the use of an optimal cylindrical mesh. The SWENSE method ensures a good incident wave, allowing the use of coarse mesh in the far-field; the cylindrical mesh is refined near the structure so that it is always fine "enough" to calculate the first harmonic amplitude correctly.

\item Higher harmonic components contain larger "relative errors" than that in the first harmonic component, since the high-order components are smaller and thus require finer discretizations to be captured accurately. Even so, the accuracy with the coarsest mesh is still rather satisfying, with a  maximum "relative error" smaller than 20\%.

\item For the finest mesh, errors are so low that the convergence seems to saturate. However, one must remember here that it is a "relative" error to an estimated solution and not to an exact one, so that errors of the order of a few percents do not represent the absolute accuracy of the method.

\end{itemize}

\subsubsection{Steepness study}

In this part, the wave force with eight wave steepnesses ($ka\in\{0.06, 0.08, 0.10, 0.13, 0.15, 0.17, 0.20, 0.24\}$) are calculated with \textit{foamStar-SWENSE}. The mesh 4 of the previous section is used, to make a compromise between the computational cost and the accuracy. Table \ref{tab:RelativeerrorsWithDifferentDiscretizations} shows that this mesh achieves a rather small "relative errors", with only 0.67 million cells (1/8 of the finest mesh).  The mesh is adjusted in the vertical direction according to the wave steepness so that the number of cells per wave amplitude is the same.

\begin{figure}[hp]
	\begin{center}
    \includegraphics[width=0.8\textwidth]{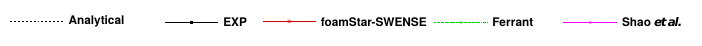}

	\begin{subfigure}{0.49\textwidth} 
		\includegraphics[width=\textwidth]{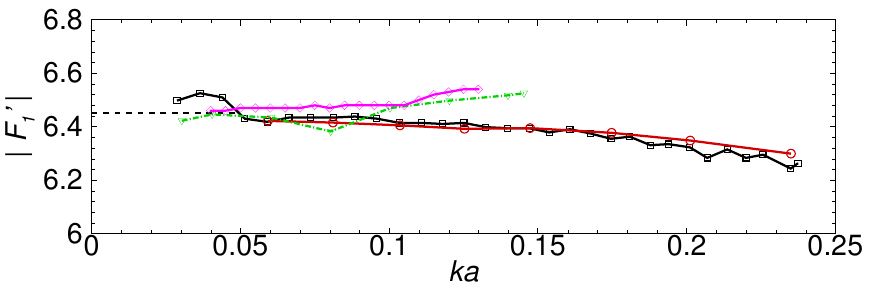}
		\caption{First harmonic amplitudes} 
    \label{fig:cylinderFirstHamronicAmplitude}
	\end{subfigure}
	\begin{subfigure}{0.49\textwidth} 
		\includegraphics[width=\textwidth]{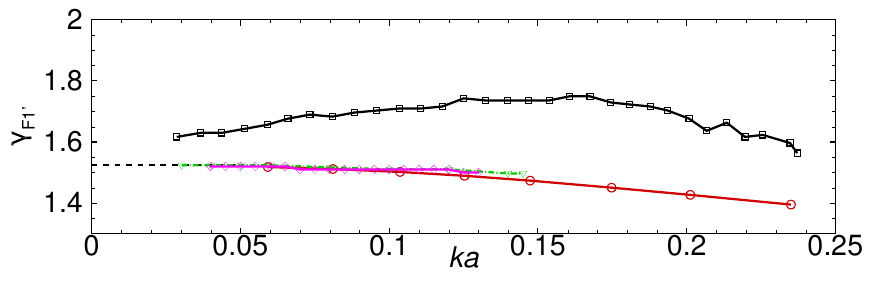}
		\caption{First harmonic phases} 
	\end{subfigure}
    \begin{subfigure}{0.49\textwidth} 
		\includegraphics[width=\textwidth]{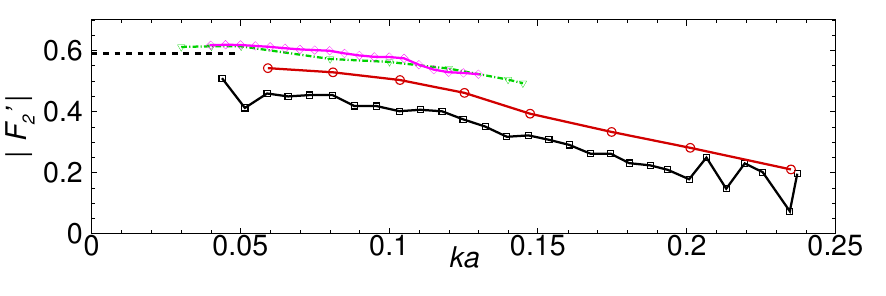}
		\caption{Second harmonic amplitudes} 
	\end{subfigure}
	\begin{subfigure}{0.49\textwidth} 
		\includegraphics[width=\textwidth]{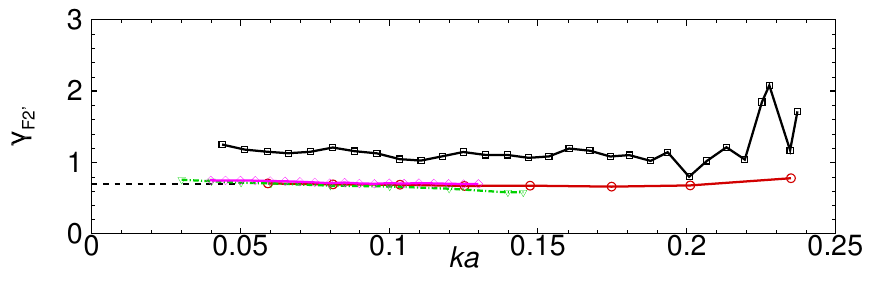}
		\caption{Second harmonic phases} 
	\end{subfigure}
    \begin{subfigure}{0.49\textwidth} 
		\includegraphics[width=\textwidth]{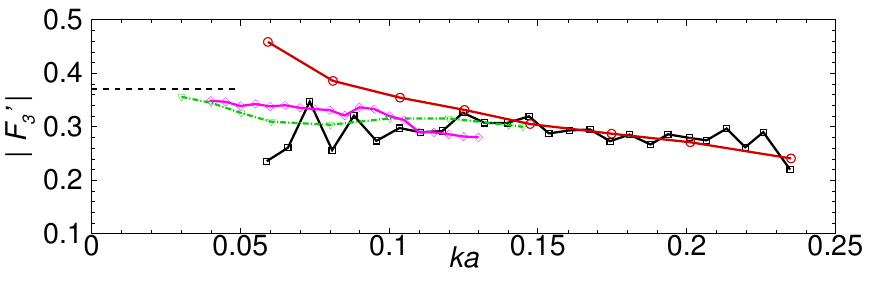}
		\caption{Third harmonic amplitudes} 
	\end{subfigure}
	\begin{subfigure}{0.49\textwidth} 
		\includegraphics[width=\textwidth]{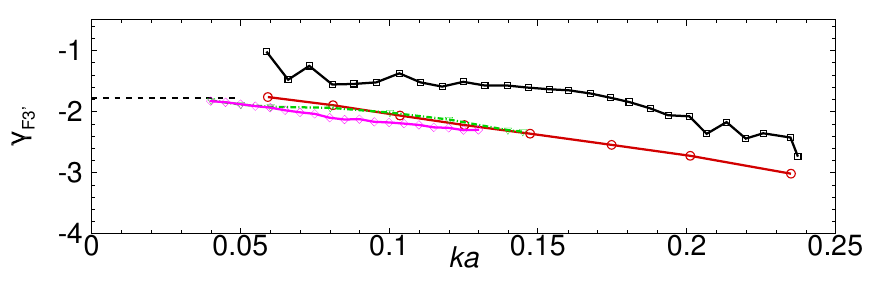}
		\caption{Third harmonic phases} 
	\end{subfigure}
    \begin{subfigure}{0.49\textwidth} 
		\includegraphics[width=\textwidth]{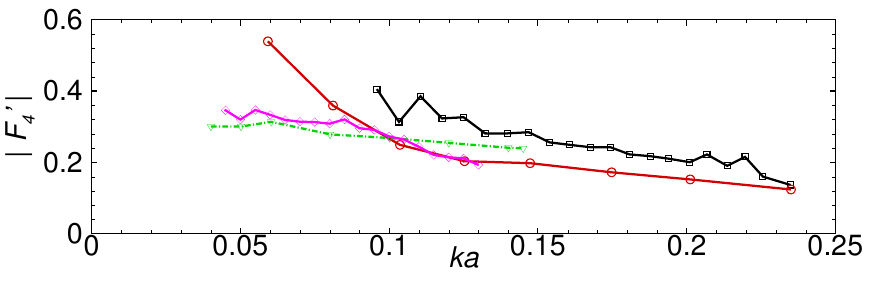}
		\caption{Fourth harmonic amplitudes} 
	\end{subfigure}
	\begin{subfigure}{0.49\textwidth} 
		\includegraphics[width=\textwidth]{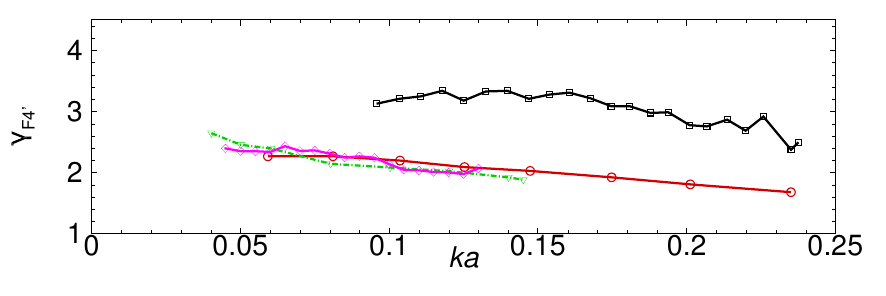}
		\caption{Fourth harmonic phases} 
	\end{subfigure}
	\caption{High-order wave forces test case: Comparison of the first to fourth harmonics of horizontal forces on vertical circular cylinder in regular waves with $kR  = 0.245$.} 
\label{fig:ComparisonForceWithExperiemnt}

\begin{subfigure}{0.24\textwidth} 
	\includegraphics[width=\textwidth]{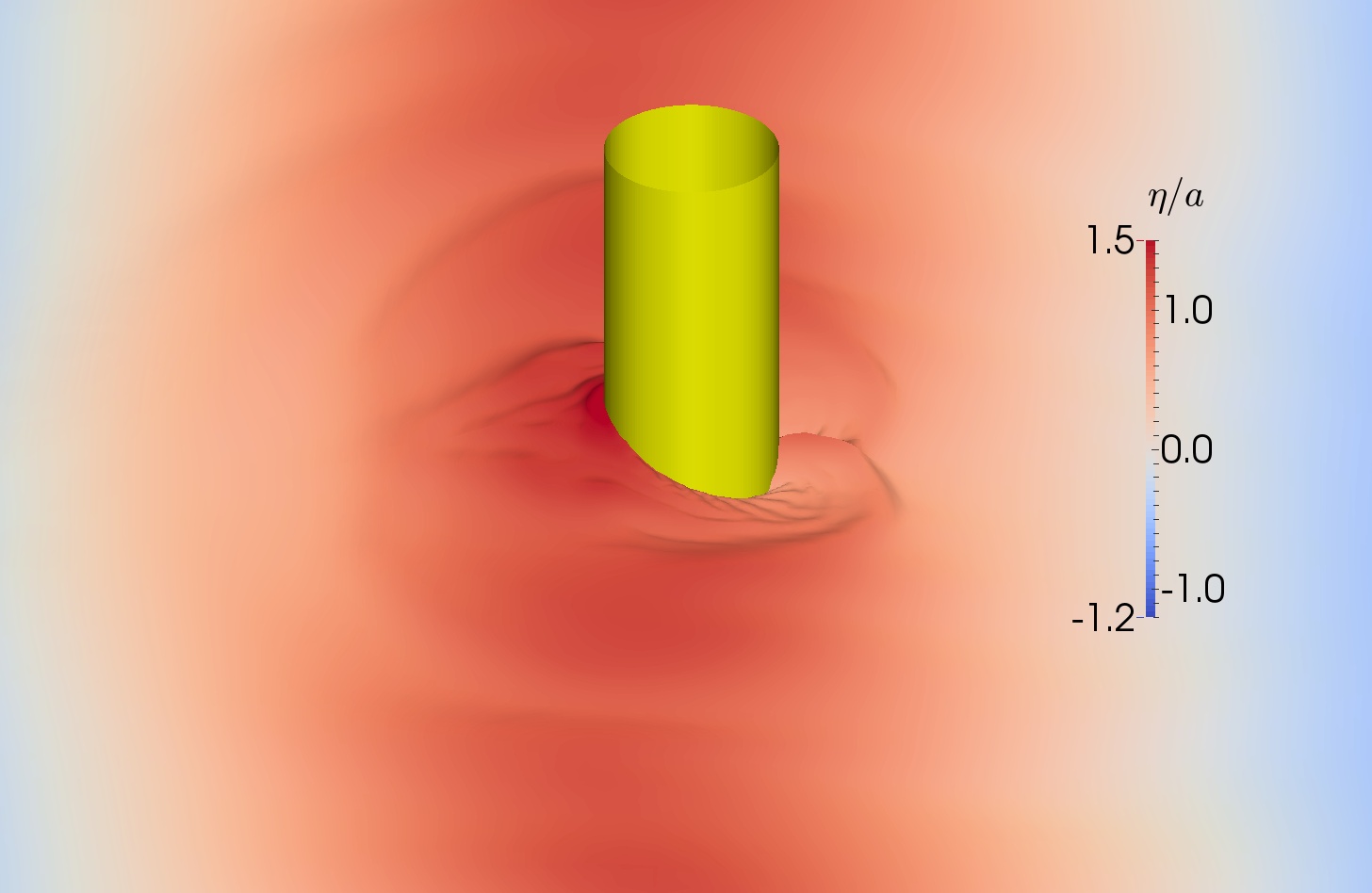}
	\caption{$t=6T$} 
\end{subfigure}
\begin{subfigure}{0.24\textwidth} 
	\includegraphics[width=\textwidth]{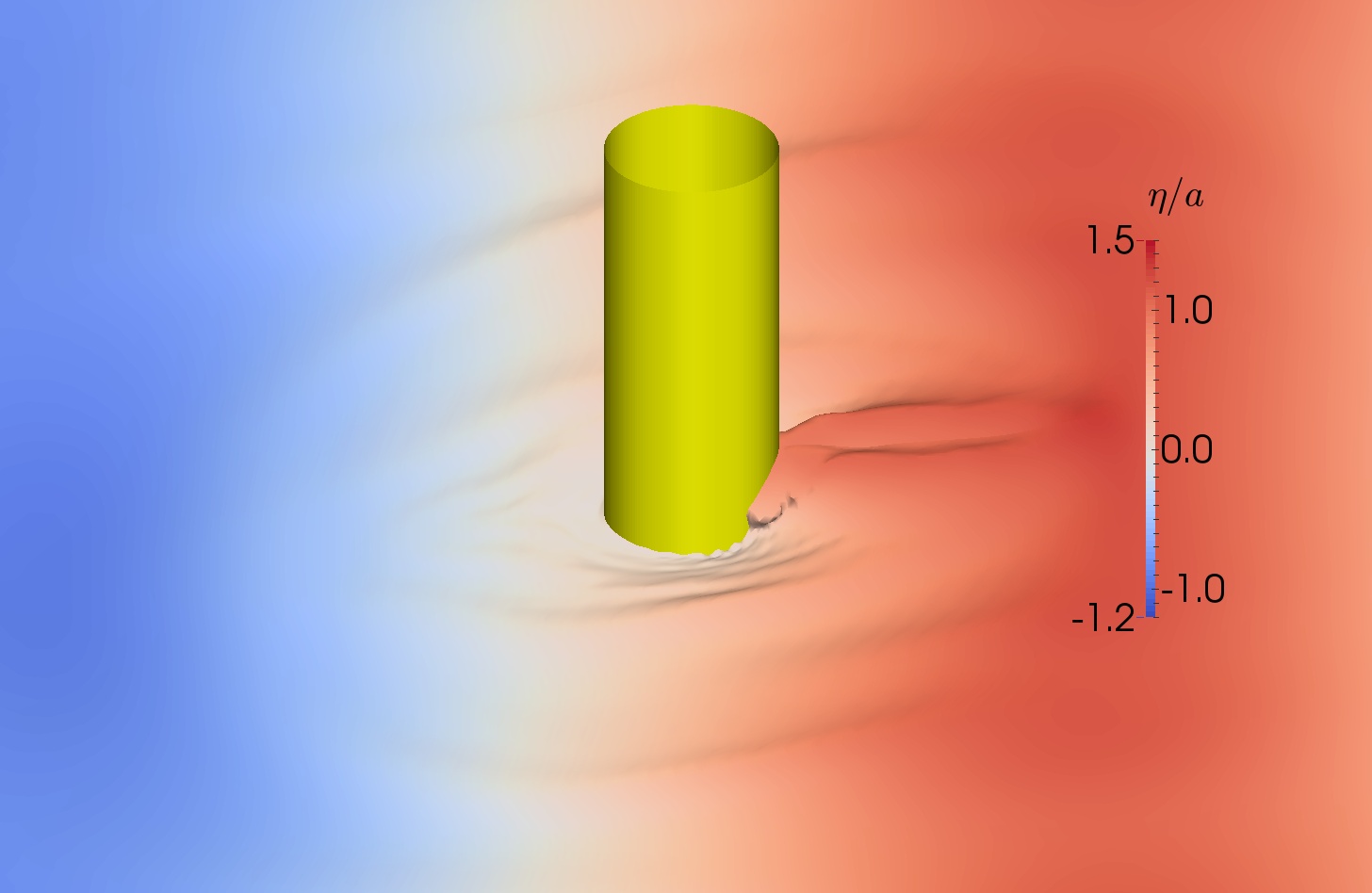}
	\caption{$t=6.2T$} 
\end{subfigure}
\begin{subfigure}{0.24\textwidth} 
	\includegraphics[width=\textwidth]{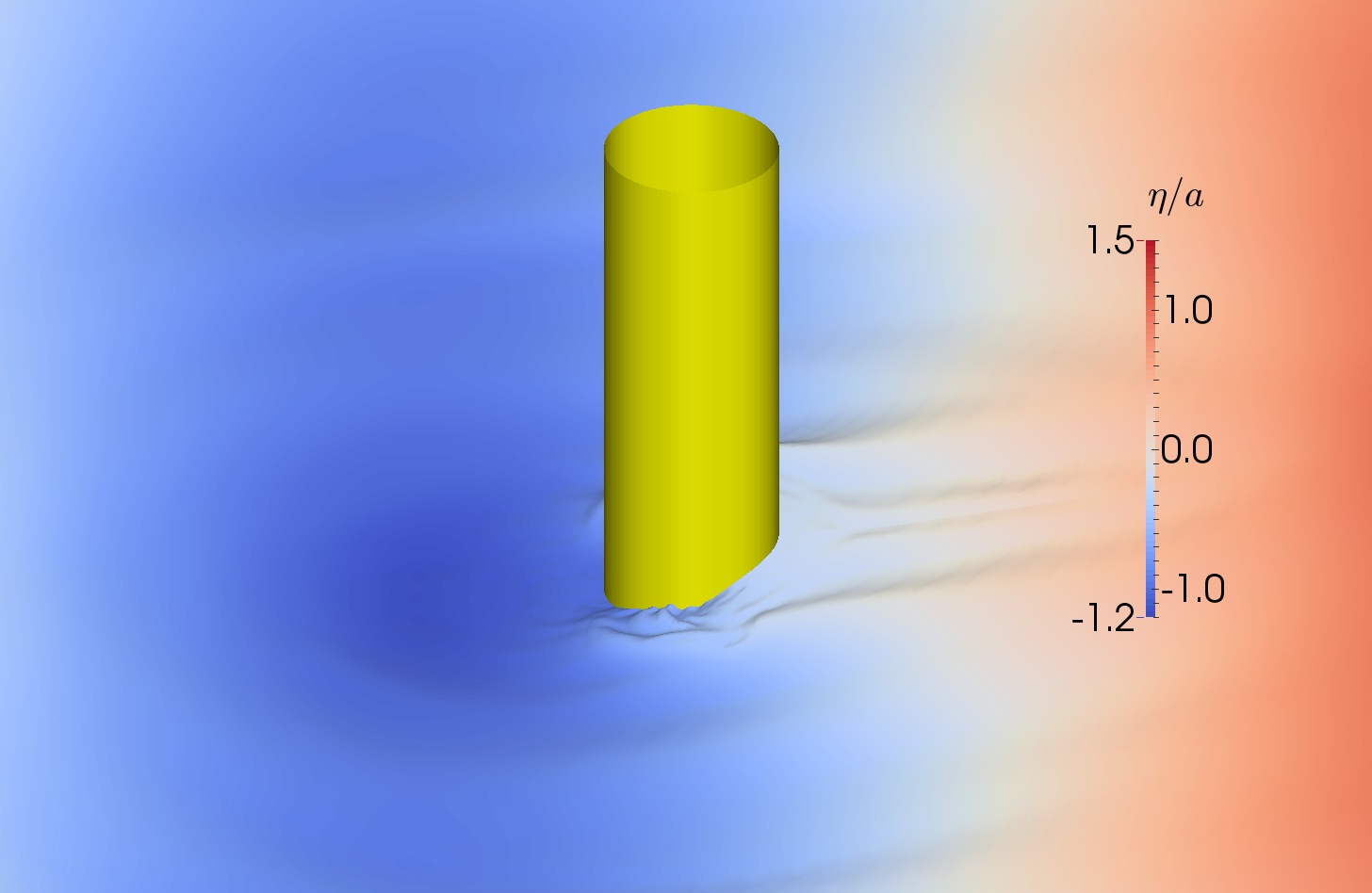}
	\caption{$t=6.4T$} 
\end{subfigure}

\begin{subfigure}{0.24\textwidth} 
	\includegraphics[width=\textwidth]{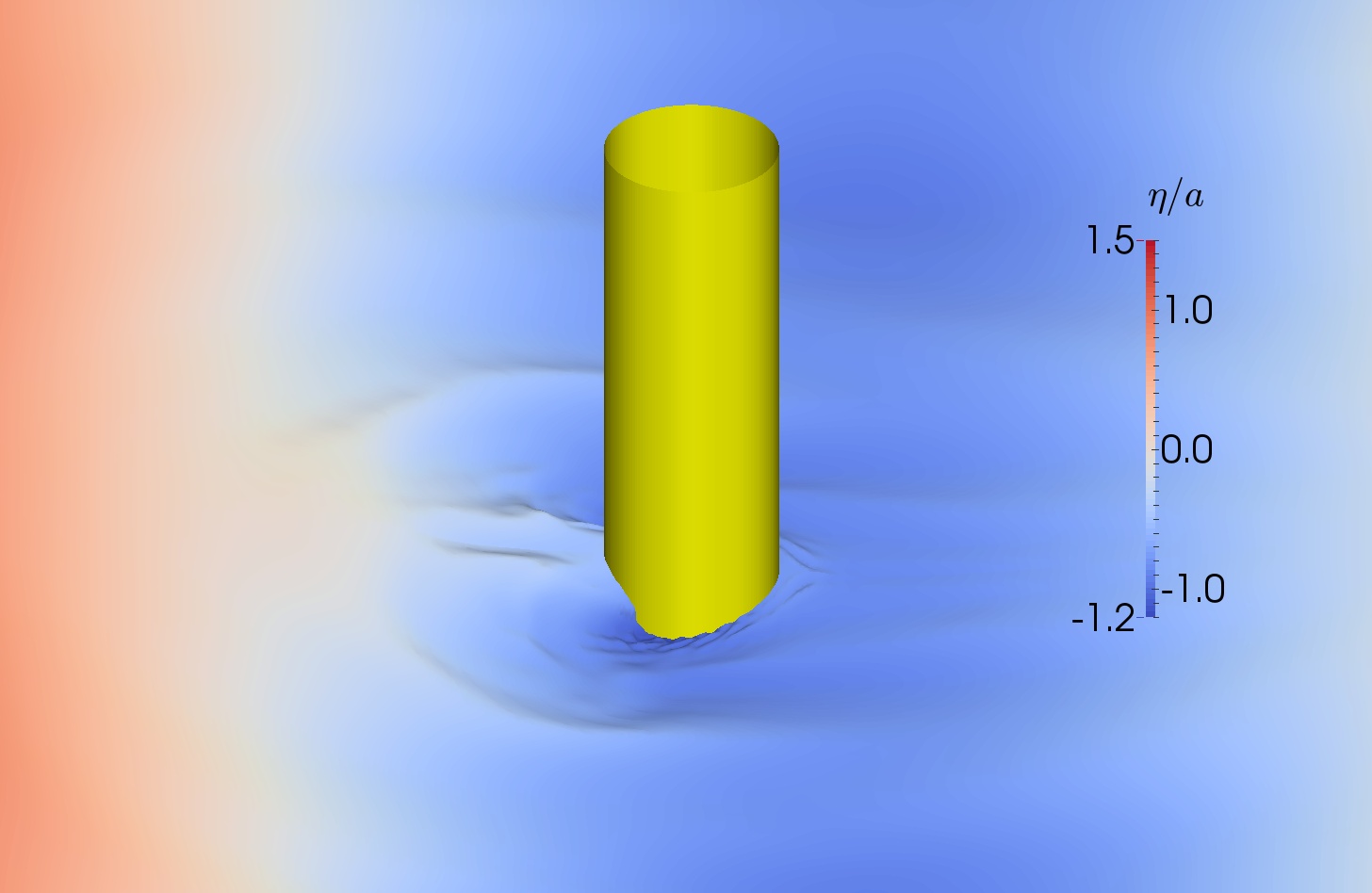}
	\caption{$t=6.6T$} 
\end{subfigure}
\begin{subfigure}{0.24\textwidth} 
	\includegraphics[width=\textwidth]{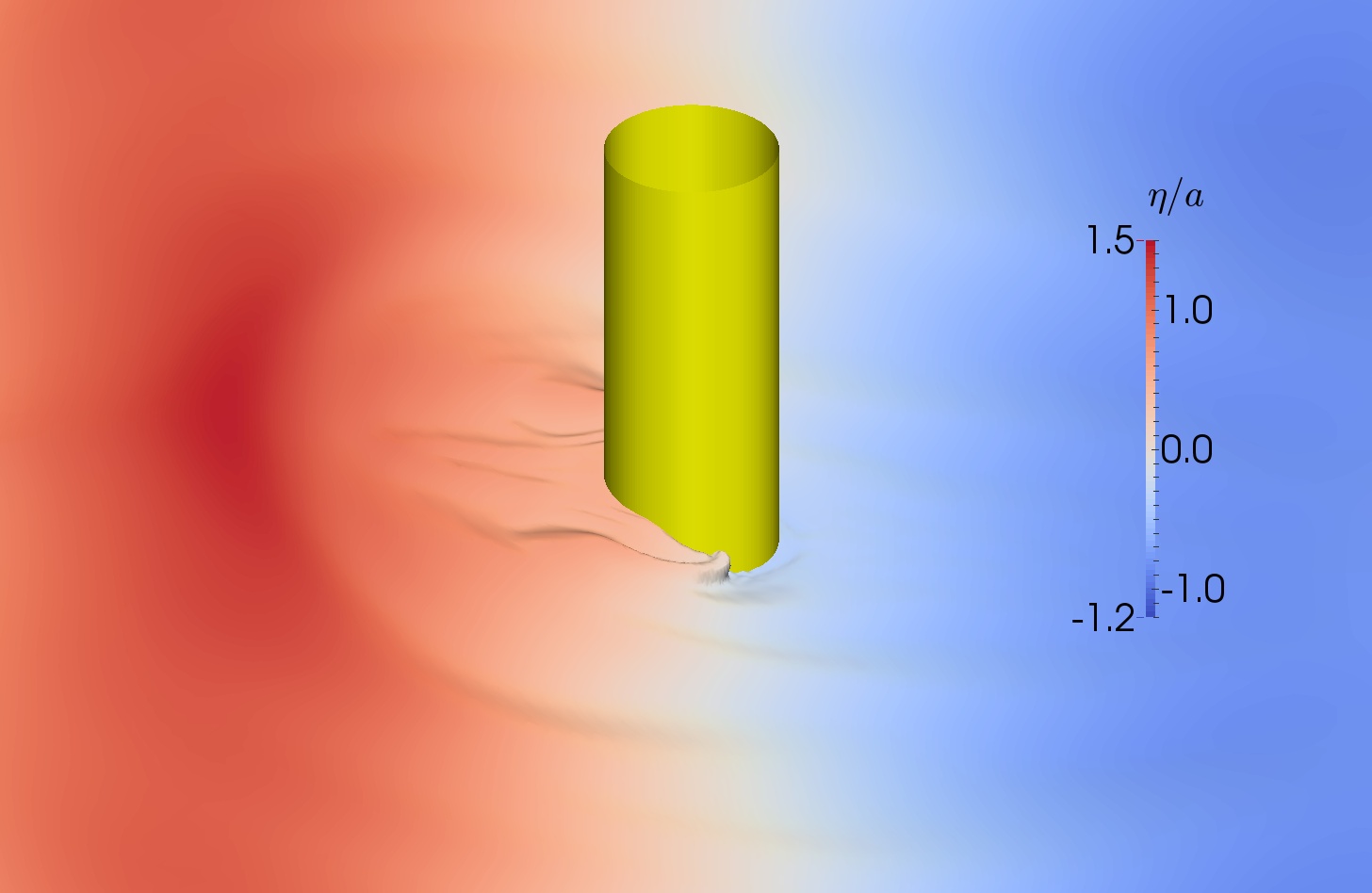}
	\caption{$t=6.8T$} 
\end{subfigure}
\begin{subfigure}{0.24\textwidth} 
	\includegraphics[width=\textwidth]{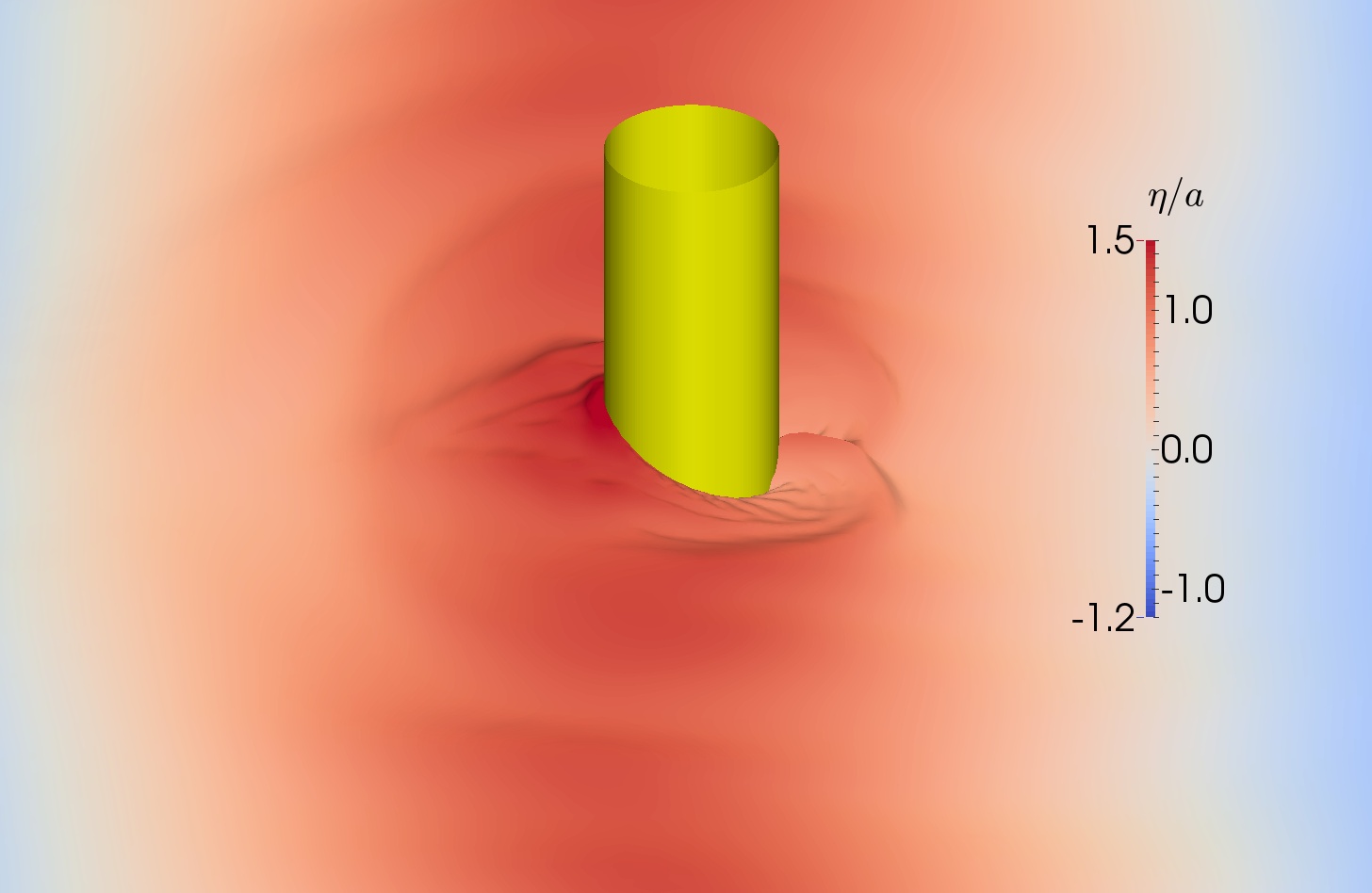}
	\caption{$t=7T$} 
\end{subfigure}

\caption{High-order wave forces test case: Free surface near the cylinder (incident wave steepness $ka=0.24$)} 
\label{fig:freeSurfaceCylinder}
	\end{center}
\end{figure}

The harmonic amplitudes and the phase shifts calculated by \textit{foamStar-SWENSE} are compared with reference data in Fig.\ \ref{fig:ComparisonForceWithExperiemnt}. From the top to the bottom, the first to the fourth harmonic components. The amplitudes are shown on the left and the phase shifts are shown on the right.  In each subplot, the horizontal axis denotes the wave steepness. The reference data include:
\begin{itemize}
  \item The third-order analytical solution of Malenica and Molin \cite{malenica1995third} (referred to as Analytical).
  \item The experiment data reference  \cite{huseby2000experimental} (referred to as EXP).
  \item Two numerical results using fully nonlinear potential flow theory (referred to as Ferrant \cite{ferrant1998fully} and Shao \etal{} \cite{ShaoHPC}).
\end{itemize}

To analyze these results, the primary focus is put on the first harmonic amplitudes (Fig.\ \ref{fig:cylinderFirstHamronicAmplitude}), since it is dominant. In general, \textit{foamStar-SWENSE} shows a very close agreement with experimental data. This agreement suggests the accuracy of the proposed method in calculating the wave load. When compared with the potential theory results of Ferrant \cite{ferrant1998fully} and Shao \etal{} \cite{ShaoHPC}, \textit{foamStar-SWENSE} shows also a good agreement at small wave steepness ($ka<0.1$). For $ka>0.1$, \textit{foamStar-SWENSE} is more accurate than the potential flow solvers, since the SWENSE method is able to capture the flow separation and violent free surface deformations (see Fig.\ \ref{fig:freeSurfaceCylinder}), where PT solvers encounter numerical difficulties \cite{licalculation}.

%
%
%
%

For the second harmonic amplitudes, both \textit{foamStar-SWENSE} and the experiment show the same decreasing trend when the wave steepness becomes larger. The results are approximately 20\% larger than the experimental data and are closer to experimental data than that of the potential flow solvers where they are available.

 For the third and fourth harmonic amplitudes, the comparison is better when the wave steepness is large. For small wave steepness, it is indeed more difficult because the magnitudes of the high-order force are very small. The oscillations appearing in the experimental data also confirm this difficulty.

For the phase shifts, a good agreement is shown among the analytical solution and all the numerical results, which are, however, different from the experimental data. The reason for this discrepancy is still not clear \cite{ShaoHPC}, but it may due to a setup slightly different in the experiment. More specifically,  in consideration of the difference getting larger while the order increases, it is possible that the beginning time of the experiment and the calculations are different.


To conclude, this comparison demonstrates that the proposed two-phase SWENSE method is able to calculate the wave force on a simple structure for a large range of wave steepness. The typical cylindrical mesh with coarse cells in the far-field can be used to reduce the computational cost without influencing the accuracy.

\subsection{CALM buoy in regular and irregular waves}

This part aims to demonstrate the ability of the two-phase SWENSE method in dealing with a more complex geometry and wave conditions. 

\begin{itemize}
  \item Geometry: The Catenary Anchor Leg Mooring (CALM) buoy \cite{JIP_2005} contains a thin heave-damping skirt. {\color{black} Violent free surface deformation and significant flow separation induced by such a geometry needs to be captured by the complementary field. }
  \item Wave condition: Both regular and irregular wave conditions are used. The irregular waves are obtained with \textit{HOS-NWT} solver and reconstructed onto the CFD mesh to validate the reconstruction method proposed in Sect.\ \ref{Section:IncidentWave}.
   \item {\color{black}Turbulence modeling: The $k-\omega$ SST turbulence model \cite{menter1994two} is used. The non-slip boundary condition is applied on the buoy with standard wall functions in OpenFOAM \cite{liu2016wallFunctions}. }
\end{itemize}

In addition, the accuracy and efficiency of \textit{foamStar-SWENSE} and \textit{foamStar} are compared in the regular wave condition. Here the results are shown in a concise version for validation purposes. The interested reader can find details about the regular wave case in a separate paper \cite{li2019comparison}.

\subsubsection{Geometry and wave parameters}

\begin{figure}[ht]
\begin{center}
\begin{subfigure}{0.4\textwidth}
\includegraphics[width=0.8\textwidth]{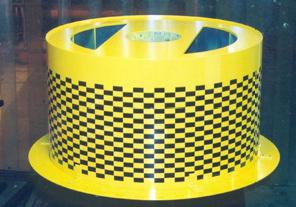}
\caption{CALM Buoy Model}
\label{fig:modelPhoto}
\end{subfigure}
\begin{subfigure}{0.4\textwidth}
\includegraphics[trim={0 1.5cm 0 0},clip,width=0.8\textwidth]{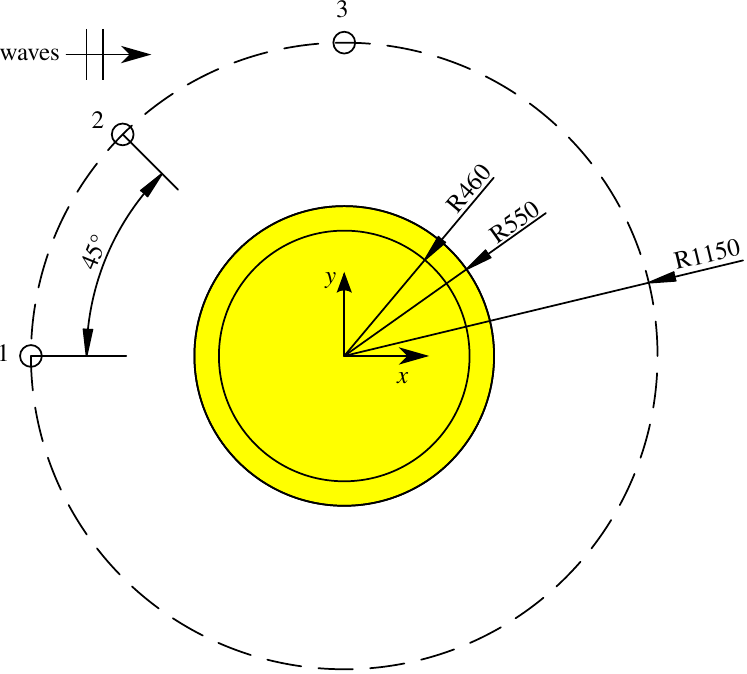}
\caption{Wave Gauge Positions}
\label{fig:probePosition}
\end{subfigure}
\caption{CALM buoy test case: Model geometry and experiment setup}
\end{center}
\end{figure}

The test case reproduces an experiment carried out in the ocean engineering basin of Ecole Centrale Nantes (50m long, 30m wide and 5m deep). The buoy is a truncated cylinder with a thin skirt near the bottom to provide additional damping forces through vortex shedding (see Fig.\ \ref{fig:modelPhoto} and Tab.\ \ref{tab:JIPParemeteres}). The horizontal and vertical wave forces and the free surface elevations at three points are measured (see Fig.\ \ref{fig:probePosition}).

\begin{table}[ht]
\begin{center}
  \caption{CALM buoy test case: Geometry characteristics}
  \label{tab:JIPParemeteres}
  \begin{tabular}{l r}
    \toprule
    Parameter & Value \\
    \midrule
    Radius   & $0.460 \,  $m \\
    Height overall   & $0.560 \,  $m \\
    Skirt radius   & $0.550 \,  $m \\
    Skirt thickness   & $0.004 \,  $m \\
    From the bottom to the mid-skirt   & $0.04 \,  $m \\
    Draft   & $0.25 \,  $m \\
    \bottomrule
  \end{tabular}

\end{center}

\centering
\caption{CALM buoy test case: Wave conditions}
\label{tab:waveConditionsForCALMBuoyInWaves}
\begin{tabular}{cccc}
\hline
\multicolumn{2}{c}{Regular waves} & \multicolumn{2}{c}{Irregular waves} \\ \hline
$T$             & 1.80 s              & $T_p$              & 2.00 s              \\
$H$             & 0.16 m             & $H_s$              & 0.12 m             \\
$kH/2$ & 0.1 & $k_pH_s/2$ & 0.06 \\ \hline
\end{tabular}

\end{table}

Regular and irregular wave conditions, chosen from the experiment, are listed in Tab.\ \ref{tab:waveConditionsForCALMBuoyInWaves}. The wave steepnesses are moderate. The irregular waves are generated with a JONSWAP spectrum with $\gamma=3$.

%

\subsubsection{Computational domain and meshes}

A cylindrical and a rectangular mesh configurations are used. The cylindrical mesh is used by \textit{foamStar-SWENSE} in both regular and irregular wave cases. However, the cylindrical mesh is unsuitable for \textit{foamStar}, because the coarse cells in the far-field deteriorate the incident waves. For this reason, a series of rectangular meshes are used in the comparative study by both \textit{foamStar} and \textit{foamStar-SWENSE}. The computational domain is defined with the regular wave's parameter ($\lambda,H$).

\noindent\textbf{Cylindrical configuration}

\begin{figure}[ht]
\begin{center}
\includegraphics[width=0.6\textwidth]{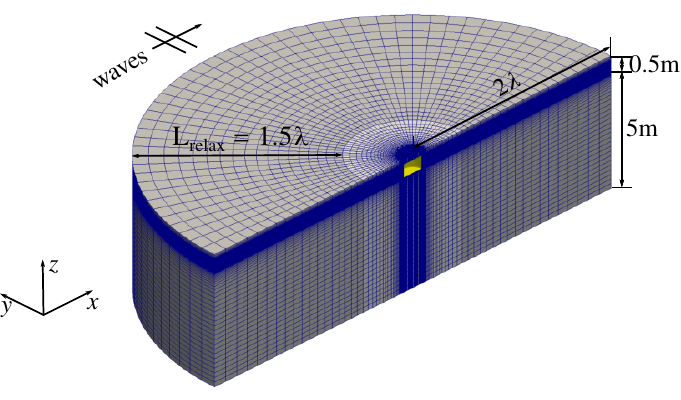}
\caption{CALM buoy test case: Cylindrical mesh layout \cite{li2019comparison}. Reprinted with permission, copyright Elsevier\textsuperscript{\textregistered}.}
\label{fig:cylindricMeshLayout}
\end{center}
\end{figure}

The domain radius is equal to $2\lambda$. The depth is equal to that of the experimental wave tank (5m). The top is 0.5m above the free surface position at rest. A longitudinal symmetry plane is used.
The cells near the structure are refined in the radial direction to capture the flow details. Cells are gradually enlarged in the far-field (see Fig.\ \ref{fig:cylindricMeshLayout}). The discretization details are listed in Tab.\ \ref{tab:JIPCylindericalMeshDetail}. In the far-field, a relaxation zone with a length of 1.5$\lambda$ is used to absorb the disturbed wave field, leaving a pure CFD zone with one wave length diameter, i.e., $r\in(-0.5\lambda,0.5\lambda)$. {\color{black} The relaxation zone setup is the same as the cylinder in wave case, following also \cite{paulsen2014efficient}.}

\begin{table}[ht]
\begin{center}
  \caption{CALM buoy test case: Cylindrical mesh information}
  \label{tab:JIPCylindericalMeshDetail}
  \begin{tabular}{l r}
    \toprule
    Parameter & Value \\
    \midrule
    $\lambda/\Delta R_{near}$    & 400 \\
    $\lambda/\Delta R_{far}$   & 10 \\
    $ 180^\circ/\Delta\theta$   & 96  \\
    $H/\Delta z $  & 16 \\
    Total Cells & 0.72 M \\
    \bottomrule
  \end{tabular}
\end{center}
\end{table}

\noindent\textbf{Rectangular configurations}

\begin{figure}[ht]
\begin{center}
\includegraphics[width=0.8\textwidth]{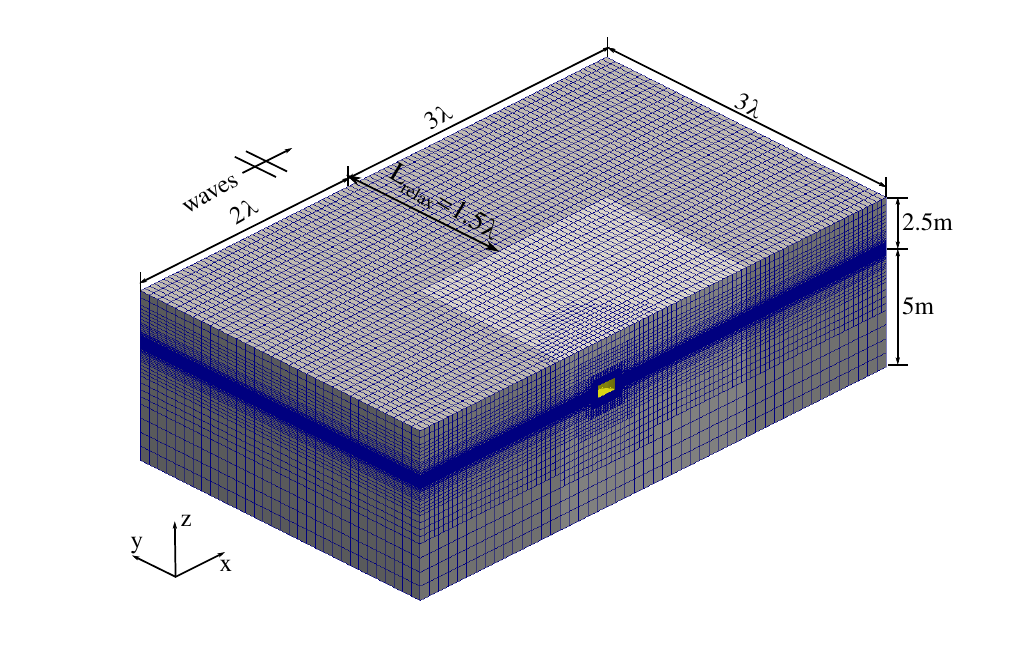}
\caption{CALM buoy test case: Rectangular mesh layout \cite{li2019comparison}. Reprinted with permission, copyright Elsevier\textsuperscript{\textregistered}.}
\label{fig:rectMeshLayout}
\end{center}

\end{figure}

The rectangular configurations use uniform Cartesian background mesh to facilitate the incident wave propagation with \textit{foamStar}. Three configurations: x20, x40, and x80 are used with 20, 40, and 80 cells per wave length in the $x$ direction. The mesh in the transverse direction is less refined than in the $x$ direction to reduce the total number of cells (see Fig.\ \ref{fig:rectMeshLayout}). In the far-field, relaxation zones with a length of 1.5$\lambda$ are set at the inlet, the outlet, and the sides. The mesh details are summarized in Tab.\ \ref{tab:meshOfcalmBuoyInWaves}. 

The local mesh refinement near the structure is invariant for all the three configurations and has a cell density similar to that of the cylindrical configuration. Thus the differences between the x20, x40, x80, and the cylindrical mesh are in the far-field only.

\begin{table}[ht]
\centering
\caption{CALM buoy test case: Rectangular mesh information}
\label{tab:meshOfcalmBuoyInWaves}
\begin{tabular}{ccccc}
\toprule
Mesh        & $\lambda/\Delta x$  & $\lambda/\Delta y$  & $H/\Delta z$  & Number of cells \\
\midrule
x20         & 20 & 10 & 16 & 1.28 M           \\
x40         & 40 & 10 & 16 & 1.47 M           \\
x80         & 80 & 20 & 16 & 2.58 M           \\
  \bottomrule

\end{tabular}
\end{table}

\subsubsection{Regular waves}

The experimental condition with regular waves ($T=1.8s, H=0.16m, kH/2=0.1$) is reproduced by both \textit{foamStar-SWENSE} and \textit{foamStar}. 

\noindent\textbf{Comparison of the accuracy}

During the simulations, the wave forces and free surface elevations are recorded during 7 periods. The time history is transformed to the frequency domain. The first and second harmonic amplitudes are extracted and compared. The amplitudes of wave force are normalized with $ak\rho g\nabla$, where $a=H/2$ is the amplitude of the waves, $k$ the wave number and $\nabla$ the displacement volume of the buoy at the designed draft. The free surface elevations are normalized by $a$. The results are shown in Tab.\ \ref{tab:compareEfdCfdResult}. The relative differences compared with the experimental data is given in percentage form.

\begin{table}[ht]
\centering
\caption{CALM buoy test case: Comparison between CFD results and experimental data for the regular wave case}
\label{tab:compareEfdCfdResult}
\resizebox{\textwidth}{!}{

\begin{tabular}{cccccccccccc}
\hline
Harmonic Amplitudes &  & $F^{ (1) }_x$ & $F^{ (2) }_x$ & $F^{ (1) }_z$ & $F^{ (2) }_z$ & $\eta^{ (1) }_1$ & $\eta^{ (2) }_1$ & $\eta^{ (1) }_2$ & $\eta^{ (2) }_2$ & $\eta^{ (1) }_3$ & $\eta^{ (2) }_3$ \\ \hline
Experiment &  & 1.390 & 0.170 & 1.180 & 0.015 & 1.220 & 0.065 & 1.210 & 0.040 & 1.040 & 0.035 \\ \hline
\multirow{6}{*}{foamStar} & \multirow{2}{*}{\begin{tabular}[c]{@{}c@{}}x80\\ (2.58M)\end{tabular}} & 1.359 & 0.168 & 1.098 & 0.010 & 1.195 & 0.060 & 1.180 & 0.036 & 1.002 & 0.045 \\
 &  & {\small -2.23\% } & {\small -1.18\% } & {\small -6.95\% } & {\small -33.33\% } & {\small -2.05\% } & {\small -7.69\% } & {\small -2.48\% } & {\small -10.00\% } & {\small -3.65\% } & {\small 28.57\% } \\ \cline{2-12}
 & \multirow{2}{*}{\begin{tabular}[c]{@{}c@{}}x40\\ (1.47M)\end{tabular}} & 1.328 & 0.165 & 1.075 & 0.011 & 1.172 & 0.057 & 1.164 & 0.035 & 0.983 & 0.041 \\
 &  & {\small -4.46\% } & {\small -2.94\% } & {\small -8.90\% } & {\small -26.67\% } & {\small -3.93\% } & {\small -12.31\% } & {\small -3.80\% } & {\small -12.50\% } & {\small -5.48\% } & {\small 17.14\% } \\ \cline{2-12}
 & \multirow{2}{*}{\begin{tabular}[c]{@{}c@{}}x20\\ (1.28M)\end{tabular}} & 1.202 & 0.130 & 1.018 & 0.017 & 1.063 & 0.057 & 1.057 & 0.037 & 0.924 & 0.039 \\
 &  & {\small -13.53\% } & {\small -23.53\% } & {\small -13.73\% } & {\small 13.33\% } & {\small -12.87\% } & {\small -12.31\% } & {\small -12.64\% } & {\small -7.50\% } & {\small -11.15\% } & {\small 11.43\% } \\ \hline
\multirow{8}{*}{foamStar-SWENSE} & \multirow{2}{*}{\begin{tabular}[c]{@{}c@{}}x80\\ (2.58M)\end{tabular}} & 1.383 & 0.182 & 1.152 & 0.014 & 1.211 & 0.060 & 1.198 & 0.032 & 1.035 & 0.051 \\
 &  & {\small -0.50\% } & {\small 7.06\% } & {\small -2.37\% } & {\small -6.67\% } & {\small -0.74\% } & {\small -7.69\% } & {\small -0.99\% } & {\small -20.00\% } & {\small -0.48\% } & {\small 45.71\% } \\ \cline{2-12}
 & \multirow{2}{*}{\begin{tabular}[c]{@{}c@{}}x40\\ (1.47M)\end{tabular}} & 1.376 & 0.181 & 1.144 & 0.012 & 1.208 & 0.060 & 1.195 & 0.032 & 1.028 & 0.051 \\
 &  & {\small -1.01\% } & {\small 6.47\% } & {\small -3.05\% } & {\small -20.00\% } & {\small -0.98\% } & {\small -7.69\% } & {\small -1.24\% } & {\small -20.00\% } & {\small -1.15\% } & {\small 45.71\% } \\ \cline{2-12}
 & \multirow{2}{*}{\begin{tabular}[c]{@{}c@{}}x20\\ (1.28M)\end{tabular}} & 1.360 & 0.183 & 1.134 & 0.011 & 1.199 & 0.059 & 1.185 & 0.039 & 1.020 & 0.051 \\
 &  & {\small -2.16\% } & {\small 7.65\% } & {\small -3.90\% } & {\small -26.67\% } & {\small -1.72\% } & {\small -9.23\% } & {\small -2.07\% } & {\small -2.50\% } & {\small -1.92\% } & {\small 45.71\% } \\ \cline{2-12}
 & \multirow{2}{*}{\begin{tabular}[c]{@{}c@{}}Cylind.\\ (0.72M)\end{tabular}} & 1.357 & 0.181 & 1.146 & 0.020 & 1.187 & 0.065 & 1.176 & 0.031 & 1.010 & 0.050 \\
 &  & {\small -2.37\% } & {\small 6.47\% } & {\small -2.88\% } & {\small 33.33\% } & {\small -2.70\% } & {\small0.00\% } & {\small -2.81\% } & {\small -22.50\% } & {\small -2.88\% } & {\small 42.86\% } \\ \hline
\end{tabular}

}
\end{table}

To analyze the results, let us focus first on the first harmonic amplitudes of the horizontal wave force:

\begin{itemize}

\item x80 (total cells 2.58M): The results of both solvers are in good agreement with the experiment. \textit{foamStar} gives slightly smaller predictions of the first harmonic amplitudes. The \textit{foamStar-SWENSE} result has a better agreement with the experiment.

\item x40 (total cells 1.47M): \textit{foamStar-SWENSE} is able to predict the wave force and elevation correctly with an accuracy of 1\%. Whereas, this difference is about 4\% for \textit{foamStar}.

\item x20 (total cells 1.28M): This discretization is known to be too coarse to simulate waves in conventional NS solvers. The coarse mesh causes excessive numerical diffusion and damps the incident waves. For this reason, \textit{foamStar} predicts a force 13.5\% smaller. However, \textit{foamStar-SWENSE}'s relative error is still within 3\%.

\item Cylindrical (total cells 0.72M): This configuration is optimal for the SWENSE method. With only 0.72 million cells, the results of \textit{foamStar-SWENSE} is almost as accurate as the result of \textit{foamStar} with x80 mesh (2.58 million cells).

\end{itemize}

\noindent\textbf{Comparison of the computational cost}

The computational times required by \textit{foamStar-SWENSE} and \textit{foamStar} are tested, the results are listed in Tab.\ \ref{tab:JIP_compareComputationalTime}. The computation is done with 24 2.5GHz processors. The wall-clock time per wave period is compared.

\begin{table}[ht]
  \centering
  \caption{CALM buoy test case: Computational time comparison of \textit{foamStar}  and \textit{foamStar-SWENSE} }
\begin{tabular}{cccc}
\hline
\multirow{2}{*}{Mesh} & \multirow{2}{*}{Number of cells} & \multicolumn{2}{c}{Computational time per wave period {\color{black}(360 timesteps)}} \\ \cline{3-4}
 &  & \textit{foamStar} & \textit{foamStar-SWENSE} \\ \hline
Cylind. & 0.72 M & - & 2417s \\
x20 & 1.28 M & 6030s & 6031s \\
x40 & 1.47 M & 6808s & 6782s \\
x80 & 2.58 M & 10364s & 9939s \\ \hline
\end{tabular}

\label{tab:JIP_compareComputationalTime}

\end{table}

Table \ref{tab:JIP_compareComputationalTime} first suggests that the computational time on the same mesh is similar for the two solvers. This information suggests that the extra terms in the governing equations of SWENSE and the associated calculations do not influence much the total computational time.

Secondly, \textit{foamStar-SWENSE} shows a clear advantage when considering the computational time at the same accuracy. 
\textit{foamStar-SWENSE} with the x20 or the cylindrical mesh obtains the same level of accuracy as \textit{foamStar} using the x80 mesh.  A speed-up between 1.71 (with x20) to 4.28 (with the cylindrical mesh) is achieved compare with \textit{foamStar} using x80 mesh. This speed-up  would be even much larger in multi-directional wave cases. In that scenario, the mesh for \textit{foamStar} could not be defined with a less resolved transversal direction, while \textit{foamStar-SWENSE} can still use the cylindrical mesh. Besides, in an irregular wave case, the mesh of \textit{foamStar} has to be defined by the smallest wave length of the irregular wave spectrum and thus contains even more cells than the present case.

{\color{black}
Please note that the SWENSE method is not designed to reduce computational cost by allowing larger timesteps. The same CFL criterion  should be set for both solvers since they convect variables with the same total velocity field. In this test, the same timestep (1/360 wave period) is used for all the simulations. The maximum CFL number is about 30 (near the refinement zone of the skirt) and the CFL number in the farfield is about $0.2$. These values are coherent with other similar simulations \cite{ransley2019blind}. 
}

\noindent\textbf{Comparison of the flow details}

\begin{figure}[ht]
\begin{center}
\begin{subfigure}{0.35\textwidth}
\includegraphics[width=\textwidth]{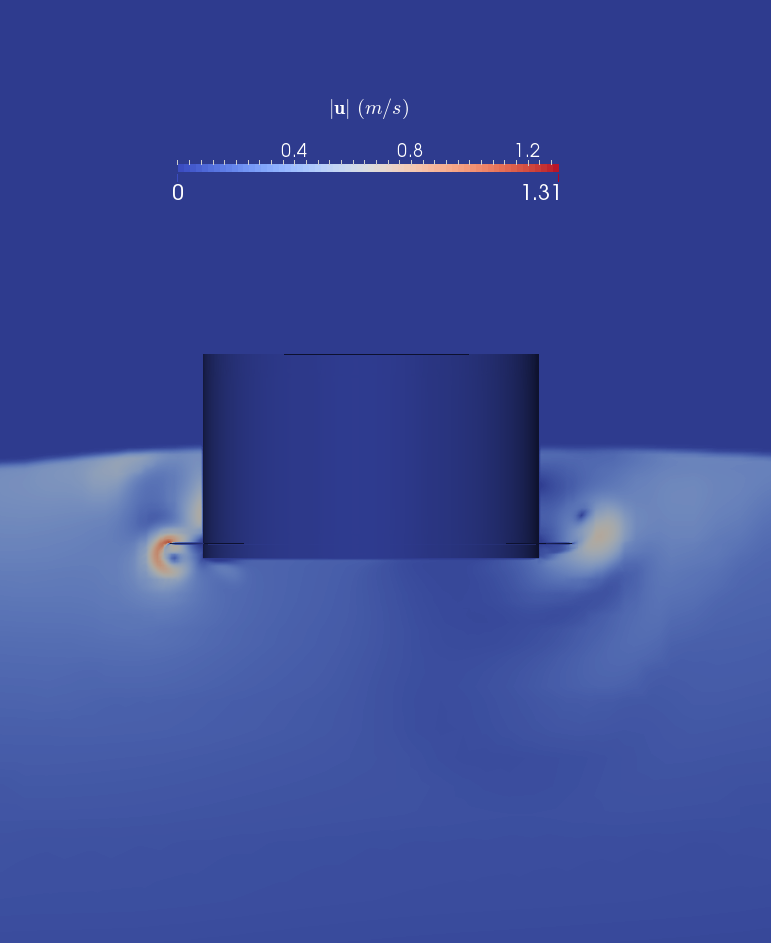}
\caption{foamStar with x80 mesh}
\end{subfigure}
\hspace{10pt}
\begin{subfigure}{0.35\textwidth}
\includegraphics[width=\textwidth]{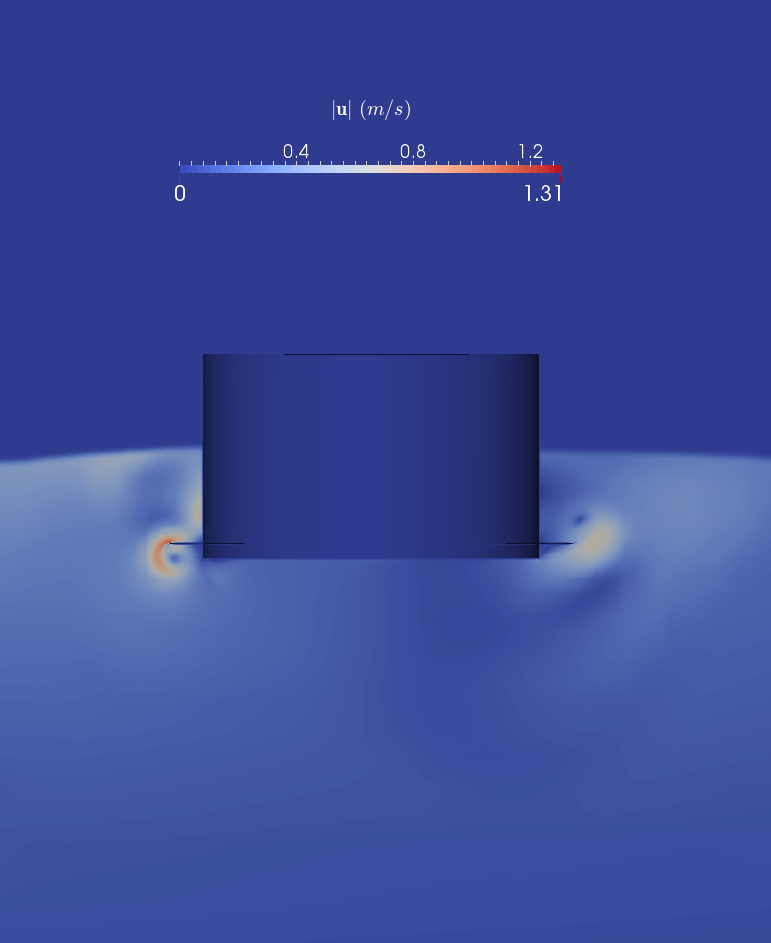}
\caption{foamStar-SWENSE with x20 mesh}
\end{subfigure}
\caption{Comparison of the velocity field in the water when a wave crest passes the buoy}
\label{fig:calmBuoyFlowDetailComparisonVeolcity}

\begin{subfigure}{0.35\textwidth}
\includegraphics[width=\textwidth]{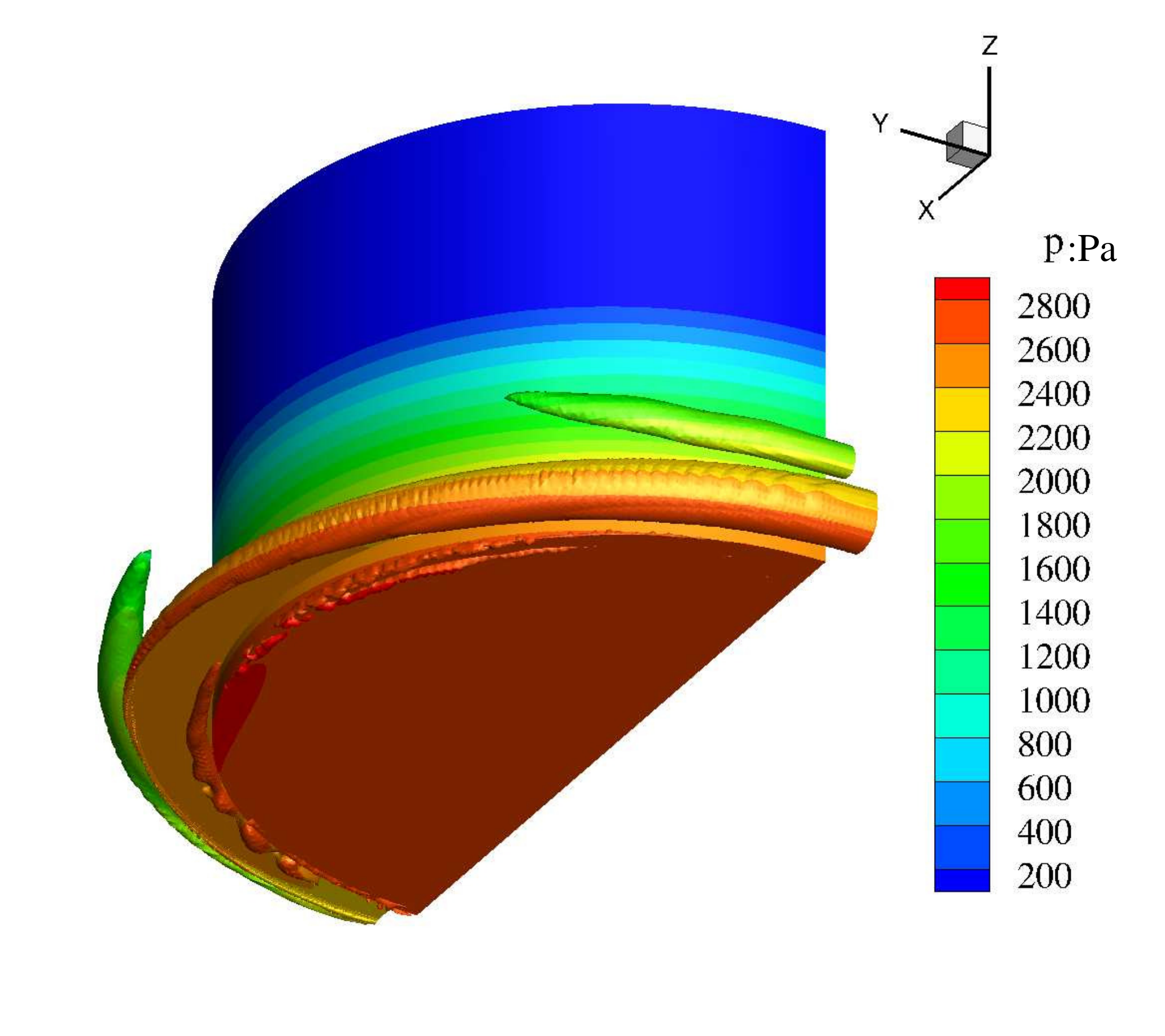}
\caption{foamStar with x80 mesh}
\end{subfigure}
\hspace{10pt}
\begin{subfigure}{0.35\textwidth}
\includegraphics[width=\textwidth]{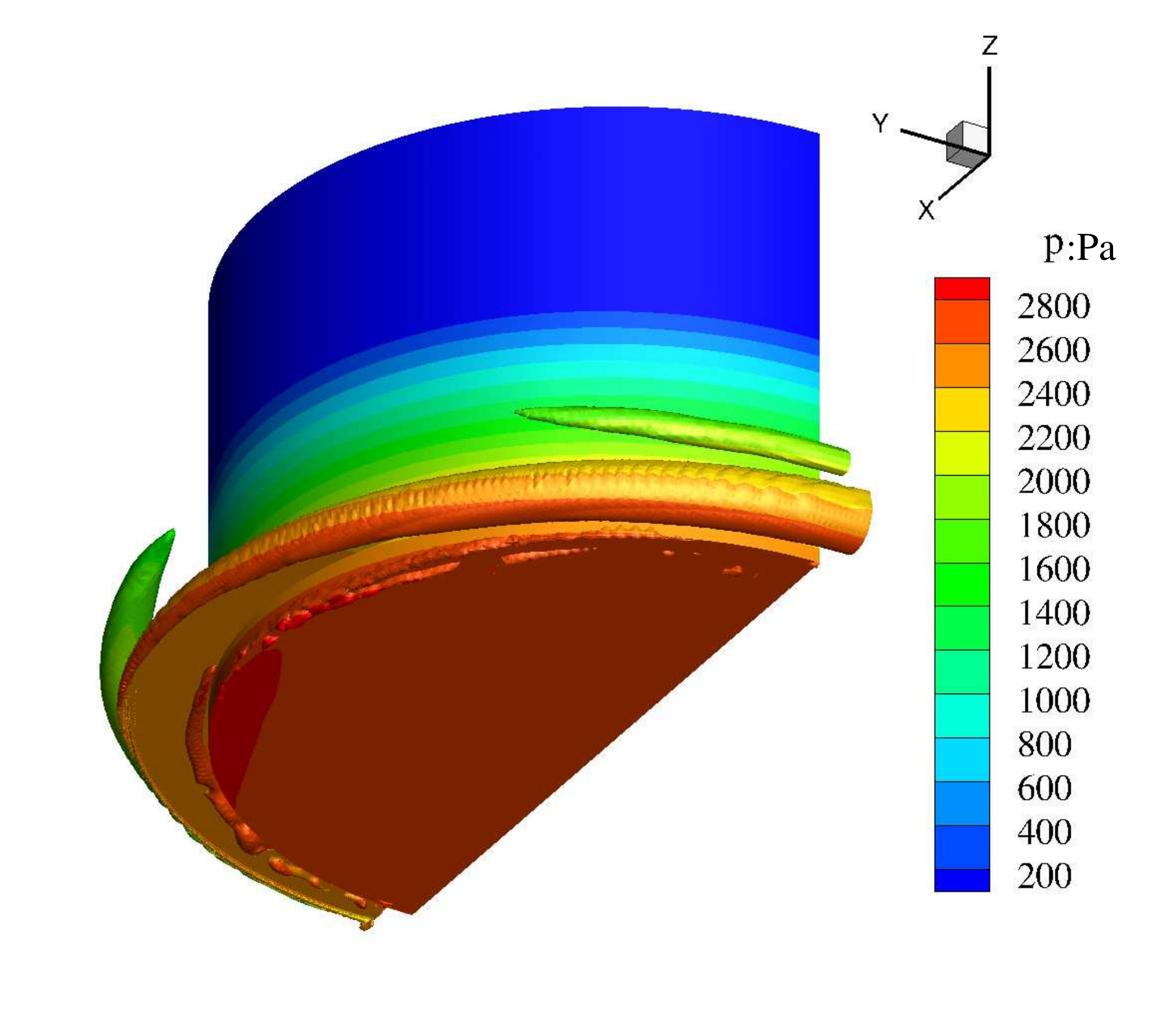}
\caption{foamStar-SWENSE with x20 mesh}
\end{subfigure}
\caption{CALM buoy test case: Comparison of the iso-surfaces of Q = 50 and pressure field when a wave crest passes the buoy \cite{li2019comparison}. Reprinted with permission, copyright Elsevier\textsuperscript{\textregistered}.}
\label{fig:calmBuoyFlowDetailComparisonQ}
\end{center}
\end{figure}

To ensure the correctness of the simulation, especially to validate the result of \textit{foamStar-SWENSE} on the coarse mesh, the flow details of the simulation are compared. Figures \ref{fig:calmBuoyFlowDetailComparisonVeolcity} and  \ref{fig:calmBuoyFlowDetailComparisonQ} plot the velocity field,  Q-criteria and the pressure fields obtained by \textit{foamStar} and \textit{foamStar-SWENSE} with x80 and x20, respectively. Good agreement is observed.

\subsubsection{Irregular waves}

This section aims to validate the SWENSE method in irregular wave cases. 
The force on the buoy is recorded and compared with the experimental data.

\noindent\textbf{Incident waves}

The irregular waves are unidirectional, generated according to a JONSWAP spectrum ($T_p = 2.0$s, $H_s=0.12m$, $\gamma=3$). The motion of the wave-maker is provided to the \textit{HOS-NWT} to calculate the incident wave field in the entire wave tank and then interpolated to the CFD mesh defined near the structure.

\noindent\textbf{Computational Domain}

\begin{figure}[h!]
\begin{center}
  \includegraphics[width=\textwidth]{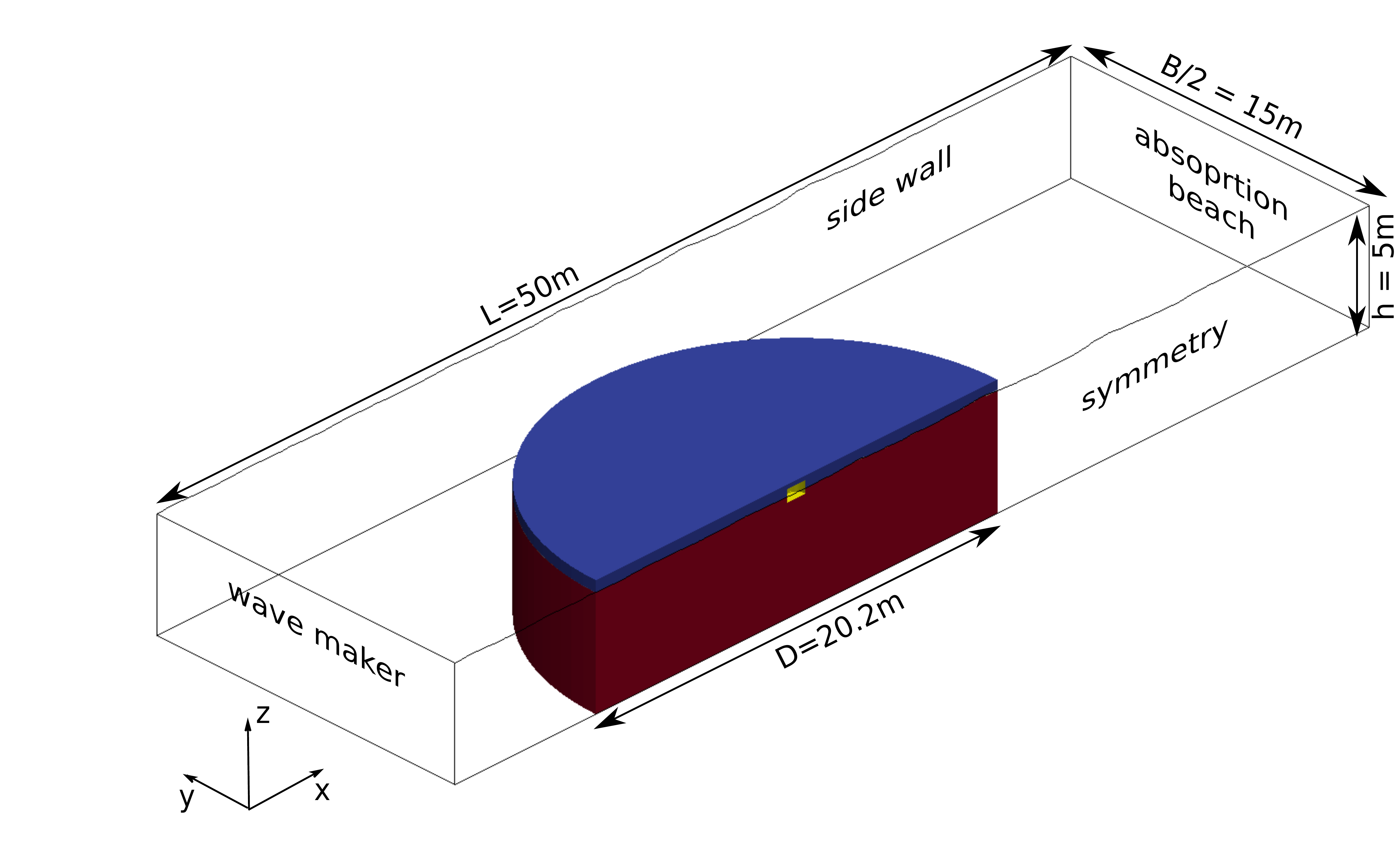}
\end{center}
\caption{CALM buoy test case: Computational domain of CALM buoy in irregular waves.}
\label{fig:computationalDomainIrregularWaves}
\end{figure}

The computational domain of this test case is illustrated by Fig.\ \ref{fig:computationalDomainIrregularWaves}. The cylindrical mesh is the same one in the regular wave case {\color{black} with the same relaxation zone.} 
The simulation thus reproduces only the first 100 s of the experiment.

\noindent\textbf{Results}

\begin{figure}[h]
\begin{center}
\begin{subfigure}{0.49\textwidth}
\includegraphics[width=\textwidth]{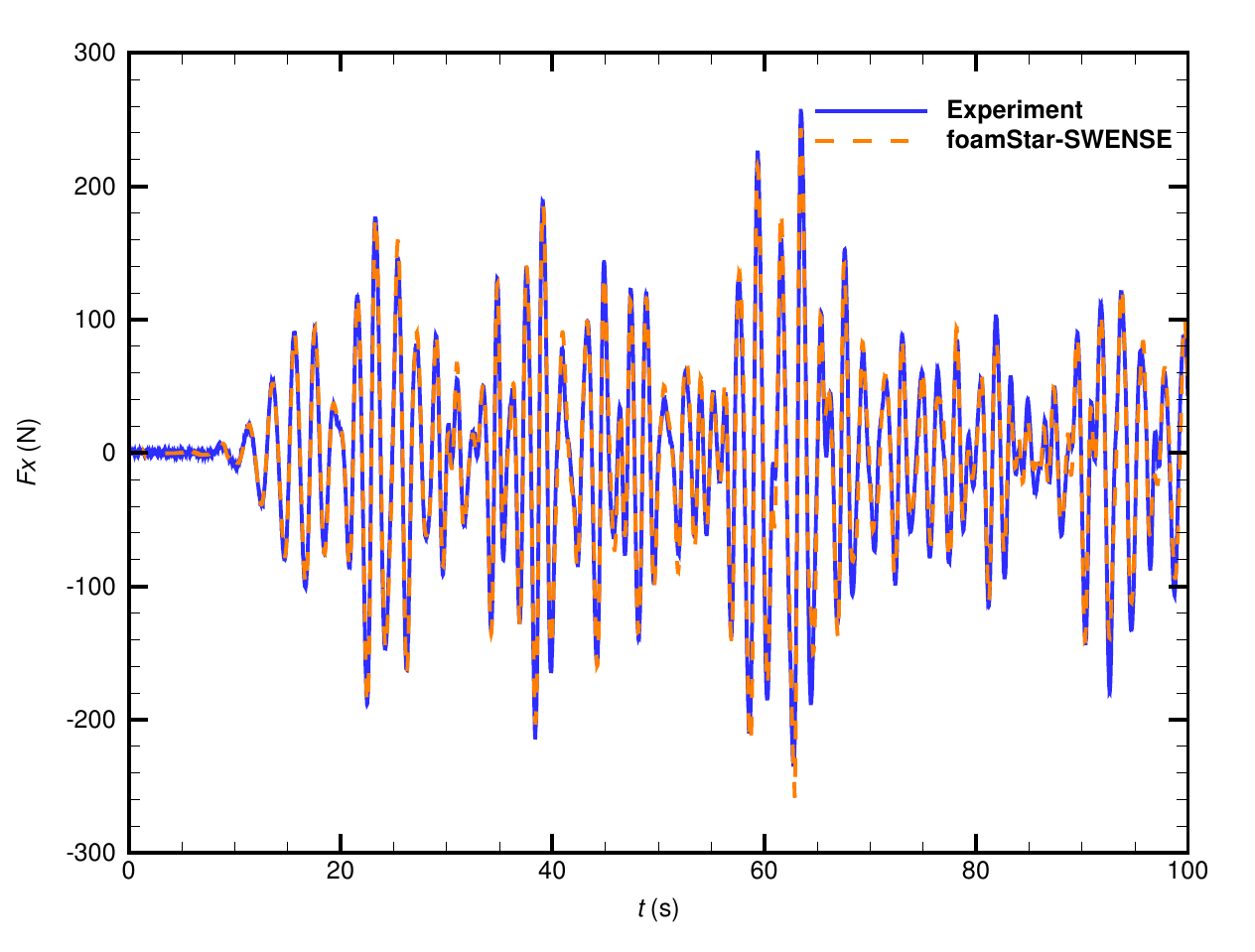}
\caption{Horizontal wave force}
\label{fig:CALMFXIrrg}
\end{subfigure}
\begin{subfigure}{0.49\textwidth}
\includegraphics[width=\textwidth]{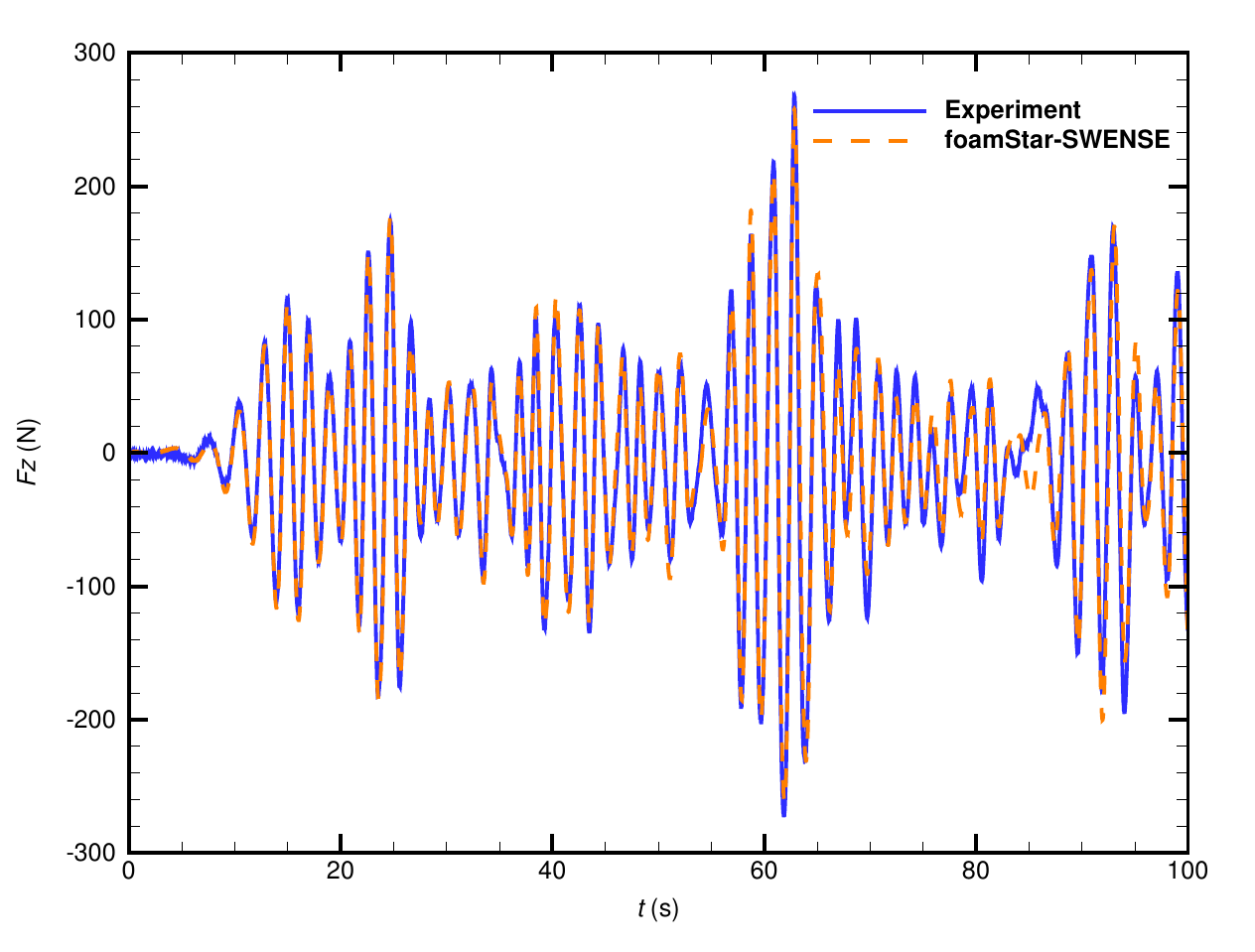}
\caption{Vertical wave force}
\label{fig:CALMFzIrrg}
\end{subfigure}
\caption{CALM buoy test case: wave forces in irregular waves}
\label{fig:CALMForceIrrg}
\end{center}
\end{figure}

The comparison of the simulation results and the experimental data, represented by the horizontal and vertical wave forces time history, is shown in Fig. \ref{fig:CALMForceIrrg}. A very good agreement is observed. At the beginning of the simulation where the wave-front does not reach the structure, the wave forces remain zero. After the wave arrival, the simulated wave force curves are very close to the experimental data during most of the simulation time. Both the amplitudes and the phases are in good agreement.

\begin{figure}[ht]
\begin{center}
\includegraphics[width=0.45\textwidth]{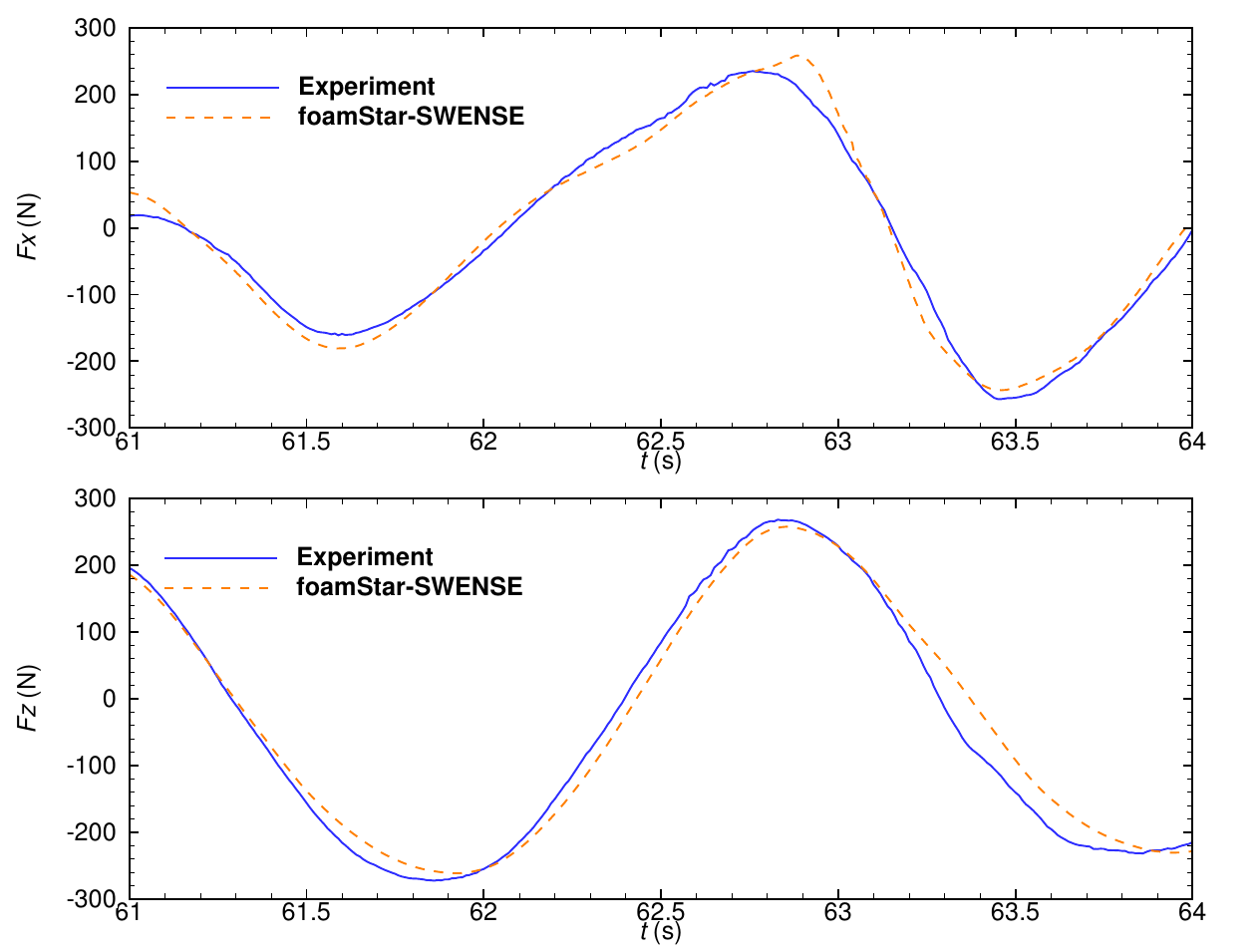}
\caption{CALM buoy test case: wave forces in irregular waves (zoomed)}
\label{fig:CALMFXIrrgZomm}
\end{center}
\end{figure}

Figure \ref{fig:CALMFXIrrgZomm} gives a focused view between $t=61s$ and $t=64s$, where a large wave group reaches the buoy. The time histories of forces, both in the horizontal and the vertical direction, are in good agreement with the experimental data. Note that the simulation results have a peak value slightly larger than the experimental data. It may be due to the overprediction of wave velocity in the crest related to the HOS method \cite{ducrozet2007modelisation}.

\begin{figure}[h]
\begin{center}
\begin{subfigure}{0.24\textwidth}
  \includegraphics[width=\textwidth]{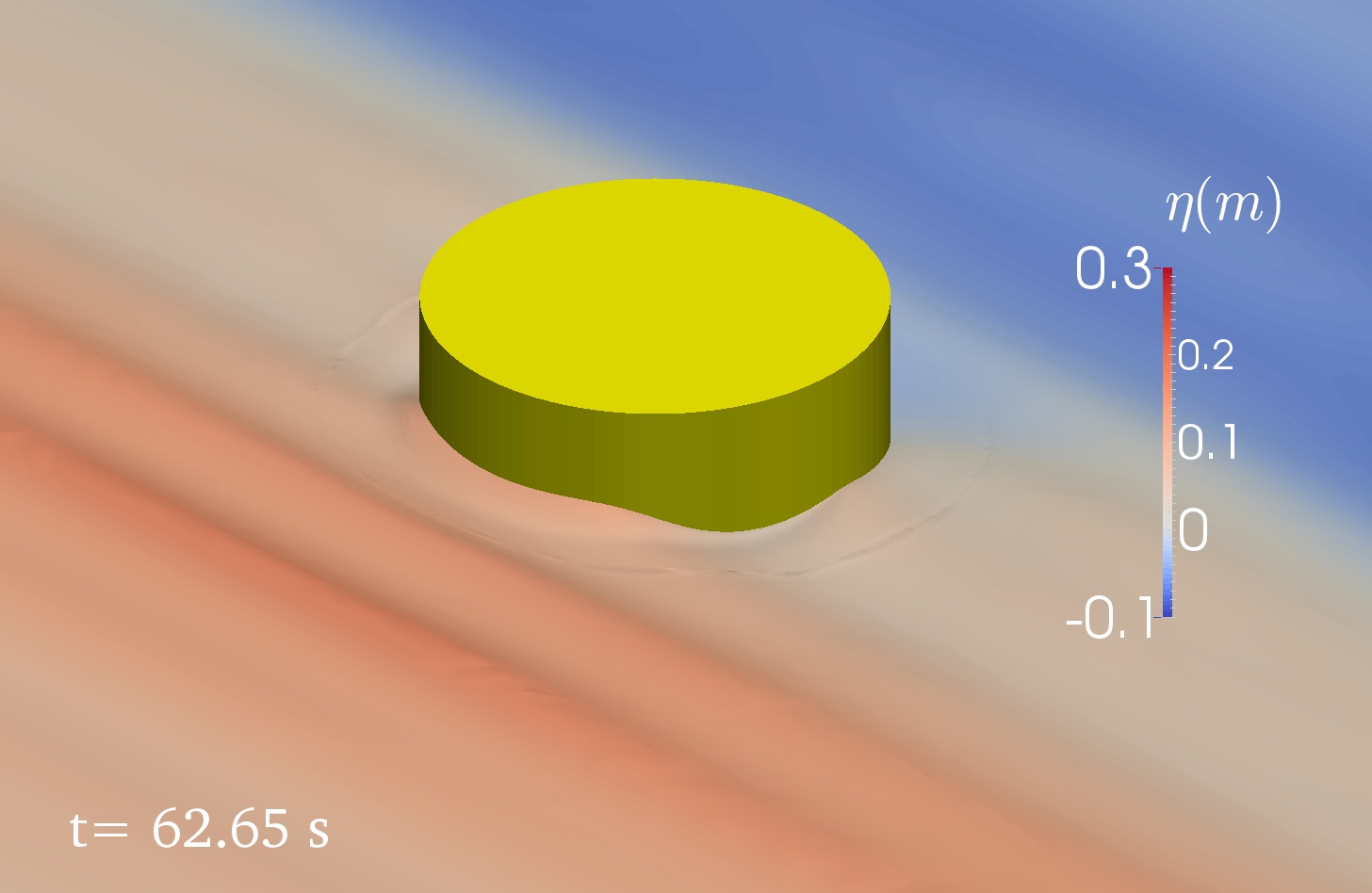}
\end{subfigure}
\begin{subfigure}{0.24\textwidth}
  \includegraphics[width=\textwidth]{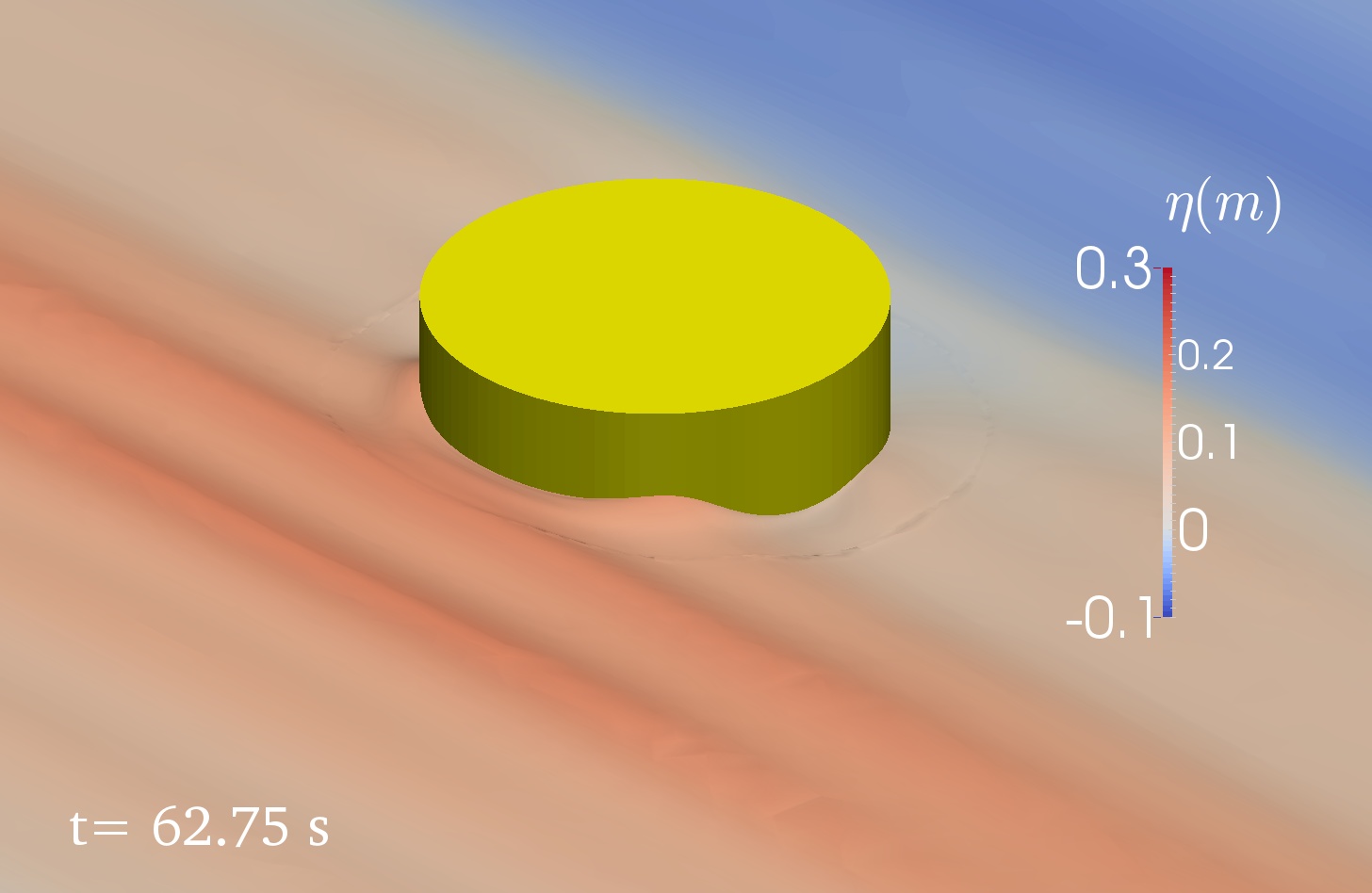}
\end{subfigure}
\begin{subfigure}{0.24\textwidth}
  \includegraphics[width=\textwidth]{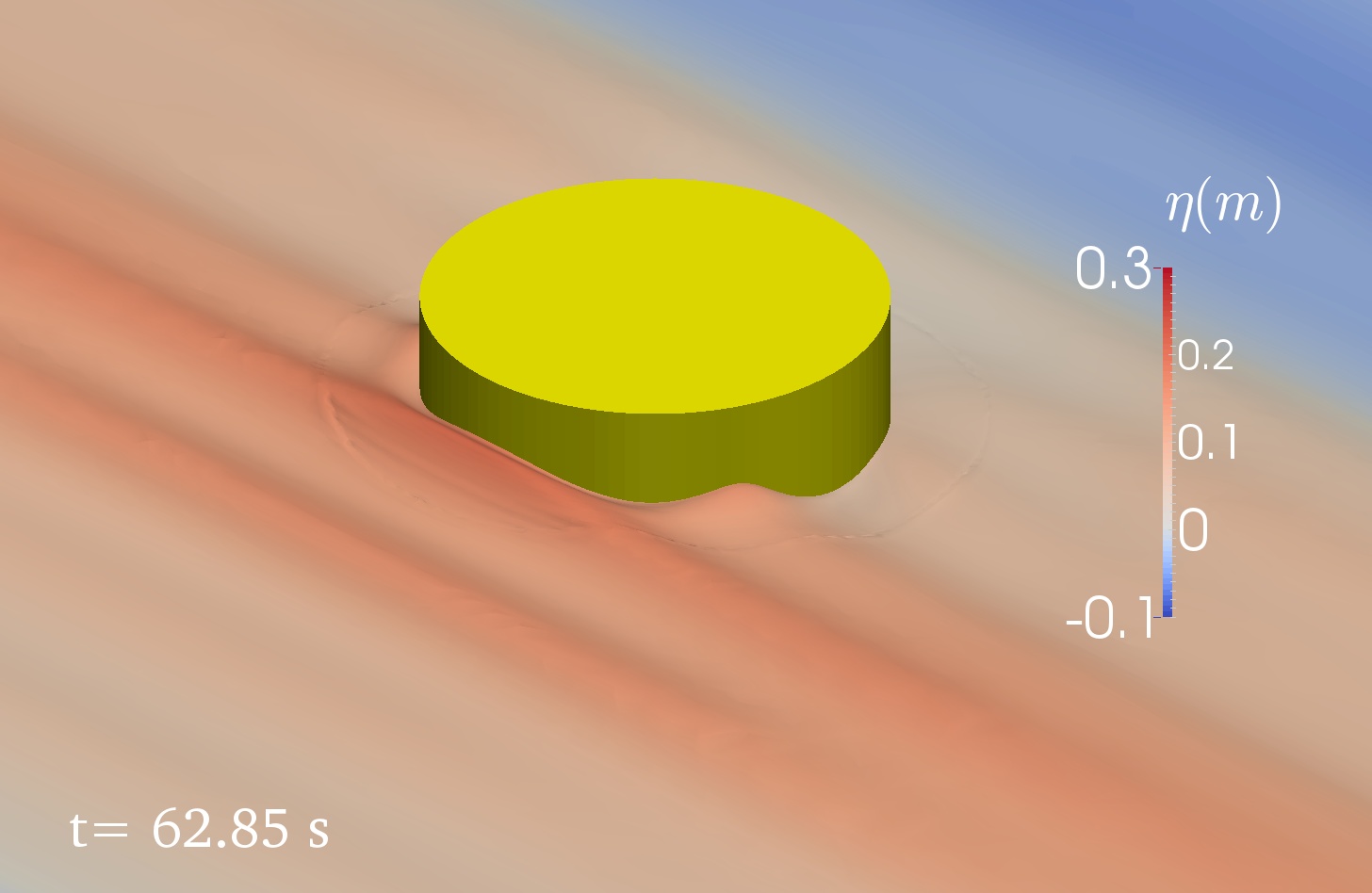}
\end{subfigure}
\begin{subfigure}{0.24\textwidth}
  \includegraphics[width=\textwidth]{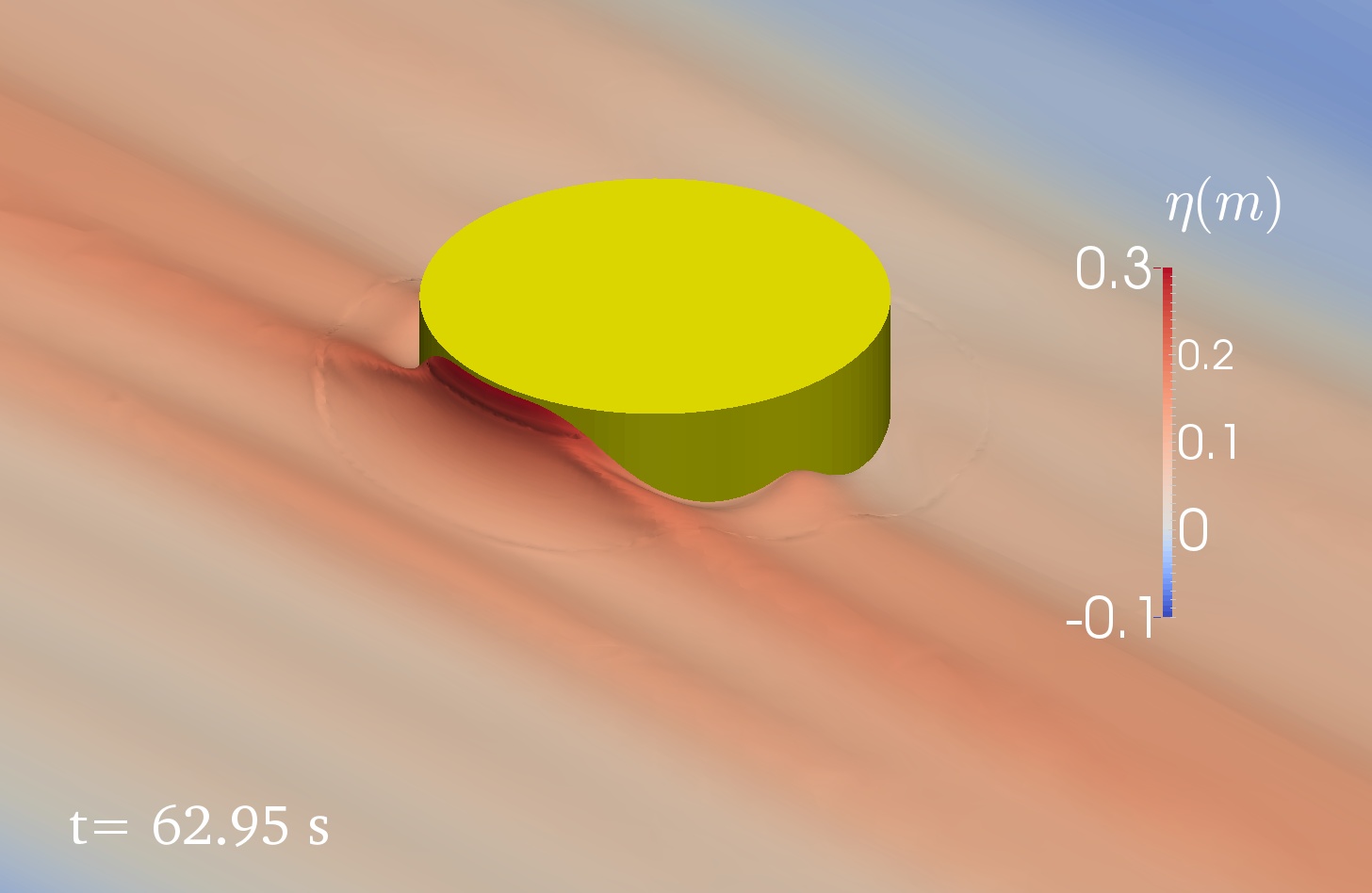}
\end{subfigure}

\begin{subfigure}{0.24\textwidth}
  \includegraphics[width=\textwidth]{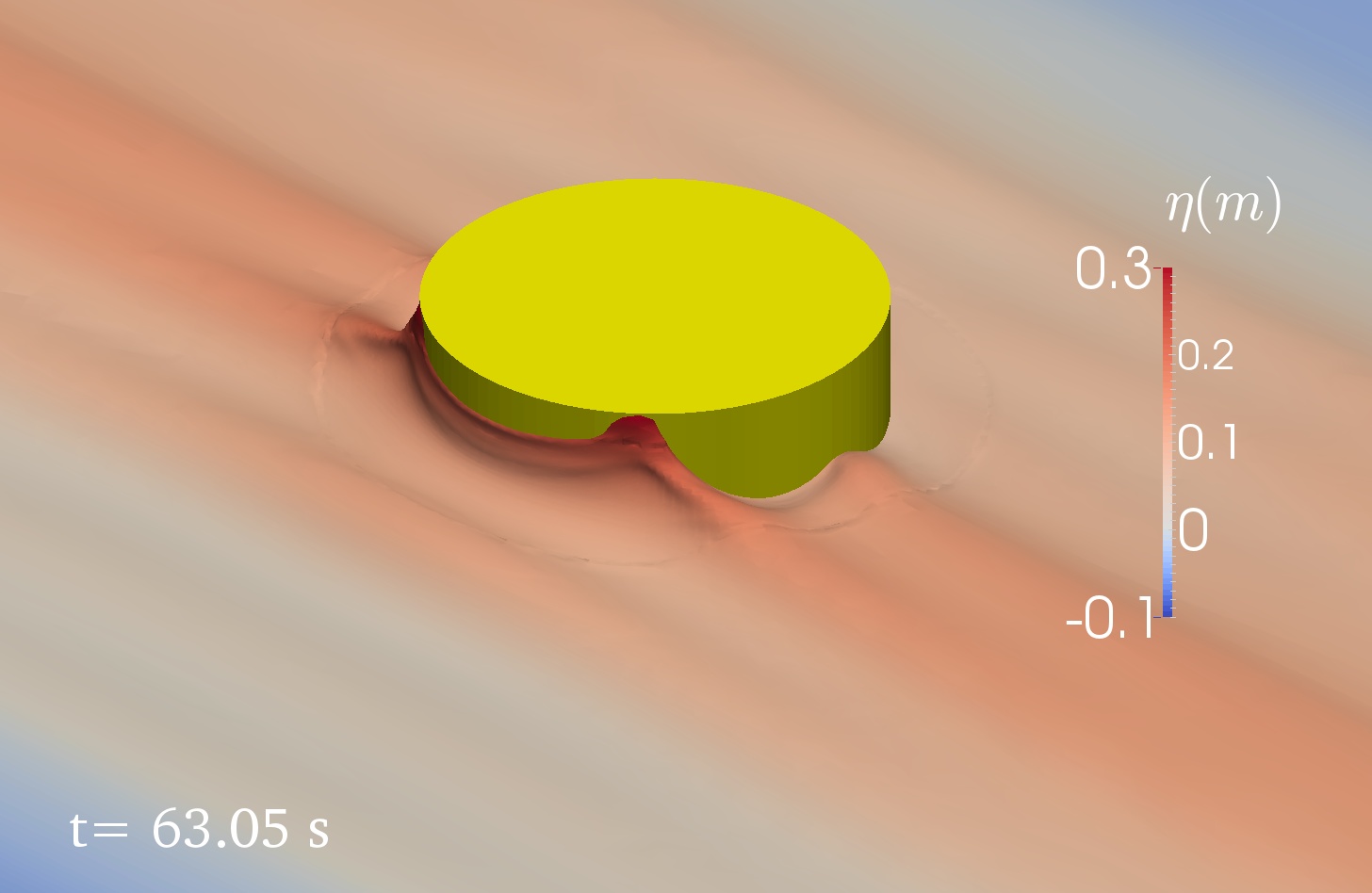}
\end{subfigure}
\begin{subfigure}{0.24\textwidth}
  \includegraphics[width=\textwidth]{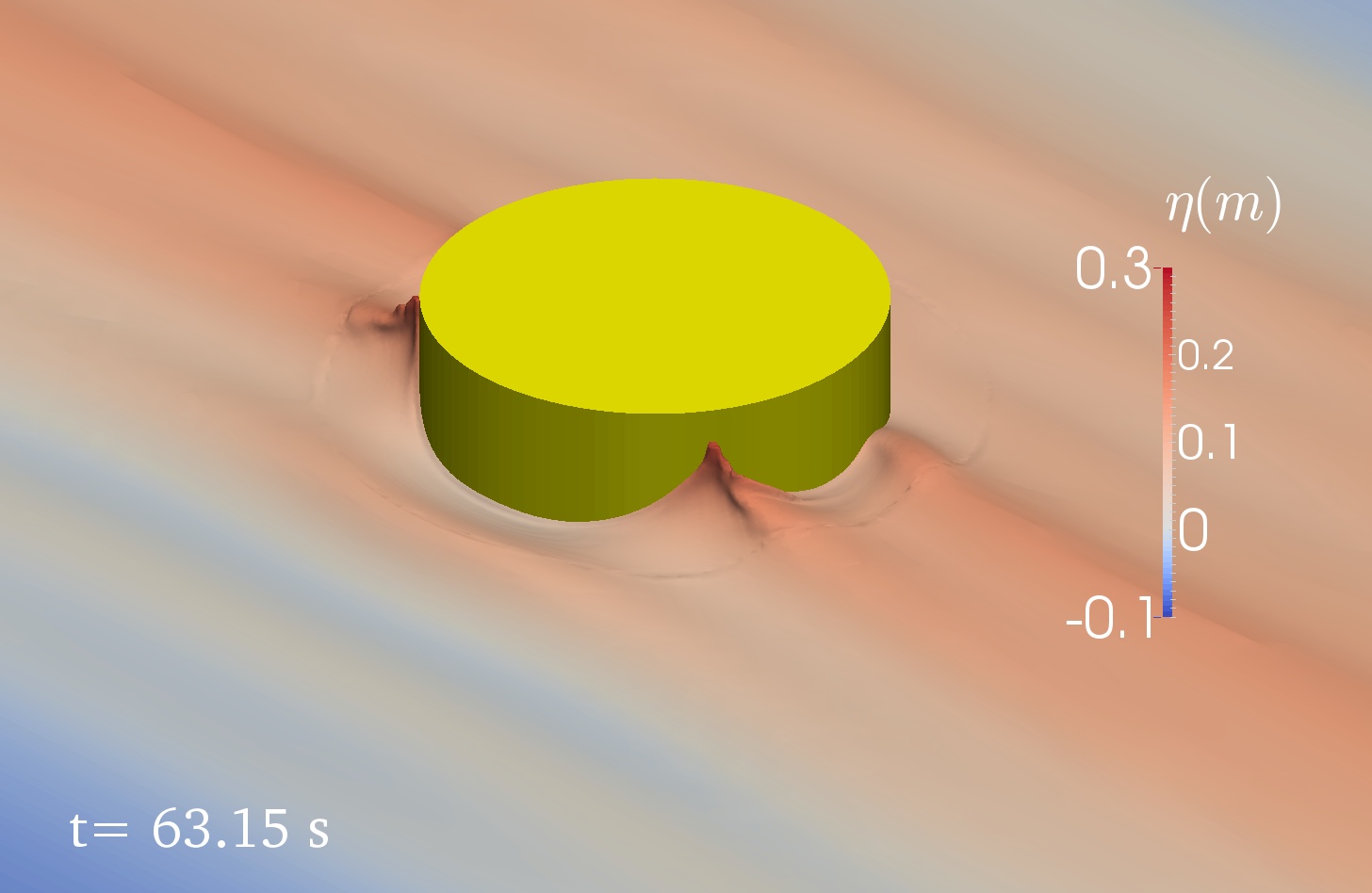}
\end{subfigure}
\begin{subfigure}{0.24\textwidth}
  \includegraphics[width=\textwidth]{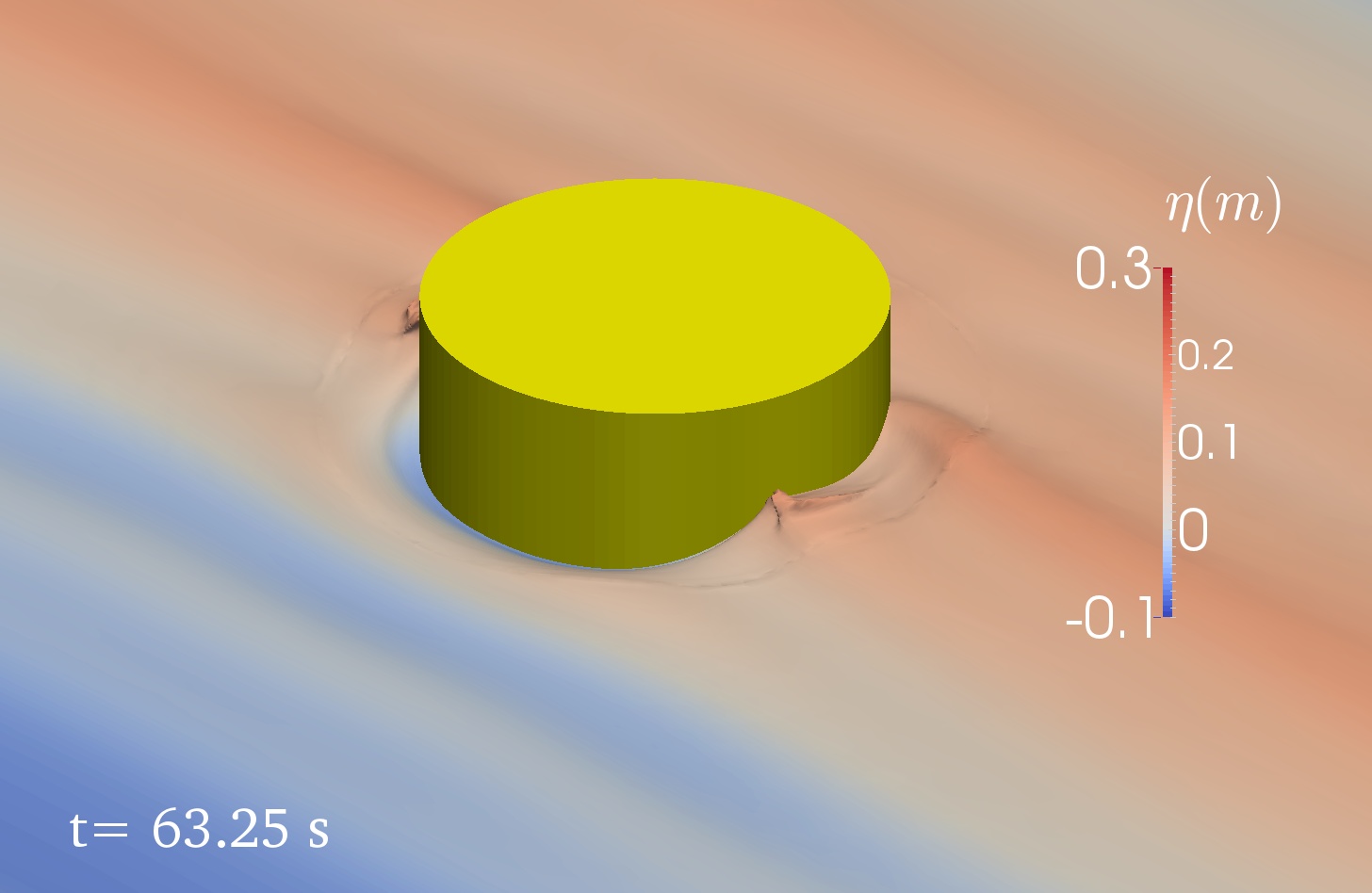}
\end{subfigure}
\begin{subfigure}{0.24\textwidth}
  \includegraphics[width=\textwidth]{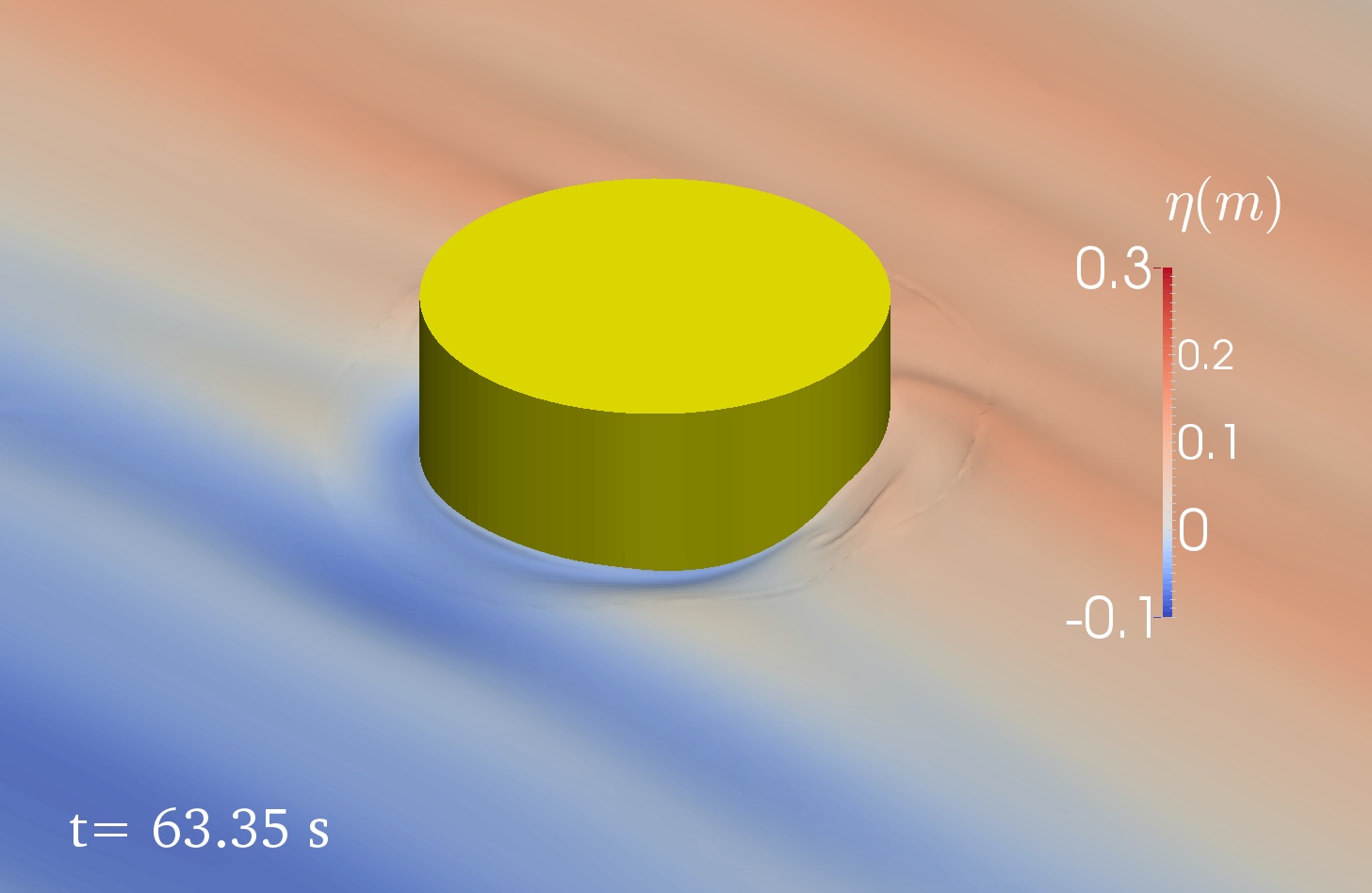}
\end{subfigure}

\caption{CALM buoy test case: The free surface (contour of $\alpha=0.5$) close to the buoy when a large wave group passes}
\label{fig:irregularVOFI}
\end{center}
\end{figure}

Flow details near the structure are provided in Fig.\ \ref{fig:irregularVOFI}. The free surface (approximated by the contour of VOF field $\alpha = 0.5$) is plotted, colored by the free surface elevation $\eta$. The time interval between each figure is $0.1s$. Violent free surface deformations are observed near the structure. At $t=62.95s$, a large runup occurs on the structure, corresponding to the peak in the time history of force. The wave breaking at $t=63.1s$ demonstrates that the present two-phase SWENSE method is able to treat violent free surface deformation.

\section{Conclusion}

In this paper, the two-phase Spectral Wave Explicit Navier-Stokes Equations (SWENSE) method is proposed to improve the efficiency and accuracy of two-phase CFD solvers when they are applied to wave-dominated phenomena. The method decomposes the total problem into an incident wave part and a complementary part. Only the latter part is solved by the CFD solver, while the incident wave solution is provided by nonlinear wave models based on spectral representation.

The governing equations of the complementary part are established from the two-phase Navier-Stokes equations and a modified version of Euler equations for incident waves. The proposed equations are mathematically equivalent to the two-phase Navier-Stokes equations and are able to keep the accuracy of the incident waves with coarse CFD mesh. The solution of the Euler equations is obtained by two spectral PT models, assuming the velocity field is irrotational: the stream function wave theory for regular waves and the HOS method for arbitrary waves in open sea or in an experimental wave tank. The definition zone and the solution of the Euler equations are extended in the air-phase.

An accurate and efficient interpolation method to map the results of HOS wave models onto the CFD mesh is proposed. The method is able to reduce divergence error of the interpolated velocity field to meet the CFD solver's need without reprojection. This interpolation method is made available for the public through an open source project \textit{Grid2Grid} \cite{2018arXiv180100026C}.

An implementation example is shown with a customized solver, \textit{foamStar-SWENSE} in OpenFOAM, based on an existing Navier-Stokes solver \textit{foamStar}. The implementation is straight-forward requiring only the modification of the governing equations.

The method is validated with three validation and application cases: {\color{black} The incident wave propagation cases prove the essential advantage of the SWENSE method, i.e., allowing the use of coarse CFD mesh to simulate incident waves: the SWENSE solver can use a grid 4 times more coarse than that of the NS solver (in each dimension) to achieve the same accuracy for the application of incident waves.} The high-order wave loads on a vertical cylinder calculated with the SWENSE method agree well with the reference experimental and numerical results.  The convergence study shows a second-order convergence behavior of the solver, which is consistent with the numerical schemes used in OpenFOAM. Note that if only the dominant order is considered, the coarsest discretization can provide a good prediction with an error of less than 10\%. This property is useful to give a fast estimation of the wave force, when the absolute accuracy is not very important, \textit{e.g.}, in the early design stage. The CALM buoy case validates the accuracy of the method with complex flow phenomena and reveals its advantage in efficiency. When comparing the computational time to achieve a same level of accuracy, the proposed two-phase SWENSE method achieves a speed-up between 1.71 to 4.28 compared with an NS solver on the regular wave  case. This speed-up would be much larger in an irregular and/or multi-directional situation. 

The present work contains the following limitations:
\begin{itemize}
  \item The incident waves are limited to non-breaking waves propagating at a constant water depth, due to the spectral wave models used here. Such limitations can be overcome in future developments by using advanced spectral wave models allowing wave breaking \cite{seiffert2018simulation,seiffert2017simulation} and variable water-depth \cite{gouin2016development}.
  \item The method is unsuitable for problems with important air effects. In the present work, the extended air velocity is nonphysical. Special treatments should be designed to correct this extended air field, when the air effects cannot be neglected.
  \item {\color{black} The present implementation approximates the interface position in the simplest manner (VOF contour $\alpha = 0.5$). In the future, VOF methods with interface reconstruction (such as PLIC \cite{scardovelli1999direct}) or sharp interface methods (such as Level-Set \cite{sethian2003level}) can be considered to improve the accuracy.}  
  \item The validation cases are limited to the calculation of the wave force on a fixed structure. However, in most marine and offshore applications, a moving structure is of interest. The next step of the work is to extend the two-phase SWENSE method to deal with moving structures.
\end{itemize}

Most of the numerical implementation is based on open source codes: OpenFOAM \cite{openfoamSite}, \textit{HOS-Ocean} \cite{HOSOceanGitHub}, \textit{HOS-NWT} \cite{HOSNWTGitHub}, and \textit{Grid2Grid} \cite{Grid2GridGitHub}. The reader is encouraged to reproduce this work and go beyond the present limitations.

\section*{Acknowledgements}
This work has been performed in the framework of the
Chaire Hydrodynamique et Structure Marines CENTRALE
NANTES - BUREAU VERITAS.
The first author acknowledges China Scholarship Council (CSC) for the financial support for his Ph.D. study. The authors are indebted to Dr.\ \vuko{}, Prof.\ Jasak  from University of Zagreb and Dr.\ Deng from Ecole Centrale Nantes for their help in this work. Three anonymous reviewers are also gratefully acknowledged. 
\begin{appendices}

\section{Problem with the direct subtraction of the Euler Equations\label{section:directSubtractionProblem}}
This section shows the numerical difficulties appearing when the Euler equation is directly subtracted from the two-phase Navier-Stokes Equations.

Recalling the NS momentum equation for two-phase incompressible fluid (Eqn. \ref{eqn:NSnonConservative})
{\color{black}
\begin{equation*}
  \frac{\partial \textbf{u}}{ \partial t} + \textbf{u} . \nabla \textbf{u} = - \frac{\nabla p }{\rho} + \textbf{g} +  \frac{\nabla.\left((\mu + \mu_t)\left(\nabla\bu+\nabla\bu^T\right)\right)}{\rho}
\end{equation*}
}
The perfect fluid Euler momentum equation (Eqn. \ref{eqn:EulerNonconservative}) reads:

\begin{equation*}
\frac{\partial \textbf{u}_{I}}{ \partial t} +\textbf{u}_{I}. \nabla \textbf{u}_{I} = - \frac{\nabla p_{I} }{\rho_I} + \textbf{g}
\end{equation*}

To demonstrate the challenge, we now subtract the two equations directly {and simplify the viscous terms as in Eqn. \eqref{eqn:twophaseSwMoment}. We obtain:}

\begin{equation}
\frac{\partial \textbf{u}_C}{ \partial t} + \textbf{u}_C . \nabla \textbf{u}_C + \textbf{u}_C . \nabla \textbf{u}_{I} + \textbf{u}_{I} .\nabla \textbf{u}_C   = - \frac{\nabla p_C }{\rho} + \underline{\underline{\frac{\nabla p_{I} }{\rho_I} - \frac{\nabla p_{I} }{\rho}}}+ {\color{black}
\frac{\nabla. \left((\mu+\mu_t)\left(\nabla\bu_C+\nabla\bu_C^T\right)\right)}{\rho}}
\end{equation}

The \underline{\underline{underlined terms}} are canceled out in the water since $\rho = \rho_I$. However, these terms have non-zero values in the air phase. They behave as source terms and affect the numerical stability.
{\color{black}
\section{Comparison of Level-Set and Volume of Fluid method\label{sect:compareLSVOF}}

The decomposed Level-Set (DLS) method \cite{vukvcevic2016decomposition} and the standard Volume of Fluid method (VOF) {\color{black}are} compared here with a pure convection case. The transporting velocity is the incident wave velocity obtained from PT (same as in Sect. \ref{sect:progressiveWaves}). The computational domain, boundary conditions are also the same as in Sect. \ref{sect:progressiveWaves}. Five meshes are used:  Three are the same as in Sect. \ref{sect:progressiveWaves}: fine ($\Delta x ,\Delta z = \lambda/200, H/40$), medium ($\Delta x ,\Delta z = \lambda/100, H/20$), coarse ($\Delta x ,\Delta z = \lambda/50, H/10$), two coarser meshes are in addition: ($\Delta x ,\Delta z = \lambda/25, H/5$) and ($\Delta x ,\Delta z = \lambda/15, H/3$).

The measurement and postprocessing procedures are the same as in Sect. \ref{sect:progressiveWaves}. Figure \ref{fig:compareLSandVOF} shows the first harmonic amplitudes of the free surface elevation (nondimensionalised by the target value). The DLS method shows no advantages against the VOF method. The VOF method creates only 1\% error with the coarsest mesh, ($\Delta x ,\Delta z = \lambda/15, H/3$), suggesting the VOF method can work together with the SWENSE method to reduce the mesh.

\begin{figure}[h!]
\centering
\begin{subfigure}[b]{0.49\textwidth}
    \includegraphics[width=\textwidth]{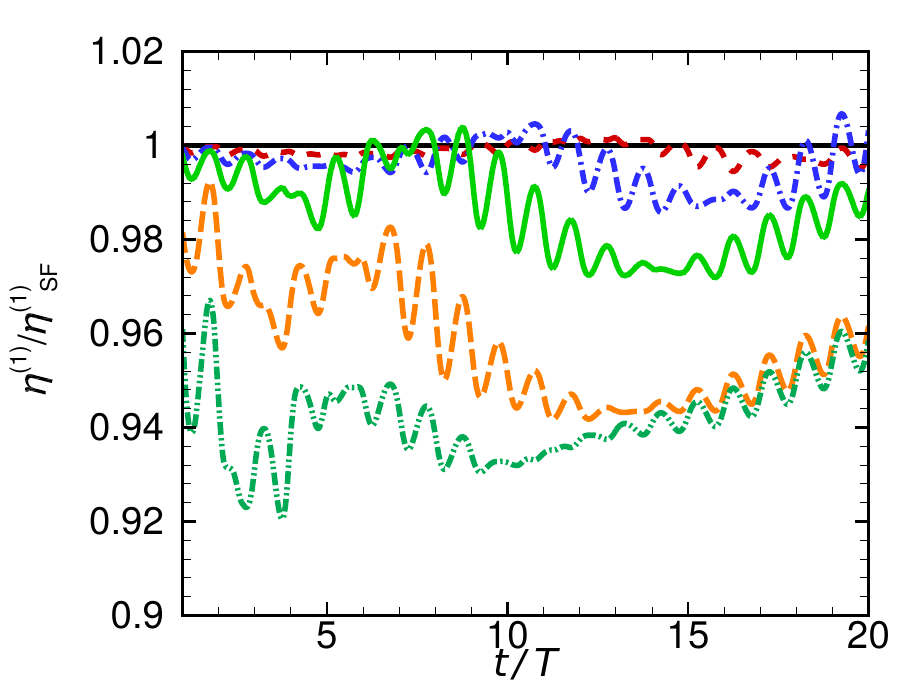}
  \caption{DLS \cite{vukvcevic2016decomposition} \label{fig:levelSetConvection}}
\end{subfigure}
\begin{subfigure}[b]{0.49\textwidth}
    \includegraphics[width=\textwidth]{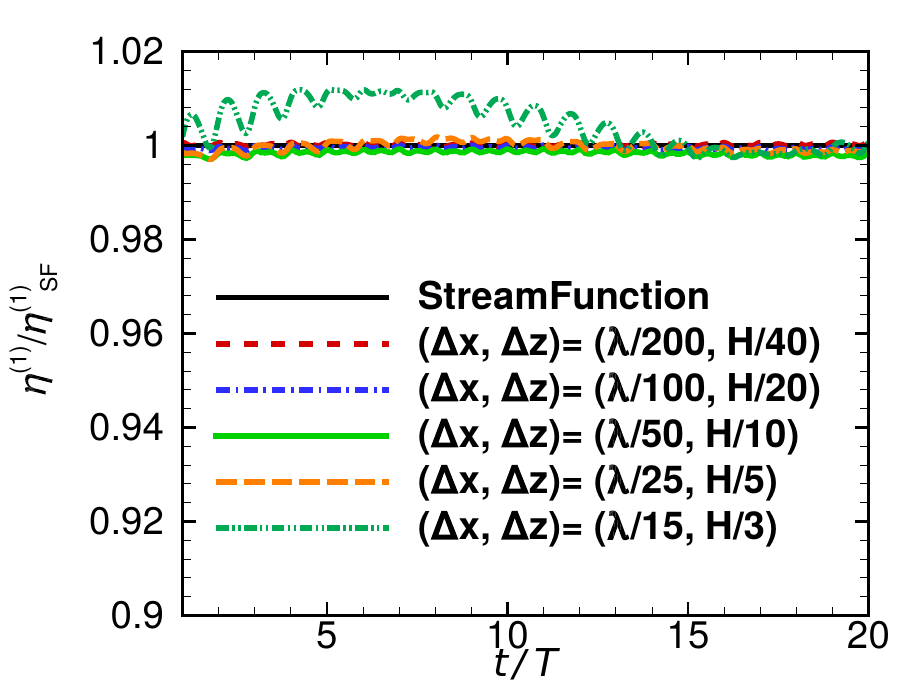}
    \caption{VOF \label{fig:VOFConvection}}

\end{subfigure}
  \caption{First hamornic amplitude of free surface elevation obtained with the passive transport of the decomposed Level-Set (DLS) and the Volume-of-Fluid (VOF) field by the incident wave velocity field. The amplitude is normalised with the target value. \label{fig:compareLSandVOF}}
\end{figure}
}

%

\end{appendices}

\bibliographystyle{abbrvnat}
\bibliography{biblio/biblio}

\end{document}